\documentstyle[epsfig,natbib2,natbibmnfix]{mn}
\newcommand{\be}{\begin{equation}}
\newcommand{\ee}{\end{equation}}
\newcommand{\bea}{\begin{eqnarray}}
\newcommand{\eea}{\end{eqnarray}}
\newcommand{\bc}{\begin{center}}
\newcommand{\ec}{\end{center}}

\renewcommand{\vec}[1]{ {\bmath #1} }

\def\gsim{ \lower .75ex \hbox{$\sim$} \llap{\raise .27ex \hbox{$>$}} }
\def\lsim{ \lower .75ex \hbox{$\sim$} \llap{\raise .27ex \hbox{$<$}} }

\renewcommand{\thefootnote}{\fnsymbol{footnote}}

\title{Growing the first bright quasars in cosmological simulations of structure formation}

\author[Sijacki et al.]
       {\parbox{18cm}{Debora~Sijacki$^{1}$\footnotemark[1],
       Volker~Springel$^{2}$ and Martin~G. Haehnelt$^{1}$}\vspace{0.3cm}\\ 
       $^1$ Kavli Institute for Cosmology, Cambridge and Institute of Astronomy, Madingley
       Road, Cambridge, CB3 0HA \\
       $^2$ Max-Planck-Institut f\"{u}r Astrophysik,
       Karl-Schwarzschild-Stra\ss{}e 1, 85740 Garching bei
       M\"{u}nchen, Germany}
       
\begin{document}

\maketitle
\begin{abstract} 
We employ cosmological hydrodynamical simulations to study the growth of
massive black holes (BHs) at high redshifts subject to BH merger recoils from
gravitational wave emission. As a promising host system of a powerful
high-redshift quasar, we select the most massive dark matter halo at $z=6$
from the Millennium simulation, and resimulate its formation at much higher
resolution including gas physics and a model for BH seeding, growth and
feedback. Assuming that the initial BH seeds are relatively massive, of the
order of $10^5\, {\rm M}_\odot$, and that seeding occurs around $z \sim 15$ in
dark matter haloes of mass $\sim 10^9 - 10^{10}\, {\rm M}_\odot$, we find that
it is possible to build up supermassive BHs (SMBHs) by $z=6$ that assemble
most of their mass during extended Eddington-limited accretion periods. The
properties of the simulated SMBHs are consistent with observations of $z=6$
quasars in terms of the estimated BH masses and bolometric luminosities, the
amount of star formation occurring within the host halo, and the presence of
highly enriched gas in the innermost regions of the host galaxy. After a peak
in the BH accretion rate at $z=6$, the most massive BH has become sufficiently
massive for the growth to enter into a much slower phase of feedback-regulated
accretion. We extend our basic BH model by incorporating prescriptions for the
BH recoils caused by gravitational wave emission during BH merger events,
taking into account the newest numerical relativity simulations of merging BH
binaries. In order to explore the full range of expected recoils and radiative
efficiencies we also consider models with spinning BHs. In the most
`pessimistic' case where BH spins are initially high, we find that the growth
of the SMBHs can be potentially hampered if they grow mostly in isolation and
experience only a small number of mergers. On the other hand, whereas BH kicks
can expel a substantial fraction of low mass BHs, they do not significantly
affect the build up of the SMBHs. On the contrary, a large number of BH
mergers has beneficial consequences for the growth of the SMBHs by
considerably reducing their spin. We also track the fate of our $z=6$ SMBH by
performing cosmological simulations all the way to $z=2$. This allows us to
study the history of BH mass assembly over a large time-span and to establish
a clear signal of `downsizing' of the BH accretion rate for the population of
BHs as a whole. We further find that the descendents of the most luminous
$z=6$ quasar correspond most likely to the most massive BHs today,
characterized by a low activity level and masses of the order of $1-2 \times
10^{10}\, {\rm M}_\odot$.
\end{abstract}


\section{Introduction}

\renewcommand{\thefootnote}{\fnsymbol{footnote}} \footnotetext[1]{E-mail:
deboras@ast.cam.ac.uk}

Black holes (BHs) are among the most remarkable objects in the Universe. They
are now widely believed to power quasars \citep{Schmidt1963, Salpeter1964,
LyndenBell1969}, which belong to the most luminous sources we know
of. Furthermore, BHs are associated with the relativistic acceleration of
particles, jet formation, and the presence of radio lobes \citep{Begelman1984,
Rees1984}, which are often accompanied by copious outflows and shocks. Given
that solid observational evidence \citep{Dressler1989, Kormendy1995,
Magorrian1998} indicates that BHs are ubiquitous inhabitants of a vast
majority of galaxies and that their properties correlate with those of their
host systems, it is inevitable to conclude that BHs should in some way
influence their host galaxies, and that in turn BHs could depend on the
properties of the galaxies they live in.

A number of well defined relationships between the BH mass and the central
stellar properties of the host galaxy, like its bulge velocity dispersion,
mass, and luminosity have been recently established
\citep[e.g.][]{Ferrarese2000, Gebhardt2000, Tremaine2002, Magorrian1998,
Marconi2003,Haring2004}. These observational findings have prompted a large
body of theoretical research \citep{Ciotti1997, Silk1998, Haehnelt1998,
Fabian1999, Kauffmann2000, King2003, Wyithe2003, DiMatteo2005, Hopkins2006,
Croton2006, Bower2006, Okamoto2008}, which focused on understanding how BHs
and galaxies coevolve. It appears clear that any successful model of cosmic
structure formation needs to incorporate BHs and their feedback effects, but
it is still far less clear which mechanisms are ultimately responsible for BH
feeding and feedback. The difficulty lies in the inherent complexity of the
problem, where a vast range of scales and an array of non-linear physical
processes need to be considered.

It is beyond the current capabilities of any numerical code to
self-consistently follow the relevant gravitational, magneto-hydrodynamical,
and star formation processes over the full dynamic range necessary for an
ab-initio treatment, from the size of the BH's sphere of influence all the way
up to cosmologically representative scales. Nonetheless, in the last few years
rapid progress has been made on three different fronts regarding direct
numerical simulations of BHs. At the smallest scales, full general relativity
simulations of BH binary mergers have become recently possible
\citep[e.g.][]{Pretorius2007, Campanelli2007, Herrmann2007, Koppitz2007,
Gonzalez2007, Baker2008}. This has greatly improved our understanding of the
interaction of BHs in these extreme situations, and of the consequences these
interactions have on much larger scales. The second breakthrough has been
accomplished in full general relativity magnetohydrodynamic simulations of BH
accretion disks \citep{Koide1999, Gammie2003, DeVilliers2003, McKinney2004,
Hawley2006, Beckwith2008}. In these studies it has been convincingly
demonstrated that BH accretion processes are indeed able of producing
relativistic jets and outflows, which also should have an impact on much
larger scales. Finally, on cosmological scales it has become possible, for the
first time, to simulate the evolution of BHs embedded in a range of host
haloes and to follow their growth and feedback over cosmic time simultaneously
with the build-up of their host galaxies and the surrounding cosmic
large-scale structure \citep{Springel2005b, Sijacki2007,
DiMatteo2008}. However, a crucial aspect that is currently still missing is a
connection between these different simulation techniques, which would give us
a much more complete and compelling picture.

One approach to overcome the present numerical limitations is to construct
sub-resolution models which encapsulate physical processes occurring on the
scales below the achievable spatial resolution of the simulation, and then to
test these sub-resolution prescriptions in full cosmological simulations
against the available observational findings. This is the approach that we
have taken already in \cite{Sijacki2007, Sijacki2008}, where we have validated
our BH model in cosmological simulations of BH-heated galaxies and galaxy
clusters, and in simulations of cosmic ray heating in clusters due to central
BH activity.

However, one of the important issues that we have not addressed in full yet is
the formation of BHs at very high redshifts, in particular in the light of the
existence of supermassive BHs (SMBHs) as early as $z\sim 6$, as observations
of the most distant quasars in the Sloan Digital Sky Survey (SDSS) seem to
indicate \citep{Fan2001, Fan2003, Fan2004, Fan2006}. This is exactly the
problem we want to address here. The formation of SMBHs with masses of the
order of $10^9 {\rm M}_\odot$ in less than a Gyr of cosmic time proves to be a
significant challenge for theoretical models despite their rareness and thus
their very low space density \citep{Fan2006rev}. The estimated bolometric
luminosities, accretion rates and masses of more than  a dozen known SDSS
quasars at $z \sim 6$ indicate that these objects have extreme properties,
accreting at a rate very close to the Eddington rate with bolometric
luminosities between $10^{47}$ and $10^{48}\,{\rm erg\,s^{-1}}$
\citep{Willott2003, Barth2003, Jiang2006, Kurk2007, Fan2006rev}.
   
Thus, the SDSS quasars represent important benchmarks for theoretical models of BH
formation and growth. Several semi-analytical and hybrid numerical approaches
have been employed to study the formation of BHs at high redshifts
\citep{Wyithe2003, Bromm2003, Volonteri2003, Volonteri2005, Volonteri2006,
Pelupessy2007, Li2007}, exploring various pathways that could lead to the
formation of sufficiently massive BHs at $z=6$. \citet{Li2007}, in particular,
used hydrodynamical simulations of multiple mergers of gas-rich disk galaxies
to study the formation of SMBHs at high redshift. Here we will take a
different approach and simulate the formation of SMBHs at $z=6$ for the first
time in fully self-consistent cosmological simulations of structure
formation. We select the most massive halo from the Millennium  simulation at
this epoch as a promising possible host of a bright high-$z$ quasar, and we
resimulate its formation with much higher mass and spatial resolution,
adopting the BH simulation model of \cite{Springel2005b} and
\cite{Sijacki2007}. In previous work we have shown that this BH model can
successfully reproduce the observed low-redshift BH mass density, the
relationships between BHs and their host galaxies, and the absence of an
overcooling problem in clusters. Here we want to investigate the important
question whether the same model is also capable of producing SMBHs at very
high redshift.

Our second aim is to extend our default BH model by exploring different
physical mechanisms which could potentially hamper the formation of massive
BHs at early times. With the advent of numerical relativity simulations of
binary BH mergers, it is now possible to explore the significant consequences
of gravitational recoils for the build up of the whole population of
BHs. Depending on the mass ratio of the merging BHs, and on the magnitudes and
orientations of their spins, the remnant BH can even be expelled entirely from
its host halo. Such ejections could happen more frequently at high redshifts,
where the potential wells of the BH host systems are expected to be
shallower. In addition, a merger of two non-spinning BHs produces in general a
spinning BH. It appears likely that theoretical models which  ignore
BH spins may miss an important aspect of the problem. We therefore take
advantage of the newest numerical relativity results and incorporate
prescriptions for BH recoils caused by asymmetric gravitational wave emission
during a BH merger, where BHs are characterized by a non-vanishing spin
value. We also allow for spin-dependent radiative efficiencies for gas
accretion. First attempts in this direction have recently been made by
\cite{Berti2008} and \cite{Tanaka2008} using semi-analytical techniques. In
this study, we employ fully cosmological hydrodynamic simulations, which
provide detailed information about the merging histories of the host haloes
and their central BHs, about the thermodynamic properties of the gas in the
vicinity of BHs, and thus about the BH accretion and feedback processes.

Finally, we also examine the fate of the most massive high redshift BHs at
lower redshifts. For this purpose we select descendents of our $z=6$ halo at
$z=4$ and $z=2$, and resimulate them once more adopting the same BH model and
numerical resolution. This permits us to track BH activity over a much larger
time span and to relate it to the dynamical history of the host halo.

The paper is organised as follows. In Section~\ref{Methodology}, we discuss
the methodology we have adopted to simulate the high redshift evolution of
BHs, while in Section~\ref{Simulations} we describe the suite of simulations
we have performed. The bulk of our results are presented in
Sections~\ref{Results1} and~\ref{Beyond_z_6}, and in Section~\ref{Discussion},
we discuss our main findings and draw our conclusions.

\section{Methodology} \label{Methodology}

\subsection{The numerical code}

In this study we use the massively parallel Tree-SPH code {\small GADGET-3}
\citep[see][for a description of an earlier version of the code]{Gadget2} in
its entropy conserving formulation \citep{SH2002}. In addition to gravity and
non-radiative hydrodynamics of the dark matter and gas components, the code
follows radiative cooling and heating of an optically thin plasma of hydrogen
and helium, subject to a time-dependent and spatially uniform UV
background. We adopt a subresolution multi-phase model for the treatment of
star formation and the associated supernovae feedback, as described in
\cite{SpringelH2003}, and include an optional extension for starburst powered
galactic winds of constant outflow speed. We follow self-consistently the
growth and feedback processes of a population of BHs embedded in the
simulations. In Section~\ref{Default_model}, we briefly summarize the main
features of our default BH model \citep[for a more detailed description
see][]{DiMatteo2005, Springel2005b, Sijacki2007}. We then discuss in more
depth several extentions of the default BH model in order to account for
recoils due to BH mergers induced by asymmetric gravitational wave
emission. We also study models with spinning BHs and explore the evolution of
the spin history in simple terms. This allows us to account for the effect of
spin-dependent radiative efficiencies.

\subsection{Default BH model} \label{Default_model}

BHs are represented by collisionless sink particles in the code that can grow
with time from an initially small mass by accreting surrounding gaseous
material, or by merging with other BHs.  Note that due to inevitable numerical
limitations we cannot track nor resolve the initial formation processes of
BHs. We assume these to be sufficiently efficient to generate a population of
BH {\em seeds} at high redshifts in all haloes above a given mass
threshold. Once such seed BH particles exist, we can then follow their
subsequent growth in mass by gas accretion. The accretion rate is
parametrized in terms of a simple spherically symmetric Bondi-Hoyle type
accretion flow \citep{Hoyle1939, Bondi1944, Bondi1952}, i.e.  \be \dot M_{\rm
BH}\,=\, \frac{4\,\pi \,\alpha\, G^2 M_{\rm BH}^2 \,\rho}{\big(c_s^2 +
v^2\big)^{3/2}}\,, \label{Bondi_eq} \ee where $\alpha$ is a dimensionless
parameter, $\rho$ is the density, $c_s$ the sound speed of the gas, and $v$ is
the velocity of the BH relative to the gas. Here we fix the value of $\alpha$
to $100$, as it has been done in all of our previous work \citep{Sijacki2007,
Sijacki2008, Puchwein2008}. Note that a value of $\alpha$ larger than unity
has been adopted because cosmological simulations typically fail to resolve
the Bondi radius, and thus underpredict the gas density used in
equation~(\ref{Bondi_eq}). Furthermore, at the highest resolved densities we
model the gas with a sub-resolution multiphase model for star formation that
gives a comparatively high mean gas temperature as a result of supernova
feedback. A volume-average of the Bondi-rates for the (spatially unresolved)
cold and hot phases of the ISM recovers a value of $\alpha$ close to
$100$. Our results, however, are not very sensitive to the adopted value for
the $\alpha$ parameter because most of the BH growth occurs during phases of
Eddington limited accretion (where the accretion rate is independent of
$\alpha$), and because BH growth is self-regulated by feedback (see also
Section~\ref{NumConvTests}). We furthermore assume that the accretion rate
cannot exceed the Eddington limit.

BHs are also allowed to grow in mass by merging with other BHs that happen to
be in the immediate vicinity (within the smoothing lengths that is used to
estimate the local gas density) and that have small relative velocities of the
order of or lower than the local gas sound speed. Note that our cosmological
simulations lack the spatial resolution to track the BH binary hardening
process at parsec and sub-parsec scales prior to the merging event \citep[see
e.g.][]{Begelman1980}. Our prescription for BH merging assumes instead that
the hardening is efficient and occurs on a short timescale. This should
justify that two BHs merge instantaneously when they are sufficiently close as
in our model. We note that whether or not a BH binary may get `hung up' for a
significant time if the hardening only has to rely on stellar dynamics is
still not completely settled. However, in the gas rich environments \citep[see
e.g.][]{Escala2004, Cuadra2009} that  we are discussing here, additional
dissipation due to the gas should help. The procedure we have adopted should
thus result in an upper bound for the possible impact of BH recoils on the
growth of early quasars.

In our model, accreting BHs affect their environment by an isotropic coupling
of a small fraction of their bolometric luminosity to the thermal energy of
the surrounding gas particles according to the expression \be \dot E_{\rm
feed} \,=\, \epsilon_{\rm f}\, L_{\rm r}\, = \, \epsilon_{\rm f}\,
\epsilon_{\rm r}\, \dot M_{\rm BH} \, c^2\,.  \ee Here $\epsilon_{\rm r}$ is
the radiative efficiency which in the default model is fixed to be $0.1$,
while for the thermal coupling efficiency, $\epsilon_{\rm f}$, a value of
$5\%$ has been adopted. With this choice of $\epsilon_{\rm f}$, the simulated
$M_{\rm BH}-\sigma_*$ relation obtained for remnants of isolated galaxy
mergers is in a good agreement with observations, as shown by
\cite{DiMatteo2005}. Moreover, it has been demonstrated that in fully
self-consistent cosmological simulations the $M_{\rm BH}-\sigma_*$ relation is
also reproduced with this feedback efficiency \citep{Sijacki2007, DiMatteo2008}.

\subsection{Choice of BH seed mass and seeding procedure} 

Two important issues in studying the build-up of a population of massive BHs
at high redshifts are the assumed BH seed mass and the selection of haloes
which contain such seeds. Both of these issues are intimately linked to poorly
understood BH formation processes in the early Universe. In fact, there are
several possible channels of BH formation \citep{Rees1978} at early times. In
this work, we are mainly interested in scenarios where relatively massive BH
seeds are produced as a consequence of central gas collapse within
proto-galactic haloes \citep[e.g.][]{Haehnelt1993, Umemura1993, Begelman2006,
Lodato2006}. In this picture, the most favourable conditions for the formation
of massive BH seeds with mass of order $10^4-10^6 \,h^{-1} {\rm M}_\odot$
occur in haloes with virial temperatures of $10^4\,{\rm K}$ and above, and
where metal line and $H_2$ cooling is be subdominant with respect to atomic
hydrogen cooling, thus preventing excessive fragmentation and star formation
\citep{Bromm2003}. Recently, there have been several promising numerical
attempts targeted to address this specific BH formation scenario
\citep{Wise2008, Levine2008, Regan2009}, suggesting that this mechanism may
indeed work.

We mimic such a scenario of BH seed formation, by assuming that the BH
particles in the code are initially characterised by a fixed mass of $10^5
\,h^{-1} {\rm M}_\odot$, and that all haloes above a given mass threshold
contain at least one such BH seed. We have, however, run an additional
simulation with a larger seed mass of $10^6 \,h^{-1} {\rm M}_\odot$, in order
to test the sensitivity to this choice (see
Section~\ref{Spinning_quasars}). As in \cite{Sijacki2007}, the actual creation
of BH seeds in the simulations is accomplished on the fly by frequently
invoking a fast and parallel friends-of-friends algorithm for halo finding; if
a halo above the mass threshold is found that does not contain any BH yet, a
seed BH at the halo centre is introduced.  For the halo threshold value we
explore two possibilities, one where the threshold value corresponds to
$10^{9} \,h^{-1} {\rm M}_\odot$, and another, more restrictive case where only
haloes with mass above $10^{10} \,h^{-1} {\rm M}_\odot$ are seeded. Our main
aim here is to understand whether and under which conditions such a population
of BH seeds produced at high redshifts will evolve into BHs as massive as
those discovered at $z=6$. Clearly, it is also important  to understand if
much smaller BH seeds -- possibly the remnants of the first Pop-III stars --
or BH seeds populating different haloes may also give rise to the formation of
luminous SDSS quasars. We defer an investigation of this question, which is
beyond the scope of this study, to forthcoming work.

\subsection{Recoils of merging BHs} \label{Method_recoils}

One of the potential hazards for early BH assembly are the recoils of the
remnant BH due to gravitational wave emission during binary BH mergers. For
certain BH binary configurations, these recoils are characterised by
relatively large kick velocities which can expel a remnant BH from its host
halo, especially at high redshifts where the host haloes typically have
shallow potential wells \citep[e.g.][]{Merritt2004, Haiman2004, Volonteri2006}.

Over the past two to three years, there has been a remarkable breakthrough in
numerical simulations of BH binaries in full general relativity \citep[for a
recent review, see][]{Pretorius2007}. These simulations provide for the first
time reliable estimates of the remnant BH kick velocities for various initial
configurations. The full parameter space of encounters of two BHs has not been
explored yet, and merger simulations involving multiple BHs are still in their
infancy \citep[see][]{Campanelli2008}. However, the results of the present day
numerical relativity BH merger calculations can already be applied to
cosmological simulations of BH growth in different interesting cases. In this
study, we focus on the following three scenarios\footnote{Note that in all
three cases explored here the BH binary orbit has been assumed to have
negligible eccentricity.}:

\subsubsection{Mass asymmetry driven BH recoils}  

In our first case we consider the recoil of a BH merger remnant that is caused
only by the mass difference of two non-spinning progenitor BHs.  The kick
velocity can then be expressed by the standard Fitchett formula
\citep{Fitchett1983} that has now been calibrated by numerical relativity
simulations \citep{Gonzalez2007}, \be v_{\rm m,\,kick} \,=\, A \eta^2 \sqrt{1
- 4 \eta} \, (1 + B \eta)\,.
\label{vmass_eq}   \ee The coefficients $A$ and $B$ are $1.2 \times 10^4$ and
$-0.93$, respectively, and $\eta \,=\, q / (1+q)^2$ is a function of the mass
ratio of the progenitors only, with $q\,=\,m_1/m_2 \le 1$. Here the maximum
kick velocity is relatively low, of order $175\,{\rm km\,s^{-1}}$, and occurs
in the case of a mass ratio of $\sim 0.36$. Note that even though the
progenitor BHs are non-spinning, the remnant BH will have non-vanishing spin
due to the angular momentum carried away by the gravitational wave
emission. The final spin of the remnant BH can be expressed as a function of
the $\eta$ parameter only, i.e. $a_{\rm fin} \,=\, 3.464\, \eta - 2.029\,
\eta^2$ \citep{Berti2007}.

\subsubsection{Recoils in configurations with arbitrary mass ratio 
and aligned/anti-aligned spins}  

In the next case we explore encounters of two BHs with arbitrary mass ratio
and non-vanishing spins\footnote{For the choice of the initial BH spin values
see Section~\ref{BHspin_evo}.}. However, we impose that the two BH spins prior
to the merger are either aligned or anti-aligned with the orbital angular
momentum vector, with equal probability. The absolute values of the BH spins
are allowed to be anywhere  between $0$ and $1$. In this case the recoil BH
remnant velocity can be computed as follows (\citet{Campanelli2007, Baker2008}
see also \citet{Herrmann2007, Koppitz2007}): \be \vec v_{\rm align,\,kick} \,=
\, v_{\rm m,\, kick} \, \hat \vec e_1 + v_{\perp}(\cos\xi \, \hat \vec e_1 +
\sin\xi \, \hat \vec e_2)\,,
\label{valign_eq}  \ee with \be v_{\perp} \,= \, H \frac{\eta^2}{1+q}(a_2 - q
a_1)\,.  \ee Here $ \hat \vec e_1$ and $ \hat \vec e_2$ are orthogonal unit
vectors in the orbital plane, and $a_1$ and $a_2$ are the dimensionless spin
vectors of the two holes. $\xi$ measures the angle between unequal mass and
spin contribution to the kick velocity. For simplicity, we assume $\xi$ to be
$90^\circ$, very similar to the value of $88 ^\circ$ found by
\cite{Campanelli2007}, although this value is intrinsically very difficult to
determine in numerical relativity simulations due to the strong precession of
the spins near the merger. We adopt a value of $\sim 7.3 \times 10^3$ for the
parameter $H$, as in \cite{Campanelli2007}, while $q$ and $\eta$ have the same
meaning as in the previous case. Note that in the case of negligible spins,
equation~(\ref{valign_eq}) reduces to~(\ref{vmass_eq}). In the case of
mutually anti-aligned and maximally spinning BHs, the maximum kick velocity of
order $460\,{\rm km\, s^{-1}}$ occurs for equal mass mergers.

\subsubsection{Recoils for arbitrary mass ratio and random spin orientations}  

In our last and most general scenario, we consider mergers of binary BHs where
the mass ratio, the spin magnitude and the spin  orientations are
arbitrary. We follow \cite{Campanelli2007} \citep[see also][for a somewhat
lower maximum kick velocity estimate]{Baker2008} and express  the recoil
velocity through a generalisation of the previous case, i.e.  \be \vec v_{\rm
rand,\,kick} \,= \, v_{\rm m,\, kick} \, \hat \vec e_1 + v_{\perp}(\cos\xi \,
\hat \vec e_1 + \sin\xi \, \hat \vec e_2) + v_{||} \, \hat \vec e_z\,.
\label{vrand_eq}  \ee Note that for calculating $v_{\perp}$ we need to
consider only the components of the spin vectors along the orbital angular
momentum. Instead, $v_{||}$ is given by \be v_{||} \,= \, K \cos(\Theta -
\Theta_0) \,\frac{\eta^2}{1+q} \,(a_2^\perp - q a_1^\perp) \,, \ee where $K=6
\times 10^4$, $\Theta_0 = 0.184$, and the symbol $\perp$ refers to the
direction perpendicular to the orbital angular momentum. $\Theta$ is defined
as the angle between $a_2^\perp - q a_1^\perp$ and the infall direction at the
merger, which we consider to be a random variable.

In the case of arbitrary spin orientations, the maximal kick velocities occur
when both spins lie in the orbital plane and are anti-aligned,  and reach up
to $4000\,{\rm km\,s^{-1}}$ for maximally spinning BHs. Thus, this case
appears most dangerous for retaining BHs in the centre of host galaxies after
a merger. Note, however, that even if such an unlikely merger configuration
occurs, the probability of receiving a kick of that order of magnitude is
further  reduced due to the parameter $\Theta$.

\begin{table*} \bc
\begin{tabular}{lrccccll} \hline \hline Simulation & $N_{\rm HR}$/$N_{\rm
gas}$ & $m_{\rm DM}$ [$\,h^{-1}{\rm M}_\odot\,$] & $m_{\rm gas}$
[$\,h^{-1}{\rm M}_\odot\,$] & $z_{\rm start}$ & $z_{\rm end}$ & $\epsilon$
[$\,h^{-1}{\rm kpc}\,$] & $R_{\rm HR}$ [$\,h^{-1}{\rm Mpc}\,$] \\ \hline
zoom$5$ & $1364500$ & $6.75\times 10^6$ & $1.32\times 10^6$ & $127$ & $6.2$ &
$1.0$ & $3.4$ \\ zoom$8$ & $5588992$ & $1.65\times 10^6$ & $0.32\times 10^6$ &
$127$ & $6.2$ & $0.625$ & $3.4$\\ zoom$10$ & $10916000$ & $0.84\times 10^6$ &
$0.16\times 10^6$ & $127$ & $6.2$ & $0.5$ & $3.4$\\ zoom$5$\_z$4$   &
$21090625$ & $6.75\times 10^6$ & $1.32\times 10^6$ & $127$ & $3.9$ & $1.0$ &
$8.4$\\ zoom$8$\_z$4$ & $86387200$ & $1.65\times 10^6$ & $0.32\times 10^6$ &
$127$ & $5.3$ & $0.625$ & $8.4$\\ zoom$5$\_z$2$ & $99712125$ & $6.75\times
10^6$ & $1.32\times 10^6$ & $127$ & $2.1$ & $1.0$ & $14.0$\\ \hline \hline
\end{tabular}
\caption{Numerical parameters of the cosmological galaxy cluster simulations
analysed in this study. The names of the various runs performed are given in
the first column, where zoom\# indicates \#$^3$ higher mass resolution than
the parent Millennium simulation. Rows four and five represent resimulations
of the main descendent of our $z=6$ halo that has been selected at $z=4$,
while in the last row we have selected for resimulation the main descendent at
$z=2$. The values listed from the second to the fourth column refer to the
numbers and masses of the high-resolution dark matter and gas particles. Note
that the actual values of $N_{\rm gas}$ and $m_{\rm gas}$ vary in time due to
star formation. The last four columns give the initial and final redshifts of
the runs, the gravitational softening length $\epsilon$, and the radius of the
roughly spherical high resolution region, respectively.
\label{tab_simpar}} \ec
\end{table*}

\subsection{Numerical issues}

Due to the nevertheless limited mass resolution of our cosmological
simulations (see Section~\ref{Simulations}), even in our highest resolution
runs the dark matter and gas particle masses are comparable to the BH seed
mass. This implies that if a low-mass BH receives a recoil kick that is lower
than the escape velocity from its host halo, the subsequent evolution of the
BH's orbit would be computed inaccurately in the simulation, due to a
misrepresentation of the dynamical friction forces. As a result, a recoiled BH
would artificially wander through its host halo for a long time.

In order to correct for this effect one could try to explicitly estimate the
dynamical friction time of a displaced BH, and then gradually reposition it
over this timescale towards the centre of the halo. However,
the orbit of the BH during the approach to the halo's centre would still have
to be chosen arbitrarily. We therefore adopt here a much simpler model. Recall
that we are primarily interested in the question whether a remnant BH will
leave its host halo or not, and to what extent such expulsions can affect the
growth of the first bright quasars. For our purposes it is therefore
sufficient to compare the BH's kick velocity at the moment of the merger to
the local escape velocity from its current host halo. If the former is larger,
we give the correct kick velocity to the BH and follow its dynamical orbit that
will carry it out of the halo (later re-accretion onto a more massive dark
halo is however possible). In the opposite case, we simply neglect the small
recoil that the BH would have received and do not alter its velocity. This
approximation hence assumes that the displaced BH will return to the halo
centre on a negligible time scale. In order to estimate the local escape
velocity from the host halo, we assume a singular isothermal
sphere profile for the halo. This effectively provides an upper limit to the
number of BHs that are actually kicked out.

\subsection{BH spin evolution due to the mergers} \label{BHspin_evo} 

Given that even a merger of non-spinning BHs produces significant final BH
spins, and that spins play an important role in determining the magnitude of
the recoil velocity, it is prudent to consider BHs characterised by a
non-vanishing spin and to account for spin evolution. Here we implement a
basic model for the treatment of BH spins. Initially, when a dark matter halo
is seeded with a small mass BH, we not only assign an initial mass to the BH
but also a certain amount of spin that can range from $0$ to $1$. For
simplicity, we assume that all BH seeds are characterised by the same initial
spin value, and we perform several simulations were we vary the choice for the
initial spin in order to understand how the subsequent evolution of the BH
population depends on this parameter. Furthermore, when two BHs merge we
estimate the final spin of the BH remnant on the basis of analytical fits to
the numerical relativity calculations of \cite{Rezzolla2007a, Rezzolla2007b,
Rezzolla2008}. This allows us to consider both the initial spin values of the
two BHs prior to the merger as well as their configuration (either
aligned/anti-aligned spins, or randomly orientated spins) to accurately
estimate a final spin value. We can also follow spin flip events and estimate
how often they occur during the cosmological BH growth. Finally, we perform
additional ``control'' runs where we keep the BH spins constant and equal to
their initial value, for comparison.

Note that in this study we do not take into account spin changes due to gas
accretion, a process that could in principle be very important.  We here
refrain from attempting to model this physical process due to its intrinsic
complexity and uncertainty, given that it is possible that gas accretion
episodes lead either to spin--up or to spin--down of BHs
\citep[e.g.][]{Volonteri2005, King2006}, depending on the nature of the
accretion flow. However, by exploring a large range of initial BH spin values
we can still gain some insight into a possible modification of typical BH mass
accretion histories if, for example, extended episodes of gas accretion would
lead to rapidly spinning BHs. An investigation of a spin evolution driven by
gas accretion through suitable simulation sub-grid models  is an interesting
subject for future work.

\subsection{Spin-dependent radiative efficiencies}

In our new BH model, we can self-consistently account for the different
efficiency of spinning BHs for turning the gravitational binding energy of an
accretion flow into radiation. Spin-dependent radiative efficiencies will both
change the amount of material that can be accreted onto a hole as well as
modify the amount of feedback energy that is released. For standard accretion
disks the radiative efficiencies of accreting BHs can vary from $0.057$ for a
non--spinning BH to $0.42$ for a maximally spinning BH, thus possibly causing
a large effect on the early growth history of BHs \citep{Shapiro2005,
Volonteri2006}. Following \cite{Bardeen1972}, we estimate the radiative
efficiency at the innermost stable co-rotating circular orbit as \be e_{\rm
max} \,=\, 1 - \frac{\tilde r - 2 + a/\sqrt{\tilde r}}{\sqrt{\tilde r ^2 - 3
\tilde r + 2 a \sqrt{\tilde r}}}\,, \label{Bardeen} \ee  where \be \tilde r
\,=\, 3 + A_2 - \sqrt{(3 - A_1)(3 + A_1 + 2 A_2)} \,, \ee \be A_1 \,=\, 1 + (1
- a^2)^{1/3} \big[ (1 + a)^{1/3} + (1 - a)^{1/3} \big] \,, \ee and \be A_2
\,=\, \sqrt{3 a^2 + A_1^2}\,.  \ee Here $a$ denotes the dimensionless BH spin,
and the radius of the innermost stable co-rotating circular orbit is given by
the product of $\tilde r$ and the BH's total gravitational mass.

\section{Numerical simulations} \label{Simulations}

\begin{figure*} \centerline{ \vbox{\hbox{
\psfig{file=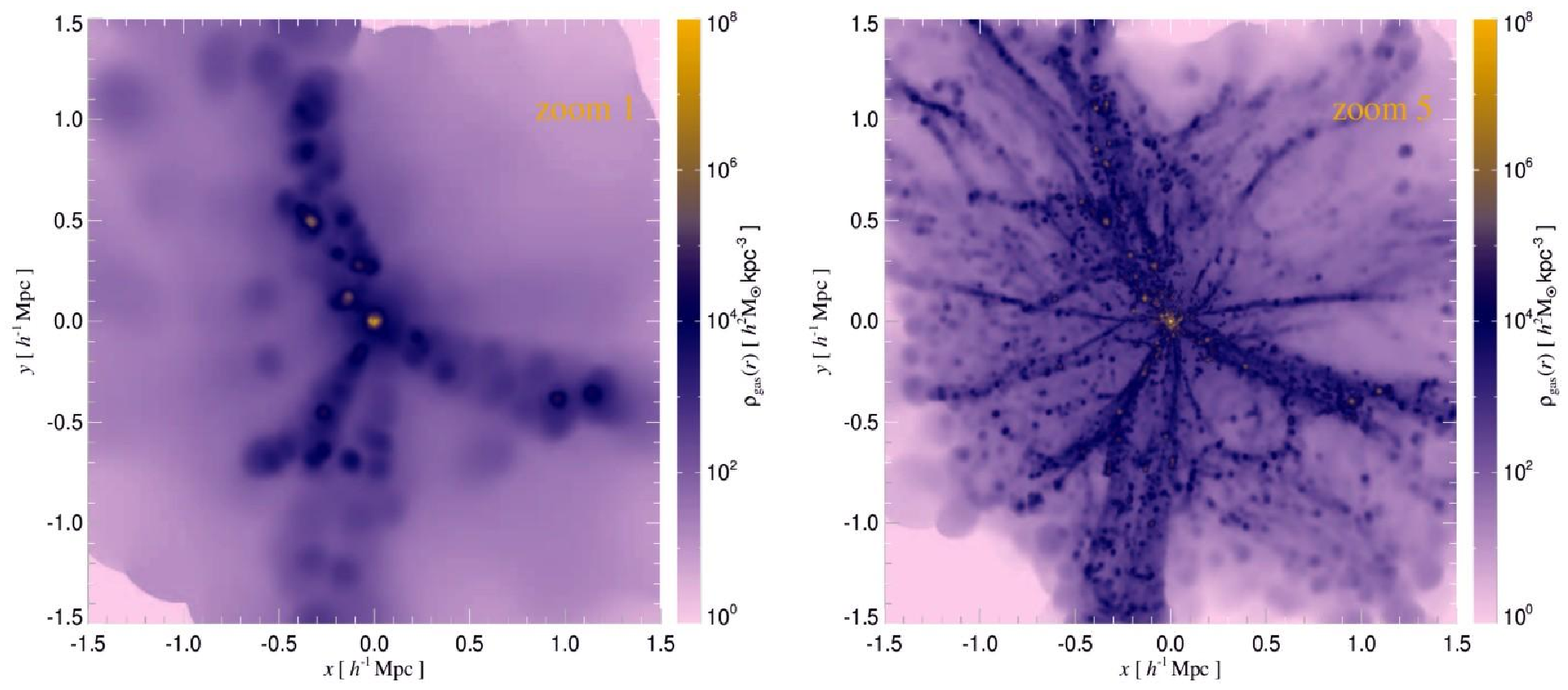,width=17.truecm,height=7.3truecm}}
\hbox{
\psfig{file=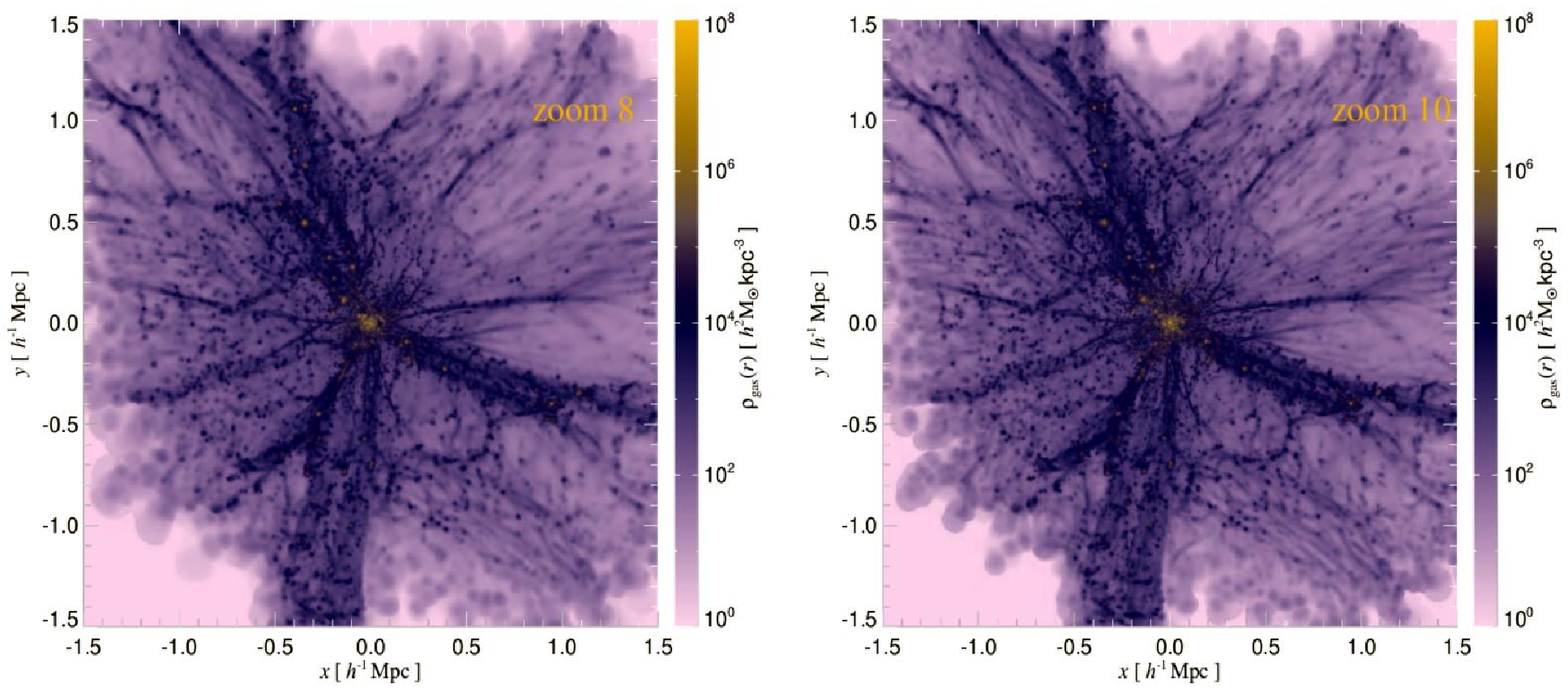,width=17.truecm,height=7.3truecm}}}}
\caption{Projected mass-weighted gas density maps of the most massive $z=6.2$
halo ($M_{\rm200}=4.86 \times 10^{12} \,h^{-1}{\rm M}_\odot$) from the
Millennium simulation, which was selected for resimulation. The upper-left
panel shows the halo simulated with the resimulation technique with the
same spatial and mass resolution as in the parent Millennium
run. The remaining panels illustrate the same halo
resimulated with $5^3$, $8^3$, and $10^3$ times higher mass resolution, as
indicated on the panels.  }
\label{zoommap}
\end{figure*}

\begin{figure*} \centerline{ \hbox{
\psfig{file=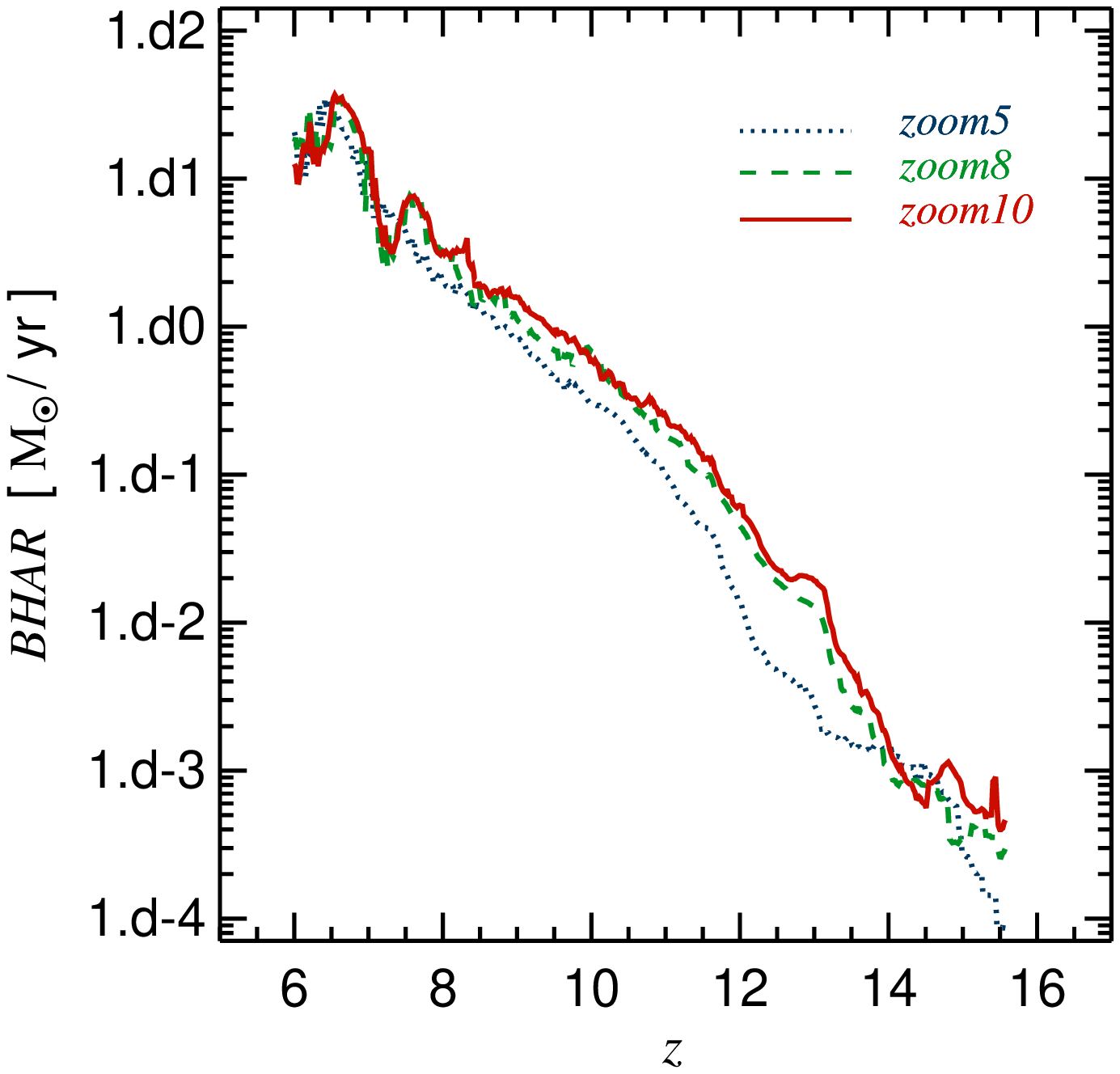,width=8.truecm,height=8truecm}
\psfig{file=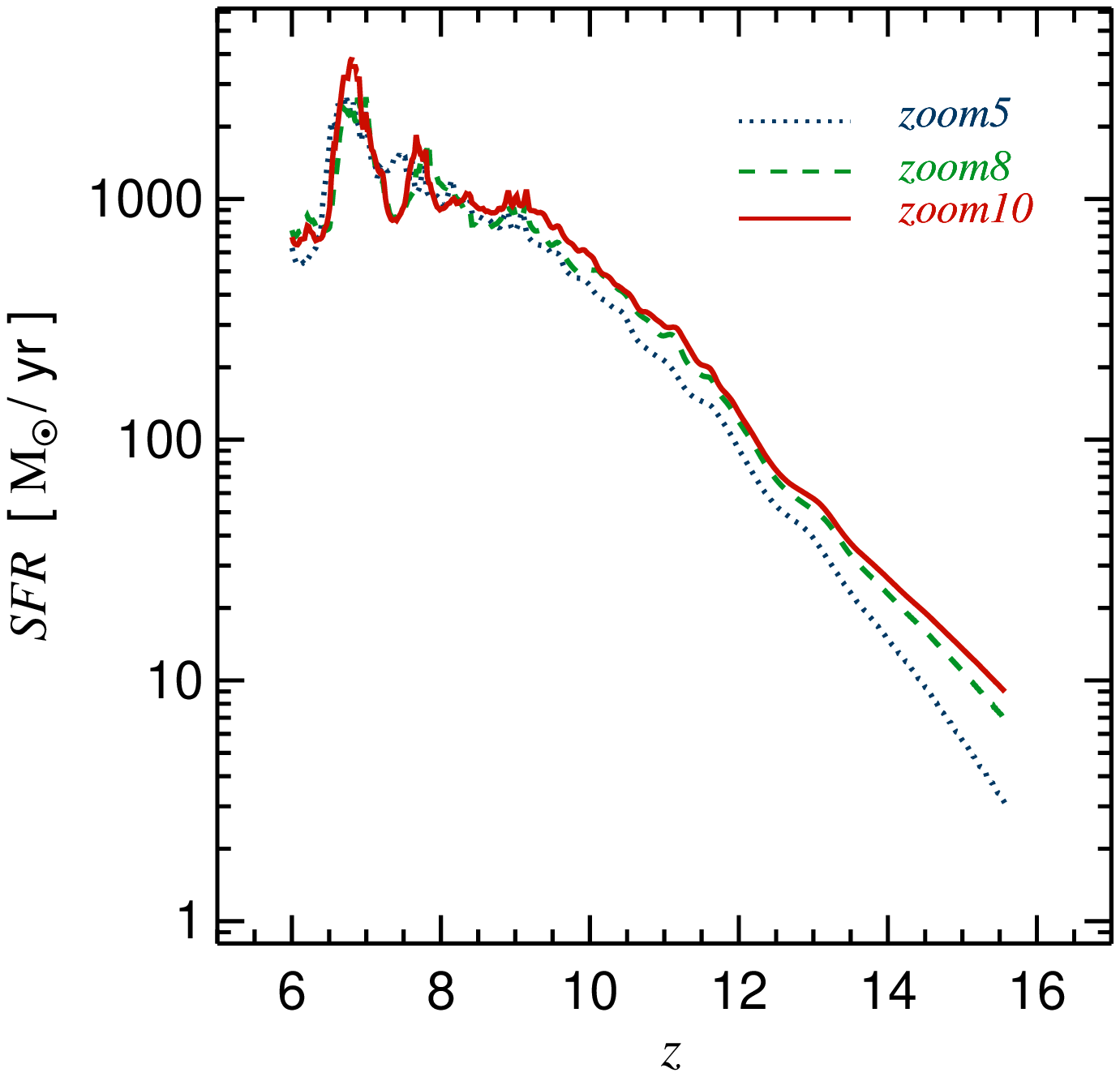,width=8.truecm,height=8.truecm}}}
\caption{BH accretion rate (left-hand panel) and SFR (right-hand panel) in the
simulated volume as a function of redshift. The different curves are for the
same simulation performed at different numerical resolutions, increasing the
mass resolution by up to a factor of 8. The BHs have been seeded here in
haloes with masses above $10^{9} \,h^{-1} {\rm M}_\odot$. Reasonable numerical
convergence in these quantities has been reached.}
\label{numconv}
\end{figure*}

We have performed a suite of numerical simulations especially designed to
address the issue of early BH growth.  For this purpose, one ideally needs to
simulate both an extremely large volume of the Universe of order of
$0.5-1\,{\rm Gpc}$ cubed, as the space density of high-redshift luminous
quasars is very low, and at the same time reach a very large dynamic  range in
order to resolve at least $\sim {\rm kpc}$ scales reliably. Simulations that
reach this resolution throughout such large volumes are currently beyond the
reach of any state-of-the-art cosmological code, especially when gas and BH
physics are included as essential ingredients.

We have therefore adopted a different approach that directly focuses on rare
high density peaks \citep[see e.g.][]{Gao2005}, where the first SMBHs are most
likely to form. By restricting the region that needs to be  resolved with high
resolution to a small fraction of the total simulated volume we can reach
sufficient resolution for numerically converged and meaningful results. The
identification of a suitable target region requires a homogeneously resolved
high-resolution parent simulation. For this purpose we use the dark
matter-only Millennium simulation \citep{Millennium}, which has a volume of
$(500\,h^{-1}{\rm Mpc})^3$, just large enough to expect about one luminous
quasar at $z=6$ in the simulation box, given the observed space density of the
high-redshift SDSS quasars, and their probably high duty cycle
\citep{Shankar2008}.  For our primary simulations, we have selected the most
massive dark matter halo formed at this epoch and resimulated it at much
higher mass and force resolution. We note that this target dark matter halo is
part of a protocluster region that collapses to a very rich cluster of
galaxies by redshift $z=0$ \citep{Millennium}.

The resimulations were performed by selecting the Lagrangian region of our
target halo in the original initial conditions and by
populating it with a larger number of lower mass particles, adding in
additional small scale power up to the new Nyquist frequency, as
appropriate. Around this high-resolution region, the mass resolution has been
progressively deteriorated by using ever more massive particles at larger
distances, but making sure that the large-scale gravitational tidal field
acting on the high-resolution region remained accurately represented. In order
to test our resimulation setup, we have performed several dark matter-only
simulations, verifying that the mass of the original target halo is reproduced
accurately to within $1-2\%$. For the simulations with gas, we have split each
high-resolution dark matter particle into a dark matter and gas particle,
displacing them by half of the original mean inter-particle separation while
keeping the centre-of-mass of each pair fixed. In this way the initial
particle loads of dark matter and gas stay spatially separate as long as
possible until the first non-linear structures form.

Based on the initial conditions constructed in this way we have performed a
number of runs including gas, stellar and BH physics. We have additionally
performed pure cooling and star formation runs, in order to gauge the impact
of BH feedback effects. We have further resimulated our $z=6$ target halo at a
number of different numerical resolutions, starting from a mass resolution of
a factor $5^3$ higher than the original Millennium run up to $10^3$ times
higher mass resolution. In Table~\ref{tab_simpar}, we summarize the main
numerical parameters of our simulation suite. Moreover, in order to continue
our simulations below $z=6$, we have identified the main descendent
of our $z=6$ halo both at $z=4$ and $z=2$ in the original Millennium run, and
resimulated those as well, starting again from a redshift of $z=127$.

\begin{figure} \centerline{ 
\psfig{file=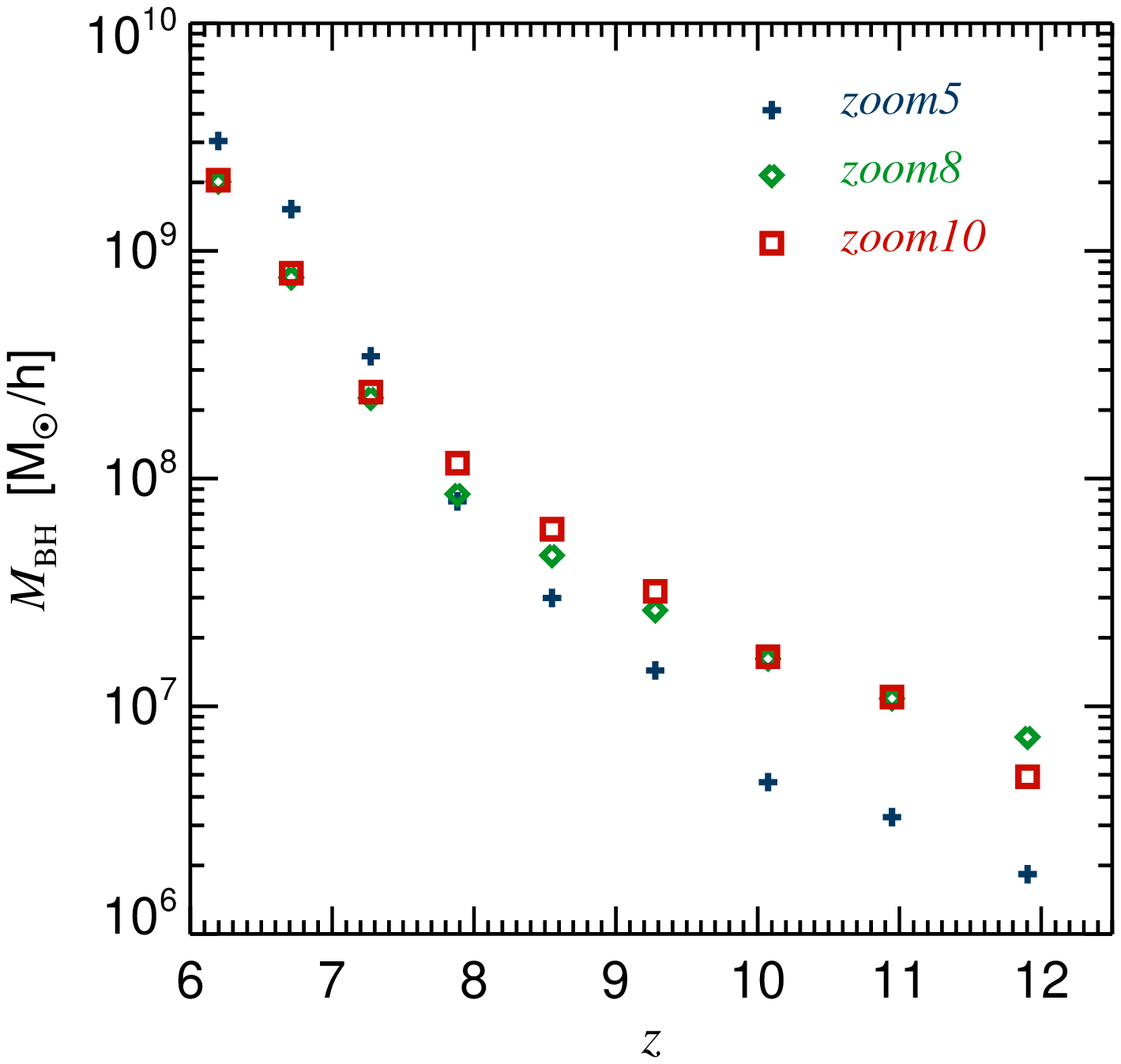,width=9.3truecm,height=9.truecm}}
\caption{The mass of the most massive BH in the simulated volume as a function
of redshift for the same runs as in Figure~\ref{numconv}. While the zoom$5$
simulation somewhat underpredicts the BH mass at high redshift, for $z < 8$ it
catches up with the higher resolution runs shown. The BH mass and accretion
rate of the zoom$8$ run are basically indistinguishable from the zoom$10$
values, indicating robustness of our results.}
\label{mbhnumconv}
\end{figure}

The cosmological parameters adopted in this study are the same as in the
Millennium simulation, namely: $\Omega_{\rm m}=0.25$, $\Omega_{\rm
\Lambda}=0.75$, $\sigma_8 = 0.9$, $h=0.73$ and $n_{\rm s} = 1.0$,
corresponding to a flat $\Lambda$CDM cosmological model. We chose $\Omega_{\rm
b} = 0.041$ in order to reproduce the cosmic baryon fraction inferred from the
most recent cosmological constraints \citep{Komatsu2008}.

\begin{figure} \centerline{ \vbox{ \hspace{0.5truecm} \vspace{-0.5truecm}
\psfig{file=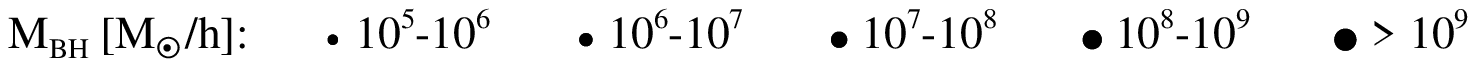,width=9.truecm,height=1.1truecm}
\vspace{-0.1truecm}
\psfig{file=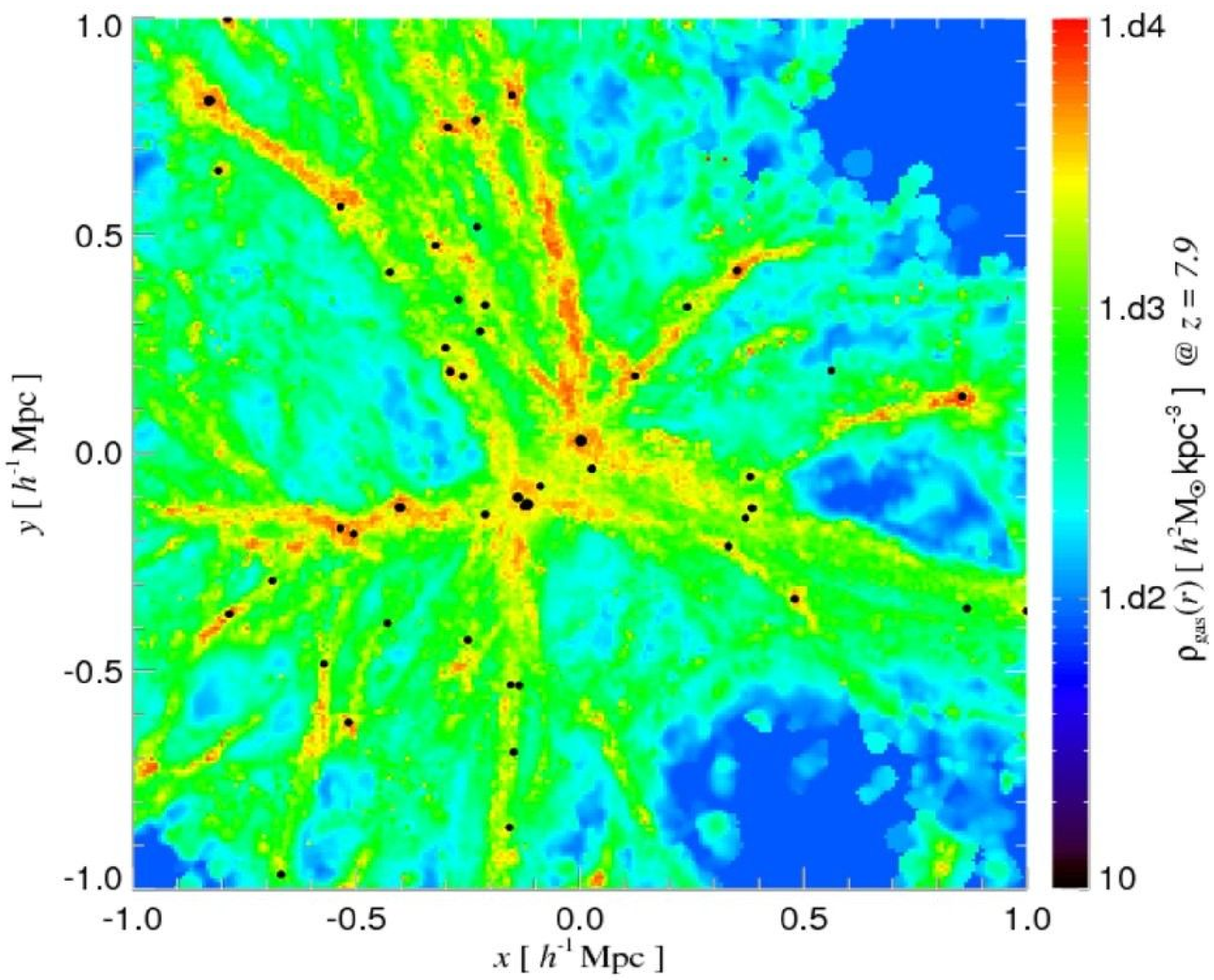,width=8.3truecm,height=6.7truecm}
\vspace{-0.1truecm}
\psfig{file=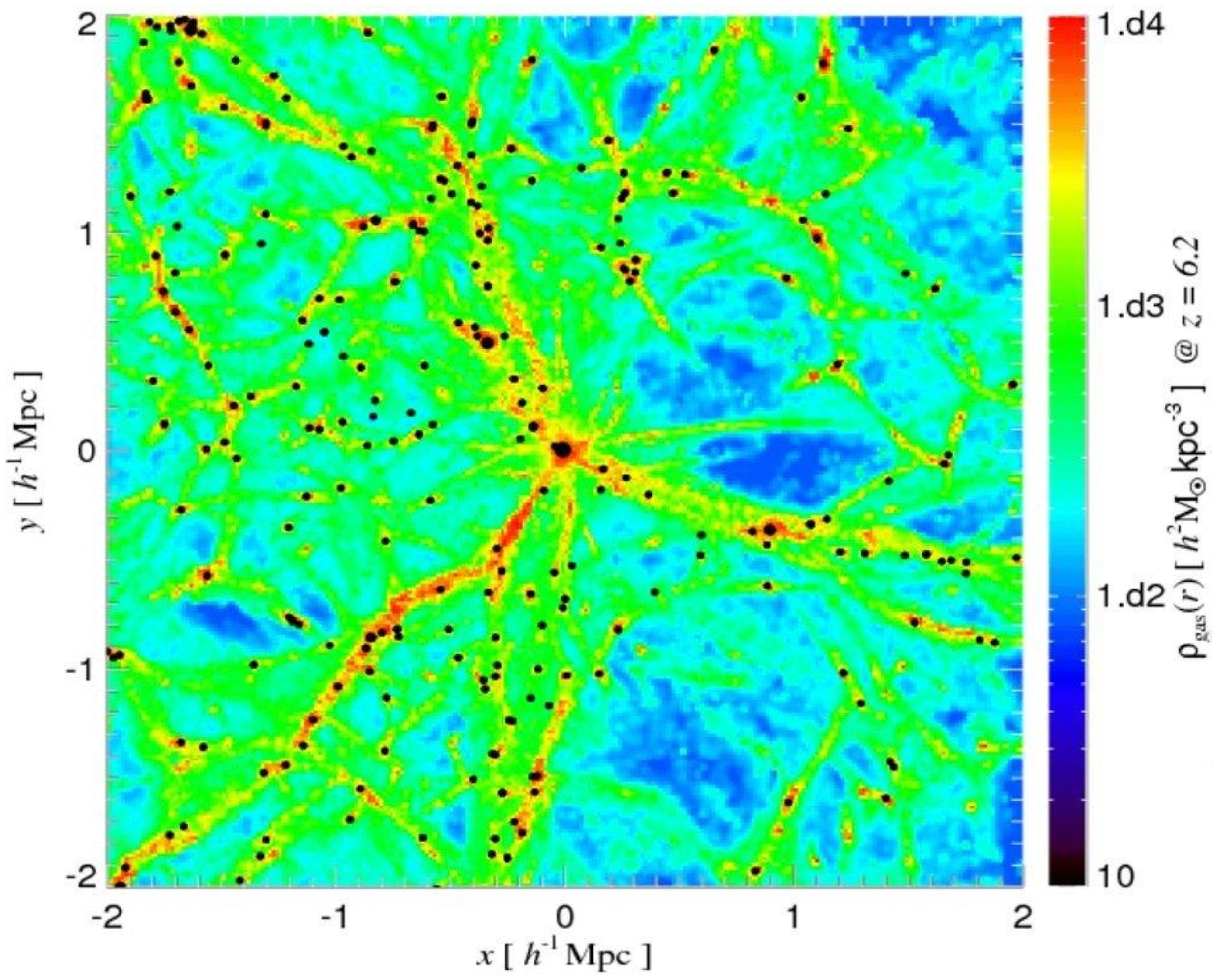,width=8.3truecm,height=6.7truecm}
\vspace{-0.1truecm}
\psfig{file=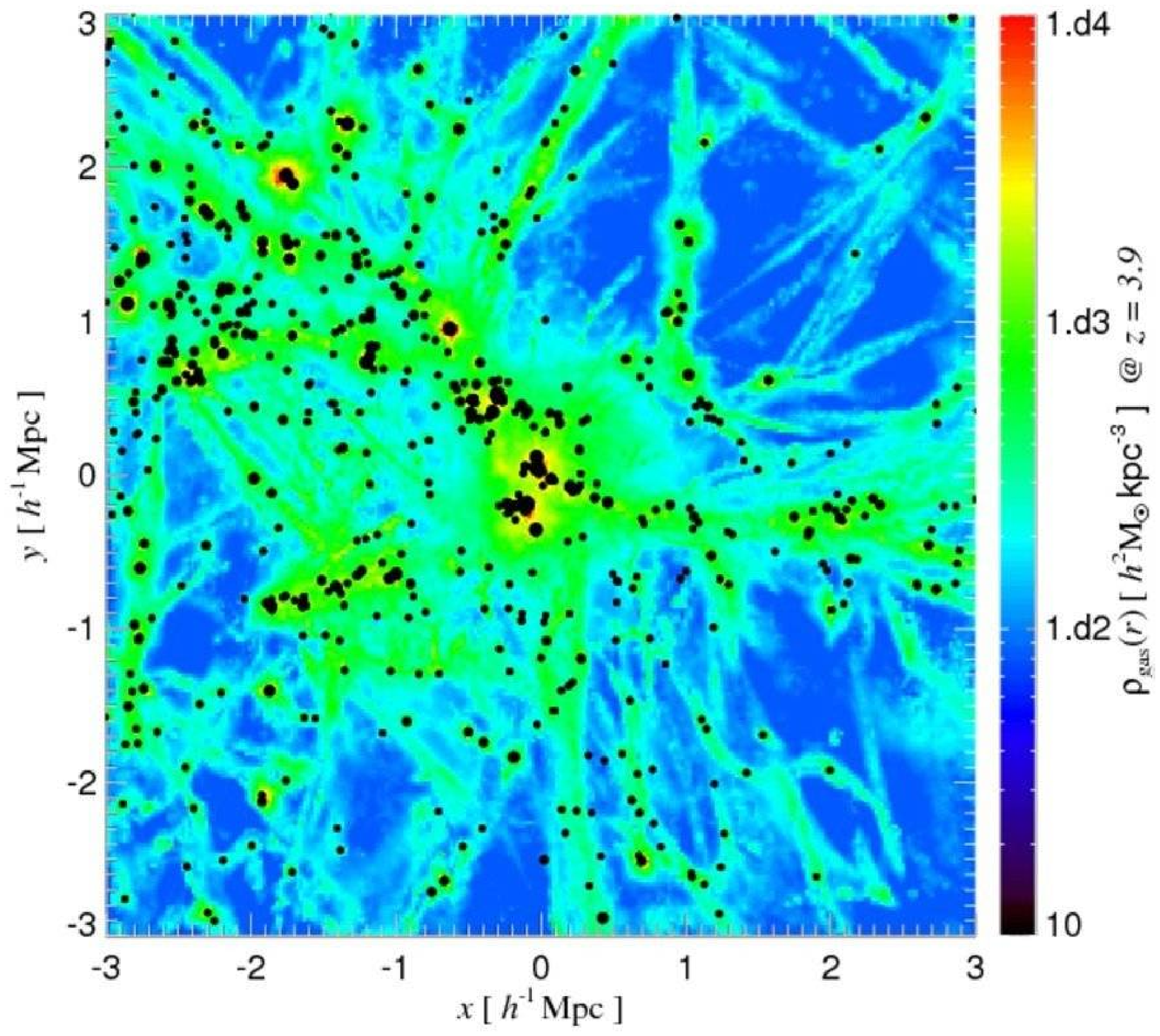,width=8.3truecm,height=6.7truecm}
\vspace{-0.25truecm} }}
\caption{Projected mass-weighted gas density maps of the most massive
Millennium halo at $z=6.2$, resimulated at much higher resolution with gas and
BHs. The middle panel shows the large-scale environment of the halo at
$z=6.2$. The upper and lower panels show the main progenitor and the main
descendent of this halo at $z=7.9$ and $z=3.9$, respectively. The black dots
denote the positions of BH particles, with their size encoding the BH mass as
indicated on the legend. Note that at $z=7.9$ two already very massive BHs in
the centre of their respective host haloes are about to undergo a merger and
thus contribute to the build-up of the SMBH at $z=6.2$.}
\label{BHmaps}
\end{figure}

\section{Reproducing high-redshift quasars} \label{Results1}

\subsection{Numerical convergence tests}\label{NumConvTests}

We start our analysis with a series of resimulations of the $z=6$ target
halo\footnote{See Table~\ref{tab_halopar} for a list of its main properties.}
using our default BH model. We systematically vary the numerical resolution by
considering progressively improved spatial and mass resolutions that are
$5^3$, $8^3$ and $10^3$ times better than the original Millennium
simulation. This allows an assessment of the numerical convergence of our
results.  In Figure~\ref{zoommap}, we show the gas density distribution of our
target halo for different resimulation runs with increasing resolution. The
global properties of the resimulated halo are very similar in all the
simulations performed (e.g. with respect to the position of the most prominent
filaments ending up in the main halo; the sites of very dense gas that is
subject to star formation, etc.). However, the richness of structures
surrounding our target halo and the fine features in its interior are clearly
much better represented in our higher resolution runs. In these simulations, a
vast number of small haloes in the outskirts of the main system and a network
of fine filaments between them is revealed. We note however that the zoom$8$
and zoom$10$ runs show almost identical gas density distributions.

Previously, we have already demonstrated good numerical convergence of our BH
model in full cosmological simulations \citep{Sijacki2007}, but properly
resolving the very early growth of BHs is particularly challenging.  This
requires that the host haloes at $z \sim 15$, when the first BHs start to be
seeded, have to be represented by a sufficient number of particles such that
the BH accretion processes can be followed accurately even at this early
time. It is important to note that if the resolution in these first host
systems is too poor this will likely lead to an underestimated BH accretion
rate, causing an artificial delay of the mass assembly of the first massive
BHs. To avoid this numerical problem it is necessary to push the numerical
resolution high enough to resolve the first BH hosts adequately. This also
implies that the actually realized resolution imposes a natural limit on the
minimum host halo mass that can reliably be seeded in our scheme.

\begin{table*} \bc
\begin{tabular}{p{0.8cm}@{\quad}p{1.cm}@{\quad}p{1.cm}@{\quad}p{1.cm}@{\quad}p{0.9cm}@{\quad}p{0.8cm}@{\qquad}p{0.8cm}@{\qquad}p{0.8cm}@{\qquad}p{0.8cm}@{\qquad}p{0.7cm}@{\qquad}p{1.cm}@{\qquad}p{0.8cm}@{\qquad}p{0.8cm}}
\hline\hline \multicolumn{13}{c}{MAIN HALO PROPERTIES AT $z=6.2$}\\ \hline Run &
$N_{\rm 200}$ & $N_{\rm 200,DM}$ & $N_{\rm 200,gas}$ & $R_{\rm 200}$ & $M_{\rm
200}$ & $M_{\rm 200,DM}$ & $M_{\rm 200,gas}$ & $M_{\rm 200,*}$ & $T_{\rm 200}$
& ${\rm SFR}$ & $M_{\rm BH}$ & $M_{\rm Edd}$\\ & & & & [${\rm kpc}/h$] & [${\rm
M}_\odot\,/h$] & [${\rm M}_\odot\,/h$] & [${\rm M}_\odot\,/h$] & [${\rm
M}_\odot\,/h$] & [${\rm K}$] & [${\rm M}_\odot\, /{\rm yr} $] & [${\rm
M}_\odot\,/h$] & \\      \hline no BHs & $6436930$ & $2475953$ &
$1047133$ & $437.1$ & $4.86 \times 10^{12}$ & $4.08 \times 10^{12}$ & $3.02
\times 10^{11}$ & $4.71 \times 10^{11}$ & $5.8 \times 10^6$ & $1820$ & -- & --
\\\\ with BHs & $6206304$ & $2491315$ & $1468049$ & $437.5$ & $4.87 \times
10^{12}$ & $4.11 \times 10^{12}$ & $3.97 \times 10^{11}$ & $3.63 \times
10^{11}$ & $6.5 \times 10^6$ & $508$ & $2.01 \times 10^9$ & 0.27 \\ \hline
\hline
\end{tabular}
\caption{The main properties of our resimulated halo at $z=6.2$. The upper row
refers to the run performed without BHs while the bottom row is for the
simulation including BH growth and feedback, where BHs were seeded in haloes
above $10^{9} \,h^{-1} {\rm M}_\odot$. Both simulations have been performed
with a zoom factor of $8$. The second to the fourth columns give the total,
dark matter, and gas particle number within the virial radius (fifth
column). The total, dark matter, gas, and stellar mass of the halo are listed
in columns six to nine. The mean mass-weighted temperature and total star
formation rate within the virial radius are given in columns ten and eleven,
respectively. The last two columns report the central BH mass and its
accretion rate in Eddington units.
\label{tab_halopar}} \ec
\end{table*}

Based on the numerical resolution experiments we have performed we find that a
mass resolution of at least $5^3$ times higher than the Millennium run is
needed when BHs are seeded at the centres of $10^{9} \,h^{-1} {\rm M}_\odot$
haloes. This is illustrated in Figure~\ref{numconv}, where we plot the total
BH accretion rate (left-hand panel) and star formation rate (SFR; right-hand
panel) as a function of time for our three high resolution runs (as indicated
in the legend). Note that while the zoom$5$ run somewhat underpredicts the BH
accretion rate and SFR at high redshifts, as expected given that a fraction of
small haloes remains poorly resolved in this run, it has very similar BH
accretion rate and SFR values for $z<9$ compared with the zoom$8$ and zoom$10$
runs. On the other hand, the BH accretion rate and SFR of the zoom$8$ and
zoom$10$ simulations are basically indistinguishable, indicating that we have
resolved the bulk of the star formation and BH accretion occurring in the
haloes within the simulated timespan.

\begin{figure}
\psfig{file=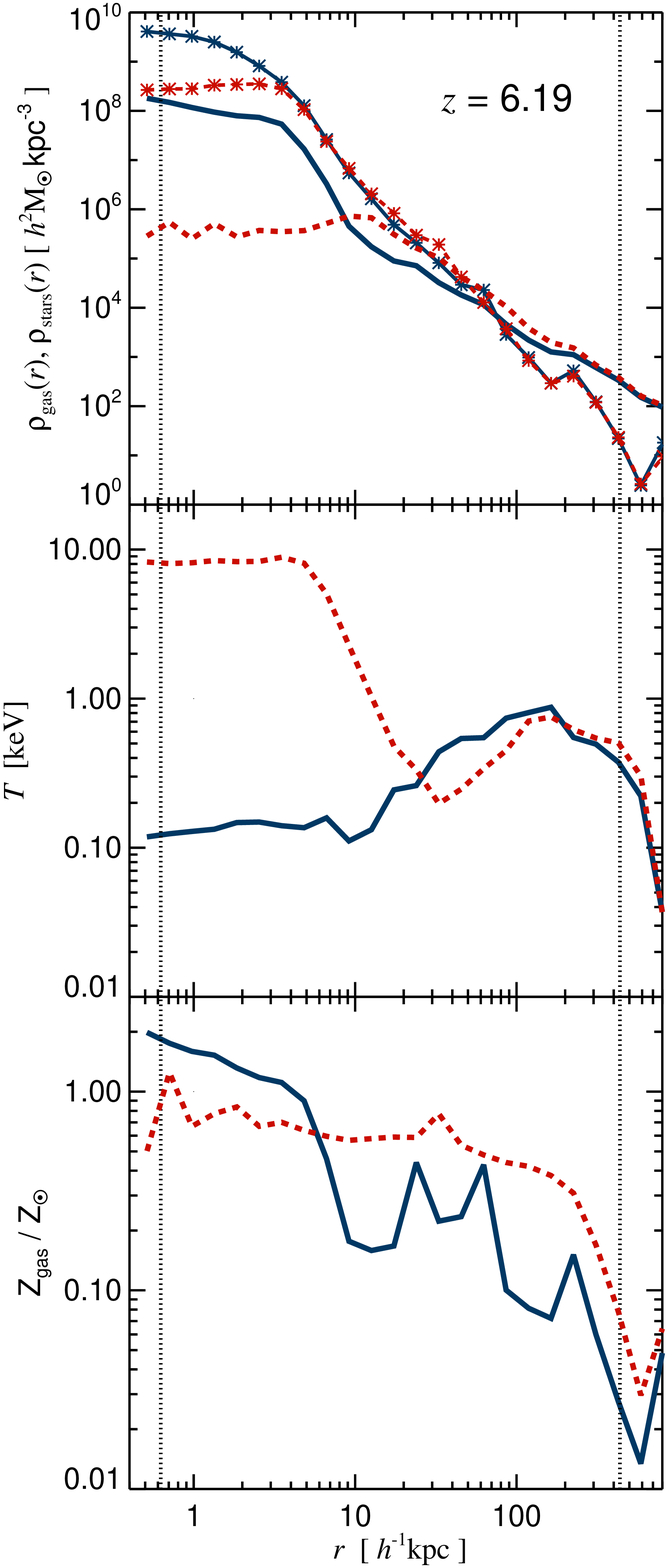,width=8.truecm,height=18truecm}
\caption{Radial profiles of gas density, mass-weighted temperature and
mass-weighted gas metallicity of the most massive halo at $z=6.2$ for a run
without BHs (blue continuous lines) and for the simulation where BH are
included (red dashed lines). In the top panel stellar density profiles are
shown as well (denoted with star symbols with the same colour coding and line
styles). Both simulations have been performed with a zoom factor of $8$, and
BHs were seeded in haloes above $10^{9} \,h^{-1} {\rm M}_\odot$. The vertical
dotted lines denote the adopted gravitational softening length and the virial
radius of the halo (in comoving units). The dramatic effect of the powerful
quasar feedback on  the properties of the host halo can be clearly seen: the
central gas density is lowered by $\sim 2$ orders of magnitude and the
temperature is increased by a similar amount.}
\label{profiles}
\end{figure}

In Figure~\ref{mbhnumconv}, we analyze the mass growth of the most massive BH
in these three runs of increasing resolution. For the zoom$5$ run the BH mass
is again somewhat underpredicted at high redshifts with respect to the higher
resolution simulations. At $z \sim 7-8$ the BH accretion rate in the zoom$5$
simulation comparatively increases, and the BH mass catches up with the values
obtained for the zoom$8$/zoom$10$ runs. Contrary to what one may naively
expect, we note that this relative BH accretion rate increase in the zoom$5$
run for $z \sim 7-8$ does not lead to an overprediction of the BH masses for
$z < 6$ (see Section~\ref{Beyond_z_6} for a detailed description of the BH
evolution at lower redshifts). The BH masses and hence BH accretion rates of
the zoom$8$ and zoom$10$ runs are essentially identical at all redshifts,
indicating that our estimates of the BH mass and the accretion rate onto our
most massive BH have converged and are reliable. We have therefore decided to
perform most of the simulations at the resolution of the zoom$8$ run for our
further analysis. The only exception are the simulations we performed to much
lower final redshifts of $z=4$ and $z=2$. For computational reasons we have
lowered for these the resolution to that corresponding to the zoom$5$
run. While this leads to a small delay in the BH growth at very high redshift,
this bias becomes negligible at lower redshift, ensuring that the BH
properties are well resolved and converged.

Our numerical tests highlight that even with the quite high resolution
achieved here, we cannot reliably track BH seeds that start out in haloes as
small as, e.g., $10^{8} \,h^{-1} {\rm M}_\odot$ haloes. Our choice for the
mass-scale of the haloes we seed is thus in part determined by numerical
limitations. In order to track BH growth in still smaller mass systems it
would be necessary to construct dedicated simulations where the required still
higher resolution can be reached. This would have to be accomplished at the
expense of either choosing a smaller mass host halo at $z=6$ for resimulation,
or by restricting the simulations to cosmic structures at $z \gg 6$. Both of
these are beyond the scope of this work. We therefore caution that our results
rely on the assumption that the BH seeding process actually occurs within the
haloes we can resolve, or that a population of BH seeds present at some
earlier redshift is compatible with the BH seed population as we introduce it
in our simulations.

Finally, Figure~\ref{mbhnumconv} shows that in our default BH model it is
possible to produce a SMBH at $z \sim 6$ with a mass of $2-3 \times 10^{9}
\,h^{-1} {\rm M}_\odot$ -- within the range of the observational estimates of
SDSS $z \sim 6$ quasars \citep[e.g.][]{Willott2003, Barth2003, Jiang2006,
Kurk2007}\footnote{We have also checked explicitly how sensitive our
results are to different choices of $\alpha$, and found that even with $\alpha=1$
(which is clearly suppressing the Bondi rate in the simulations), a SMBH of
$9.9 \times 10^{8} \,h^{-1} {\rm M}_\odot$ is produced by $z=6.2$, which is a
factor of $\sim 2$ lower than the mass of the SMBH with $\alpha = 100$,
adopting the same numerical resolution.}. Even though the BH growth starts
from relatively massive seeds in our simulations this result is {\em
non-trivial}. This is the first time that it is verified in full cosmological
simulations of structure formation that there is sufficient gas available to
fuel BH accretion and thus to build up such extremely massive BH in less than
a Gyr of cosmic time. In previous work, \citet{Li2007} had employed a sequence
of multiple mergers of isolated gas-rich spiral galaxies to show that such
rapid growth appears possible in principle. We here confirm that this is
indeed the case when the full cosmological context is modelled
self-consistently. Moreover we are finding in our simulations that the SMBH
is produced only in the centre of the most massive halo, thus further
corroborating our assumption that such rare and massive haloes are appropriate
hosts for the SDSS quasars. Our formation scenario is also in good agreement
with the estimated comoving space density of the SDSS quasars.

\subsection{The supermassive BH and its host at $z=6$}

In Table~\ref{tab_halopar} we list the main properties of our target halo at
$z=6.2$ for simulations with and without BHs. Comparison reveals the
significant effect of the AGN feedback on the halo properties. In the last two
columns we give the central BH mass  and the accretion rate in Eddington
units. The star formation rate is reduced by a factor of $\sim 3$ due to the
AGN feedback. AGN heating starts to reduce the amount of stars formed from $z
\sim 12$. Hence, the stellar mass of the host halo is decreased by $\sim
23\%$, while the amount of gas within the virial radius is larger by $\sim
31\%$. We note however that the total baryonic mass within the virial radius
up to $z=6.2$ is very similar in both runs. At $z \ge 6.2$ quasar activity
mostly affects how much gas will cool and form stars, but it is not capable of
expelling large quantities of baryons from the host
halo. Table~\ref{tab_halopar} further shows that the mean gas temperature
within the virial radius is higher in the run with BHs, a clear signature of
the thermal AGN feedback which generates very hot gas in the central regions
of the halo, as we will discuss in more detail later on.

In Figure~\ref{BHmaps}, we show projected mass-weighted density maps of our
resimulated halo at three different redshifts, $z= 7.9$ (top panel), $z=6.2$
(middle panel), and $z=3.9$ (bottom panel). The black dots
indicate the positions of BH particles. The size of the symbols encodes
information about their mass, as indicated in the legend. In the top panel the
main progenitor of our $z=6$ halo experiences a merger with a halo of similar
size which contains another relatively massive BH in its core. By $z=6.2$,
the most massive halo appears fairly relaxed. It is the only one to contain a
BH with mass above $10^{9} \,h^{-1} {\rm M}_\odot$ in its centre and is
surrounded by an intricate web of filaments containing many smaller haloes and
actively growing BHs. By $z=3.9$, our target halo has grown considerably in
mass and is once more strongly perturbed due to several merging events. There
are now also several smaller mass haloes with BHs with masses above $10^{9}
\,h^{-1} {\rm M}_\odot$ in the vicinity. One of them is just in the process of
merging with our target halo.

\begin{figure*} \centerline{ \hbox{
\psfig{file=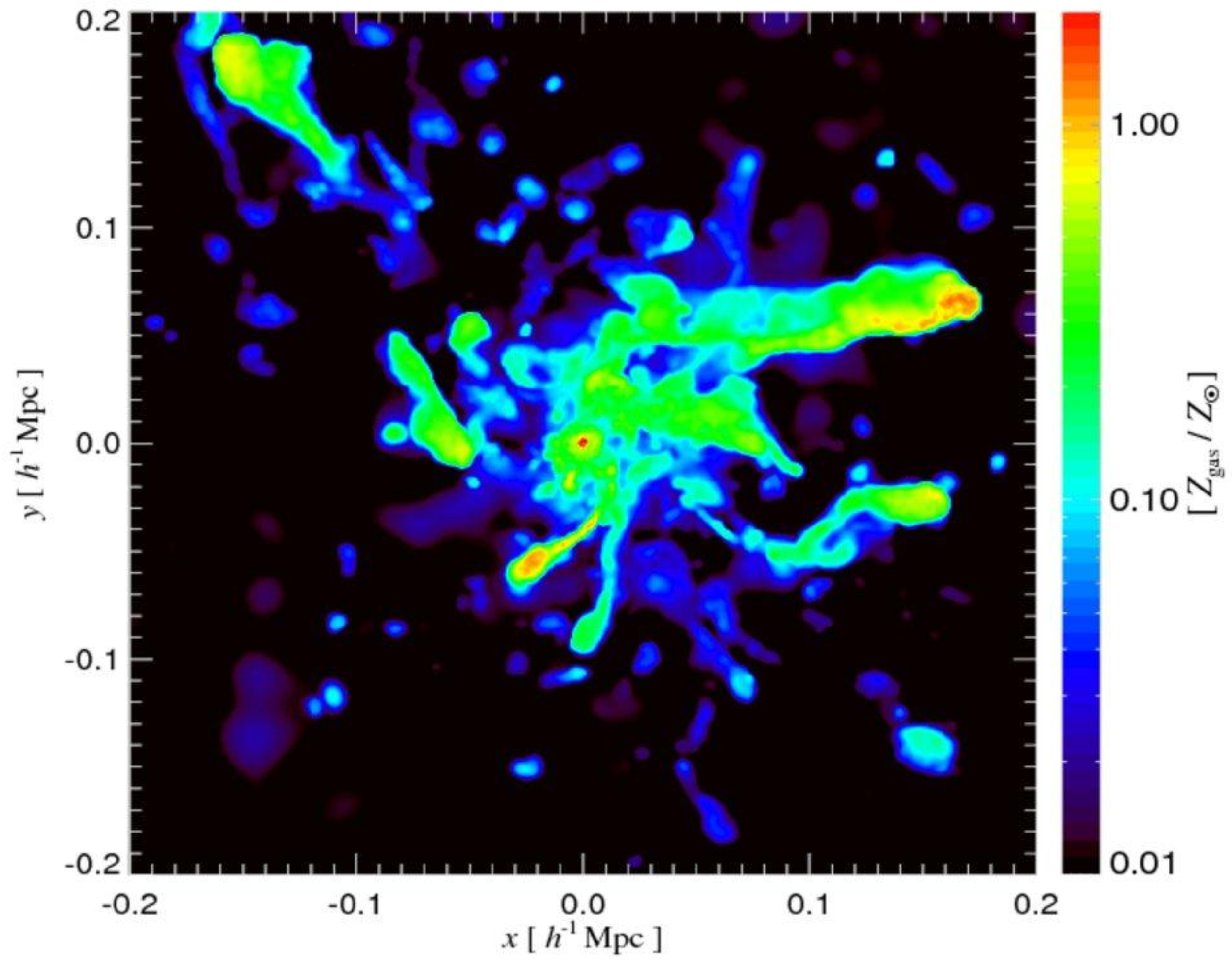,width=8.8truecm,height=7.5truecm}
\psfig{file=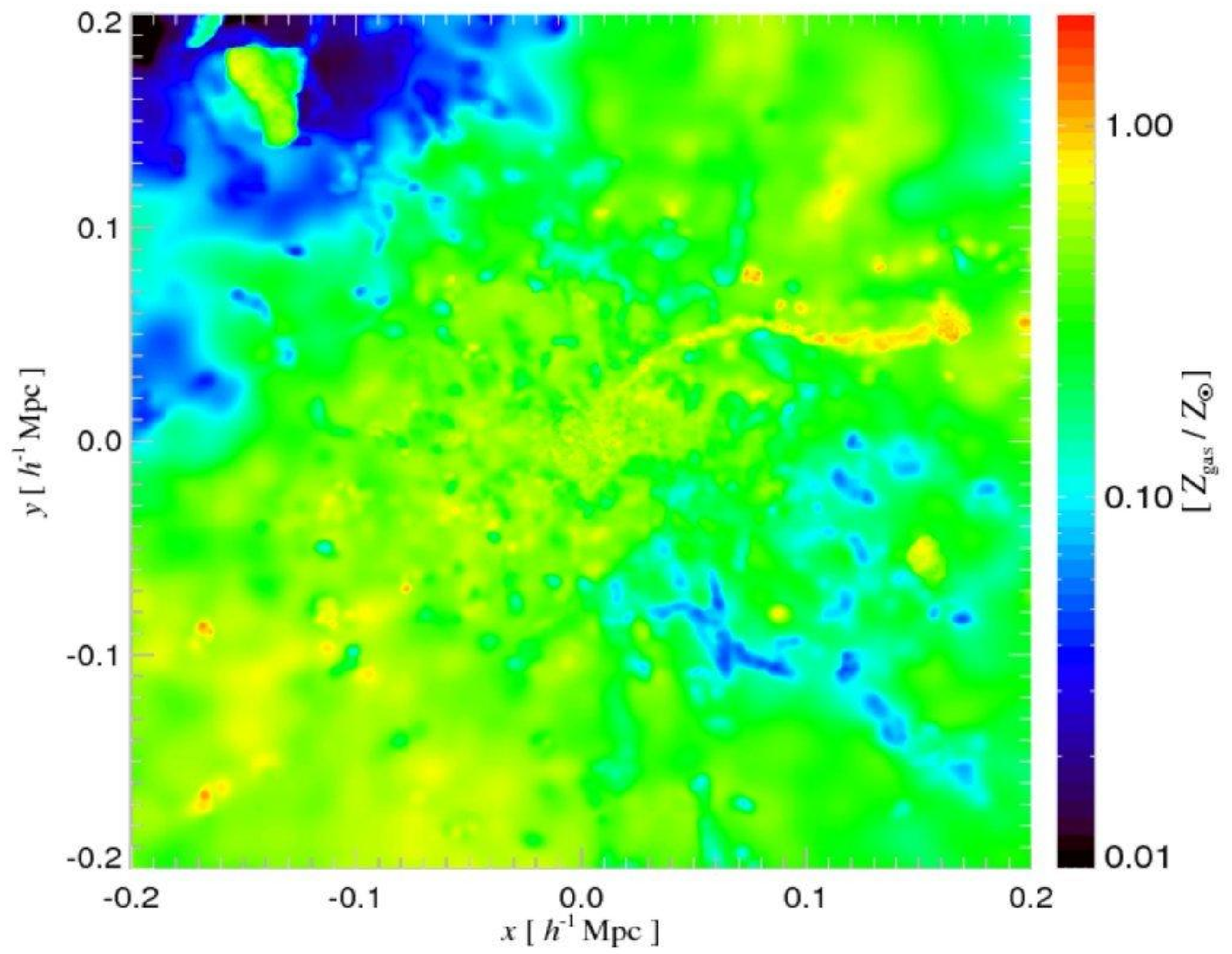,width=8.8truecm,height=7.5truecm}}}
\caption{Projected mass-weighted metallicity maps of the central region of the
most massive halo at $z=6.2$. The left-hand panel shows the results of a
simulation computed without BHs, while the right-hand panel is for a
simulation with our default BH model. The metallicity distribution is
significantly affected by quasar feedback. In the run with BHs, highly metal
enriched gas is expelled from dense star forming regions, and spread out even
out to distances of $400 h^{-1}{\rm kpc}$ from the central quasar. Note in
particular the high metallicity trail, visible to the right. It is caused by a
fly-by of a substructure and persists even in the presence of powerful AGN
activity.}
\label{metmaps}
\end{figure*}

The effect of the AGN feedback on the properties of the host halo at $z \sim
6$ becomes apparent in Figure~\ref{profiles}. We here plot radial profiles of
the gas and stellar density, the mass-weighted temperature and the gas
metallicity for the run without BHs (blue continuous lines) and with BHs (red
dashed  lines). The powerful quasar feedback significantly alters the gas and
stellar properties, especially in the central regions of the halo. With AGN
heating the gas temperature in the inner $20\,h^{-1} {\rm kpc}$ is increased
by almost two orders of magnitude, while the gas density is reduced by a
similar amount. At the same time, the distribution of metals becomes much more
uniform throughout the host halo. The right-hand panel of Figure~\ref{metmaps}
gives a striking visual impression of the effect of AGN feedback on the gas
metallicity distribution, which is shown for a simulation without BHs in the
left-hand panel. During the epoch when the SMBH is growing rapidly the thermal
feedback generates hot gas outflows which eventually shut down the BH accretion,
resulting in self-regulated growth.

\subsection{The influence of different halo threshold 
masses for BH seeding} \label{Seeding}

\begin{figure*}\centerline{ \hbox{
\psfig{file=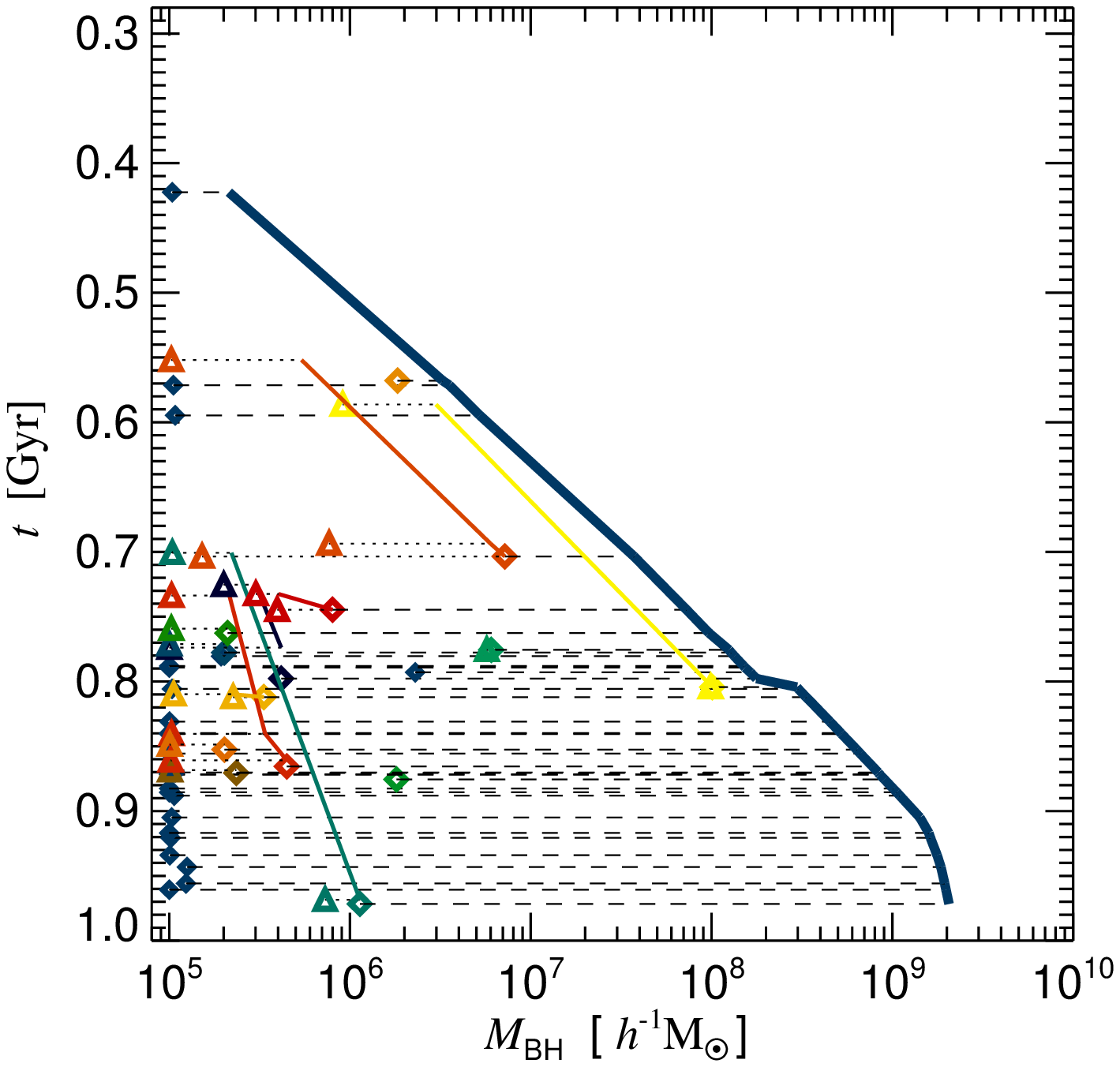,width=8.8truecm,height=8.8truecm}
\psfig{file=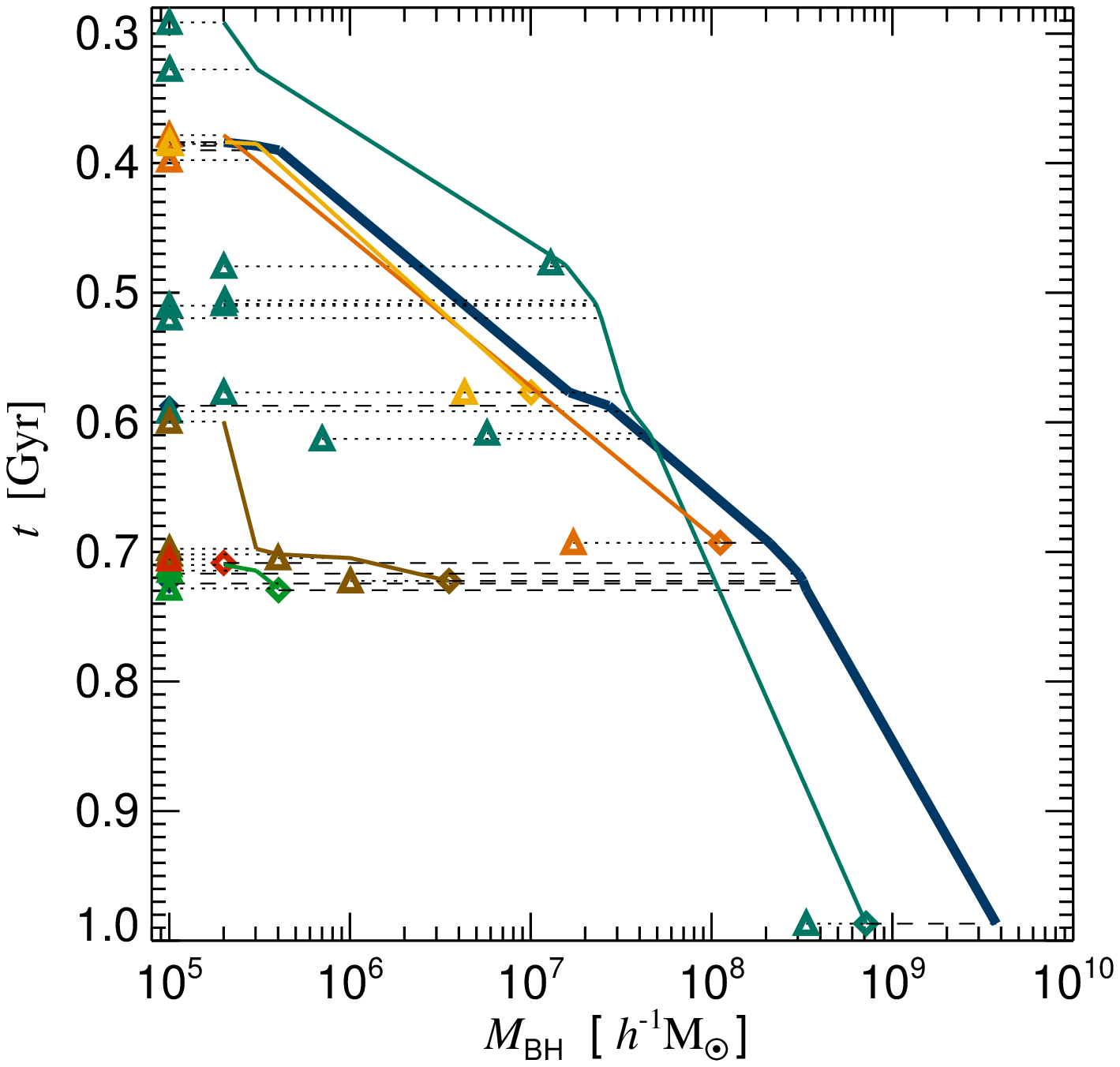,width=8.8truecm,height=8.8truecm}}}
\caption{Merger tree of the most massive BH in the simulation at $z=6$. The
left-hand panel shows the merger tree for a simulation where BHs were seeded
in $10^{10} \,h^{-1} {\rm M}_\odot$ haloes, while the right-hand panel
illustrates the case where the BHs were seeded in smaller haloes of $10^{9}
\,h^{-1} {\rm M}_\odot$. Both runs have been performed with a zoom factor of
$8$. The blue thick line represents the main progenitor of the most massive BH
at $z=6$. The diamond symbols of different colour denote the mass of the
second most massive progenitor, which merges with the main progenitor as
indicated by the dashed lines. Blue coloured diamonds denote BHs that have
never merged before. Thin continuous lines of different colour (matching the
colour of the diamonds) show instead how the second progenitors of our main BH
evolve with time. Finally, the triangles with matching colour indicate BHs
that merge with the second progenitors of our main BH. BH seeding in smaller
haloes causes significant changes in the BH merger history: most notably, BHs
seeded earlier tend to be more massive at a given redshift. This is due to the
combined effect of accretion and mergers.}
\label{mthaloes}
\end{figure*}

\begin{figure*}\centerline{ \hbox{
\psfig{file=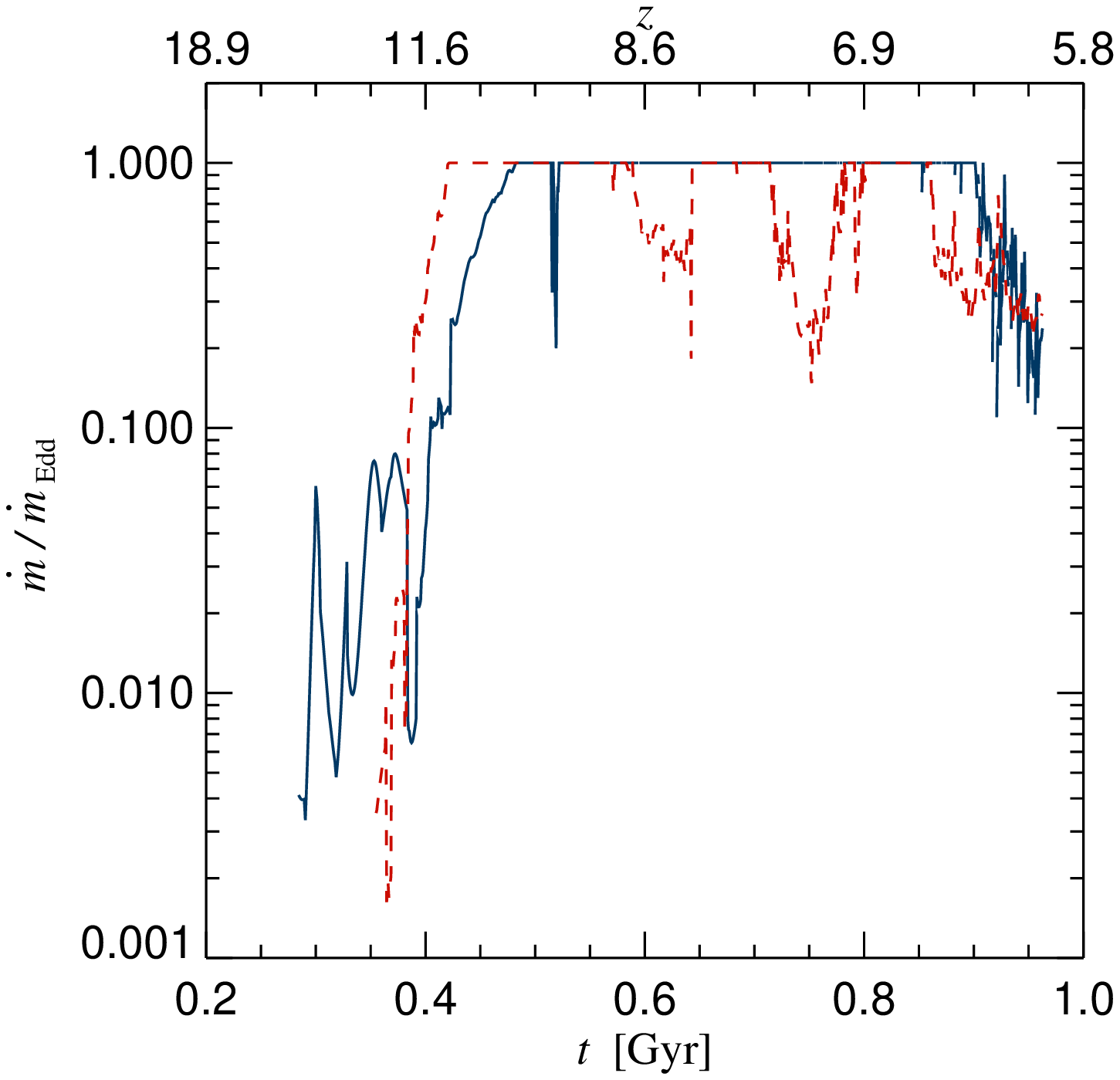,width=8.8truecm,height=8.8truecm}
\psfig{file=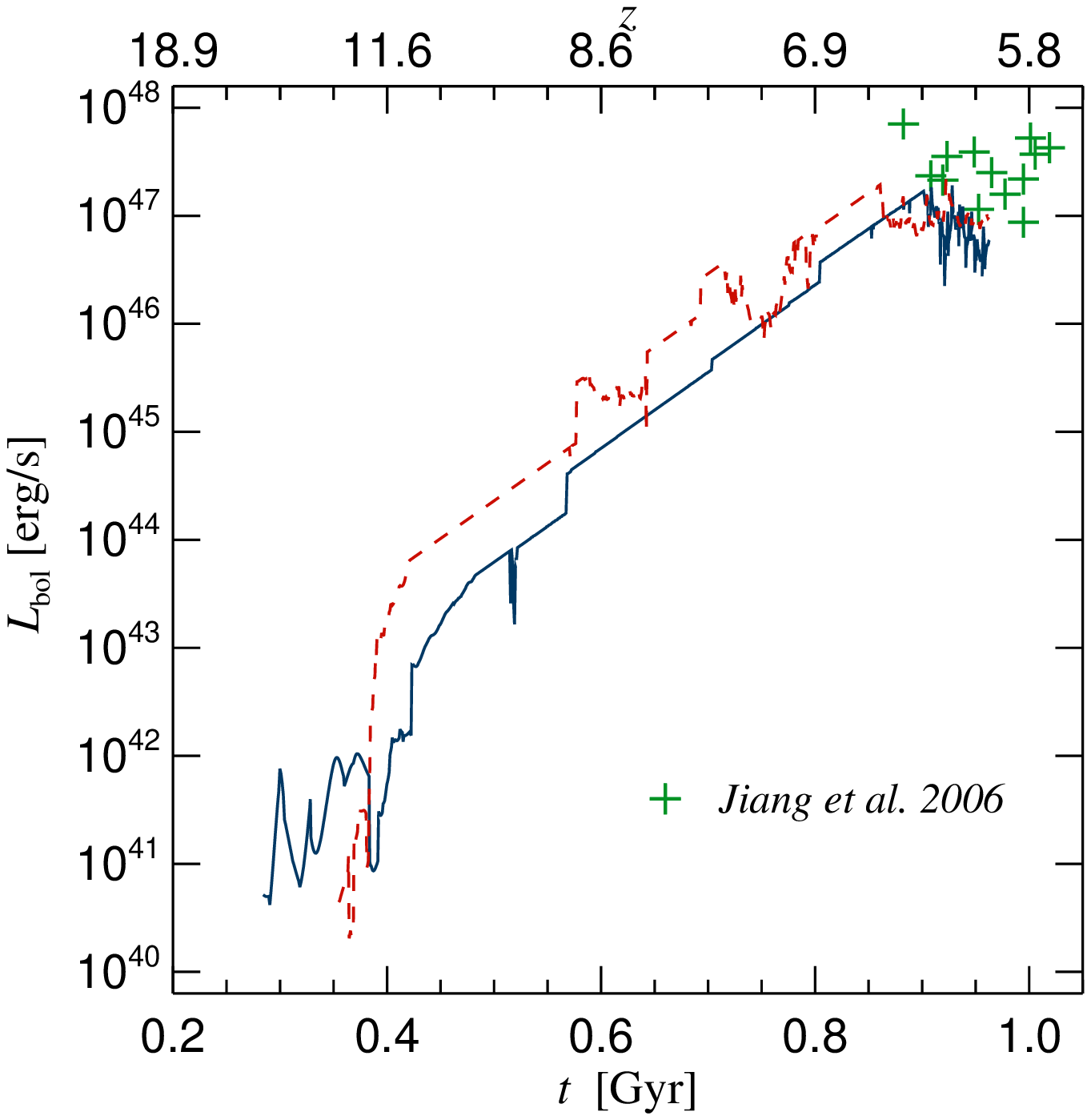,width=8.8truecm,height=8.8truecm}}}
\caption{BH accretion rate in Eddington units (left-hand panel) and bolometric
luminosity (right-hand panel) of the main progenitor of the most massive $z=6$
BH. The blue continuous lines denote the run where BHs were seeded in $10^{10} \,h^{-1}
{\rm M}_\odot$ haloes, while red dashed lines are for the run where this threshold
value has been reduced to $10^{9} \,h^{-1} {\rm M}_\odot$. The green cross
symbols indicate observational estimates of $L_{\rm bol}$ for a sample of high
redshift quasars by \cite{Jiang2006}.}
\label{medd_mainprog}
\end{figure*}

In order to investigate how sensitive the mass growth of the most massive BH
at $z=6$ is to details of the adopted seeding procedure we compare two
particular runs. In one, BHs were seeded within $10^{9} \,h^{-1} {\rm
M}_\odot$ haloes, while in the other the BH seeds were introduced in more
massive haloes of $10^{10} \,h^{-1} {\rm M}_\odot$. All other parameters of
these two simulations were kept  the same. The simulations were run with a
zoom factor of $8$. Obviously, these tests do not cover the whole parameter
space of possible seeding prescriptions. They provide us nevertheless with a
good idea of how much, for example, mergers with other small mass BHs
contribute to the build up of the most massive BH.

In Figure~\ref{mthaloes}, we plot with the thick blue line the evolution of
the main progenitor of the most massive BH at $z=6$ in the run where BHs were
seeded in $10^{10} \,h^{-1} {\rm M}_\odot$ haloes (left-hand panel), and in
the run where instead $10^{9} \,h^{-1} {\rm M}_\odot$ haloes were seeded
(right-hand panel). The symbols and lines with different colours indicate the
merging history of the secondary progenitors that merge with the main
progenitor (see the caption of Figure~\ref{mthaloes} for further explanations).

The main progenitor in the simulation where the BH seeds are placed in smaller
mass haloes is always more massive. At $z=6$, this difference translates into
the SMBH reaching $\sim 3 \times 10^{9} \,h^{-1} {\rm M}_\odot$ instead of
$\sim 2 \times 10^{9} \,h^{-1} {\rm M}_\odot$. In order to better understand
the cause of this mass difference, we have calculated the total mass of all
BHs that have merged onto the main progenitor until $z=6$. This number turns
out to be $\sim 28\%$ and $\sim 6\%$, respectively. This means that in total
the SMBH at $z=6$ has increased its mass by $\sim 7 \times 10^{8} \,h^{-1}
{\rm M}_\odot$ more due to BH mergers and by $\sim 3 \times 10^{8} \,h^{-1}
{\rm M}_\odot$ more due to the gas accretion compared with the SMBH that has
been seeded in the $10^{10} \,h^{-1} {\rm M}_\odot$ halo. It should be
stressed however that the mass difference of $\sim 7 \times 10^{8} \,h^{-1}
{\rm M}_\odot$ due to BH mergers still for the most part comes from gas that
has been accreted by secondary progenitors prior to the merger with the main
progenitor. The total mass of all seeds that eventually end up in the SMBH is
still very small with respect to its final mass, of the order of few
$\%$. This finding therefore supports the important conclusion that
irrespective of the details of our seeding prescription, the main channel of
mass growth of SMBHs is {\it gas accretion}.

In order to get a handle on the  average efficiency of the BH  accretion
luminosity, $\epsilon_{\rm l}$ (or equivalently, on the average Eddington
ratio assuming constant radiative efficiency $\epsilon_{\rm r}$) we adopt the
following equation which characterises the BH growth due to accretion: \be
M_{\rm BH}(t) = M_{\rm BH} (t_0) \,{\rm exp}\bigg( \epsilon_{\rm l} \frac{1 -
\epsilon_{\rm r}}{\epsilon_{\rm r}} \frac{(t - t_0)}{t_{\rm S}}\bigg) \,.
\label{eqnavgaccr}\ee Here $t_0$ and $t$ are the initial and final epochs that
we identify with $0.35$ and $1\,{\rm Gyr}$, respectively, for this analysis,
while $M_{\rm BH}(t_0)$ and $M_{\rm BH}(t)$ are the BH masses at these two
epochs, taken directly from our simulations (for $M_{\rm BH}(t)$ we consider
the total BH mass at epoch $t$ accumulated by accretion only). The
characteristic accretion timescale is given by the Salpeter time, $t_{\rm S}$,
that we assume to be equal to $\sim 0.45\,{\rm Gyr}$. Substituting these
values in equation~(\ref{eqnavgaccr}), we find that the average $\epsilon_l$
value is of the order of $0.8$. This highlights that even though BHs
experience extended episodes of Eddington-limited accretion in our model, it
is still unlikely that BHs can accrete always close to the Eddington rate from
very high redshift to the epoch of $z=6$ quasars, as is often assumed in
simple treatments.

In Figure~\ref{medd_mainprog} we illustrate how the accretion rate (left-hand
panel) and the bolometric luminosity (right-hand panel) of our main progenitor
evolve with time in the simulations where BHs were seeded in $10^{9} \,h^{-1}
{\rm M}_\odot$ haloes (red dashed lines) and $10^{10} \,h^{-1} {\rm M}_\odot$
haloes (blue continuous lines). At $t < 0.4\,{\rm Gyr}$, the BH's main
progenitor in the run with the lower halo threshold mass first undergoes three
mergers with other BH seeds. Afterwards it starts accreting rapidly, soon
reaching the Eddington limit. This leads to a head-start in the assembly of
the main progenitor with respect to the simulation with the higher value of
the halo mass threshold for BH seeding. For $t > 0.5\,{\rm Gyr}$, the BH which
started rapid growth earlier (red dashed lines) experiences several extended
sub-Eddington accretion episodes compared with the other run (blue continuous
lines). However, in absolute terms its accretion rate is higher (as reflected
by the bolometric luminosity). These sub-Eddington accretion episodes are
related to a temporary exhaustion of the local gas reservoir available for
accretion.

Finally, by $z \sim 6$, the most massive BHs in the two runs performed are
characterized by similar bolometric luminosities that we compare to
observations (indicated by the green crosses). The observed values are taken
from a recent paper by \cite{Jiang2006}, who have performed Spitzer
observations of 13 high redshift quasars and combined those results with
observations ranging from X-ray to radio to get a more realistic estimate of
the bolometric luminosities. Having simulated only one $z=6$ quasar we cannot
perform a detailed statistical comparison with observations; however, while
our bolometric luminosity $L_{\rm bol}$ lies at the lower end of the observed
range (note that observational error bars were not available), it is clearly
consistent with the observed values.

\subsection{Can galactic winds stall the initial growth of BHs?}

\begin{figure}   
\psfig{file=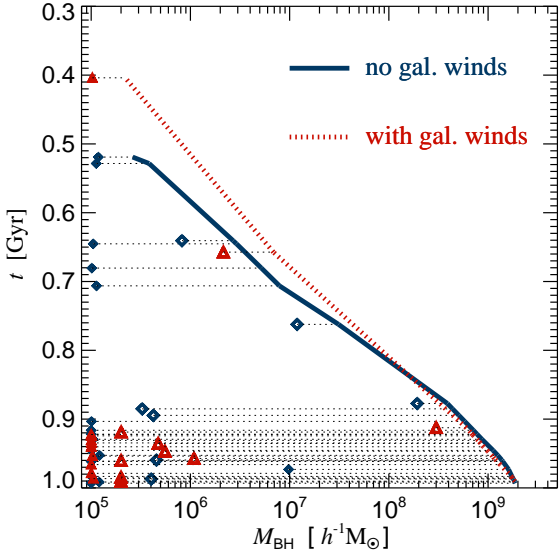,width=8.truecm,height=8.truecm}
\caption{Merger tree of the most massive BH at $z=6$ in simulations with (red
dashed line and triangles) and without (blue continuous line and diamonds)
galactic winds. Both simulations have been performed with a zoom factor of
$5$. The continuous lines show how the mass of the BH's first progenitor grows
with time while the symbols represent secondary progenitors that merge onto
it, as indicated by the dotted lines. Galactic winds have no significant
impact on the merger tree of the most massive BH throughout the whole
simulated time span.}
\label{mtwinds}
\end{figure}

Given that gas accretion drives the growth of massive BHs at $z=6$, it is
interesting to consider possible physical mechanisms that we have not taken
into account so far and that could in principle impede BH accretion and thus
stall BH growth. Note that the BH feedback itself, which is already part of
our default model, does not prevent BHs from becoming supermassive. However,
the situation may be different if the BH feedback effects are in reality much
stronger than what we have assumed in our simulations thus far, a possibility
that we examine further in Section~\ref{Spinning_quasars}, when we discuss
rapidly spinning BHs. Another possibility is stronger feedback associated
with star formation, which could also deprive BHs of the gas for accretion. We
here examine galactic winds from star formation as one possibility of such
strong feedback.

There is good observational evidence that many host galaxies of SDSS quasars
experience a very high level of star formation of the order of $\sim 1000\,
{\rm M}_\odot \,{\rm yr}^{-1}$ \citep[e.g.][]{Bertoldi2003a, Carilli2004,
Wang2008}. Starburst-driven galactic winds are therefore an obvious physical
process that could significantly deplete the amount of gas available for
accretion onto massive BHs in these systems. In order to explore such a
scenario, we have performed an additional test run (selecting zoom factor $5$
this time) where we have `switched-on' galactic winds as implemented by
\cite{SpringelH2003}. The properties of the galactic outflows are determined
by two parameters. The first is the wind velocity, which is constant regardless of
the host galaxies' mass. We select it to be rather high of order of $\sim 480
\,{\rm km}\,{\rm s}^{-1}$. The second parameter regulates the mass loading
factor of the wind, according to $\dot M_{\rm w} \, = \, \eta \dot M_{\rm
SFR}$, where we choose $\eta$ to be $2$.

In Figure~\ref{mtwinds}, we show the redshift evolution of the mass of the
most massive progenitor of the $z=6$ target BH, both without (blue continuous line) and
with galactic winds (red dashed line). Note that apart from the inclusion of
galactic winds, all other simulation parameters and the numerical resolution
were kept exactly the same in both simulations. Figure~\ref{mtwinds}
demonstrates that galactic winds have overall very little effect on the mass
growth of the most massive BH. The rather limited efficiency of galactic winds
to expel gas from the BH host halo (or to change gas thermodynamical
properties significantly) is due to the combination of two factors.
First, even though the chosen wind velocity from supernova feedback is rather
high, at $z \sim 6 - 8$ it is comparable or even lower than the escape
velocity of the host halo of the most massive BH, and thus galactic winds
cannot unbind the gas from the halo. Second, regardless of the high wind
efficiency parameter $\eta$ we have adopted, the amount of star formation
occurring in the most massive BH host halo is not particularly high at high
redshift. It only reaches the intense levels of $\sim 1000 {\rm M}_\odot \,
{\rm yr}^{-1}$ at $z < 9$, when the BH host halo has become sufficiently
massive to prevent galactic winds from leaving the halo.

This result is in line with the findings of \cite{Li2007}, who used the same
wind model with similar wind speed (but lower wind efficiency parameter
combined with higher SFRs of their simulated galaxies at high $z$). Hence, in
order to prevent or considerably reduce massive BH growth at these high
redshifts by means of starburst driven winds one would need to invoke a much
more extreme kinetic feedback. Either the wind velocity would need to be
significantly higher, e.g.~$\sim 1000 {\rm km}\,{\rm s}^{-1}$ (possibly
combined with even higher wind efficiencies and/or higher SFRs of the host
haloes). Or the nature of the wind would need to be more violent than assumed
in our model, with much larger mass-loading factors that blow away a
considerable fraction of the ISM around massive BHs. In any case the final
wind velocities would need to be high in order to prevent that the blown away
material falls back and feeds the BH after some delay. In other words, only a
substantially larger kinetic luminosity of the winds may change the above
conclusion.

\subsection{Gravitational wave recoils and the growth of SMBHs at high z}

Asymmetric gravitational wave emission during a BH merger carries away linear
momentum, which in turn imparts a velocity kick on the remnant BH. Given that
such induced BH kick velocities can have significant values amounting to
thousands of kilometres per second, it is extremely interesting to consider
their possible impact on the BH assembly in cosmological simulations, where we
have direct knowledge when BHs are merging, which properties they have at the
time they merge, and which haloes they are embedded in. If gravitational wave
recoils frequently displace or expel BHs from the centres of their host
haloes, this may have dire consequences for the build up of massive BHs at
high redshift, possibly even obstructing the formation of SMBHs by $z=6$ in
the co-evolution scenario that we study.

It is one of the important goals of this study to better understand this
question. For this purpose, we have extended our BH model and incorporated
three scenarios for BH kicks, as described earlier. These can be labelled as
the `mass asymmetry', the `parallel spins' and the `random spins'
parametrization of the kicks (see Section~\ref{Method_recoils} for the
detailed definitions).

Given that the most interesting range of kick velocities is obtained when BHs
are spinning, we first assume, for definitiveness, that all BHs are
characterized by the same initial spin parameter. We keep the BH spins
constant with cosmic time and do not change the radiative efficiency as a
function of the spin, in order to gauge more straightforwardly the impact of
BH recoils on the BH assembly. For spin-dependent radiative efficiencies, see
Section~\ref{Spinning_quasars} below, while for the evolution of BH spins due
to the mergers, see Section~\ref{Spin_evolution}. As for the magnitude of the
BH spins, we fix it at a rather high value of 0.9, but we also consider lower
values of 0.3 and 0.5 to explore the parameter space. Obviously, in the case
of mass-asymmetry kicks, we assume that the BHs are spinless (as in our
default model) and we do not change the spin magnitudes after a merger (as
should be appropriate). This case serves as a lower limit for the effect of BH
recoils.

For our three different scenarios for BH remnant recoils, we have performed a
number of resimulations of our target $z=6$ halo. In particular, we have both
considered the case when the halo mass threshold value for BH seeding is
$10^{10} \,h^{-1} {\rm M}_\odot$ and when it is $10^{9} \,h^{-1} {\rm
M}_\odot$. In the latter case, more seed BHs are introduced and hence more
mergers with the main progenitor of the most massive BH occur (there are also
more BH mergers overall), making this case more favourable for effects due to
BH kicks. Nevertheless, regardless of the seeding prescription adopted, we
find that BH kicks {\it cannot} prevent the build-up of the most massive BH,
which reaches a mass of a few times $10^{9} \,h^{-1} {\rm M}_\odot$ even when
we assume that all BHs are rapidly spinning (with a spin value of 0.9), and
irrespective of the assumed orientation of BH spins at the merger. This is
illustrated in the right-hand panel of Figure~\ref{mbh_spin}, where for the
case of randomly oriented spins with $a=0.9$ we plot the mass of the most
massive BH formed in the simulated volume (magenta star symbols connected with
dashed line).

\begin{figure*}\centerline{ \hbox{
\psfig{file=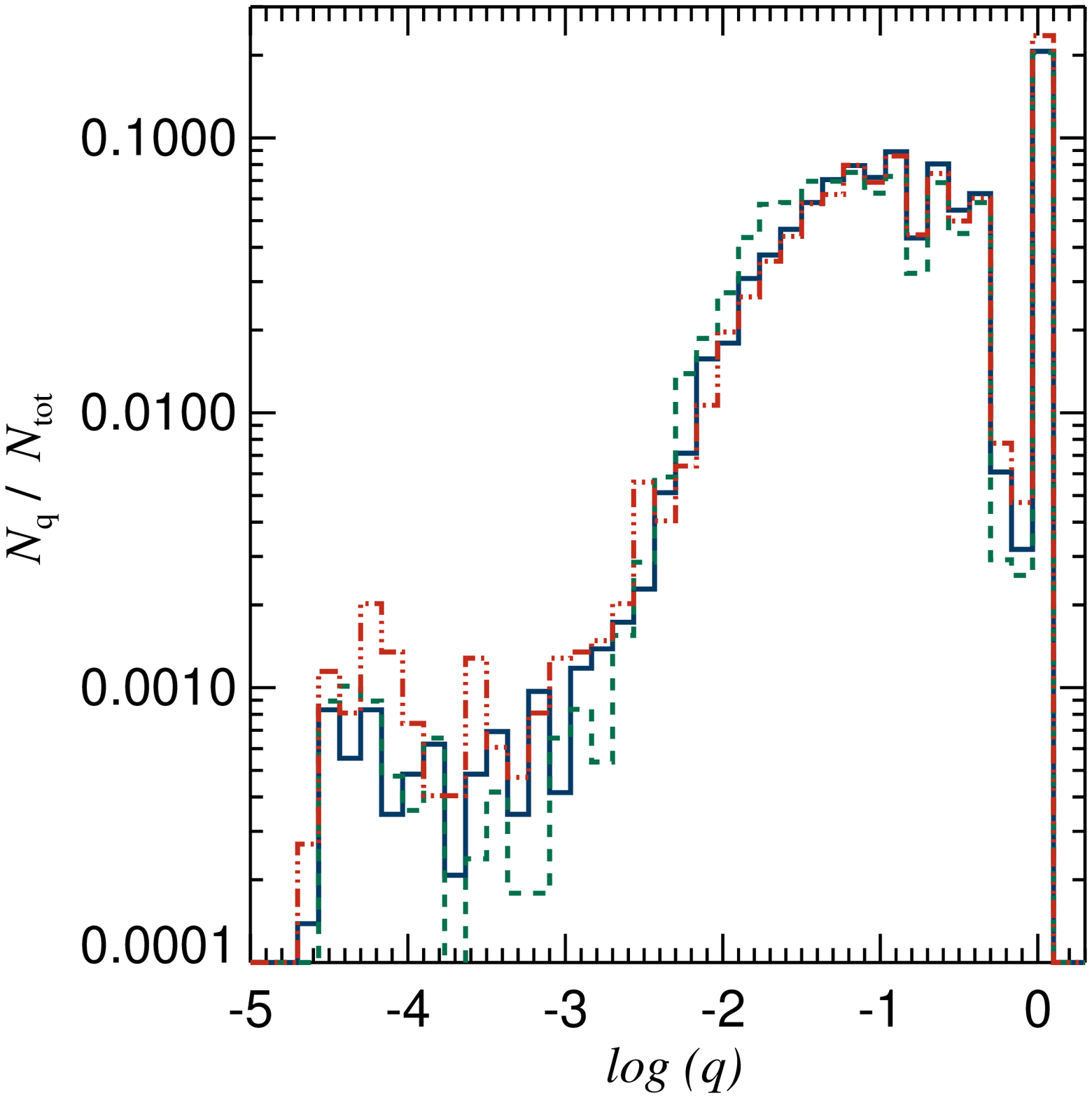,width=6.3truecm,height=6.2truecm}
\hspace{-0.5truecm}
\psfig{file=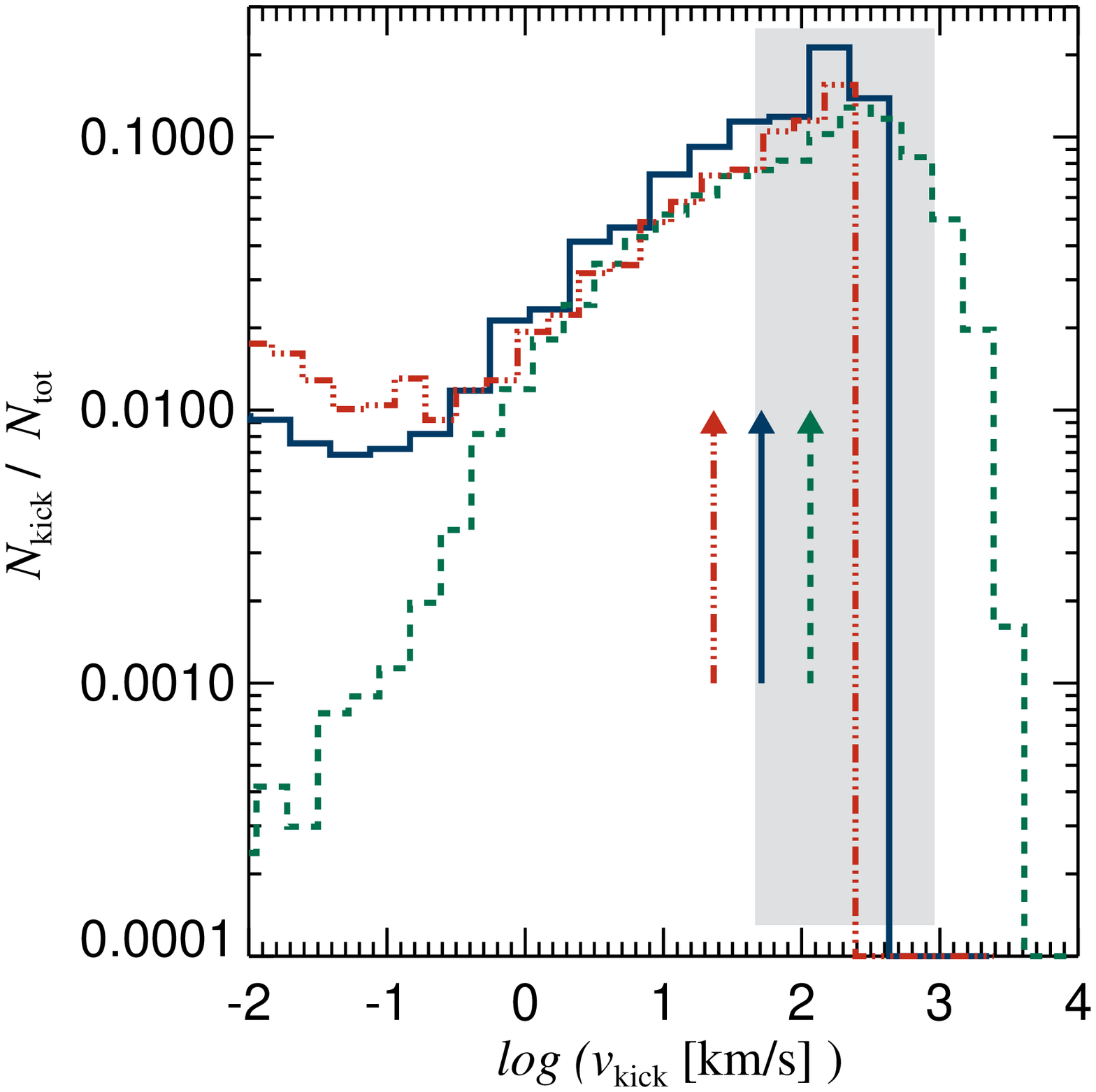,width=6.3truecm,height=6.2truecm}
\hspace{-0.5truecm}
\psfig{file=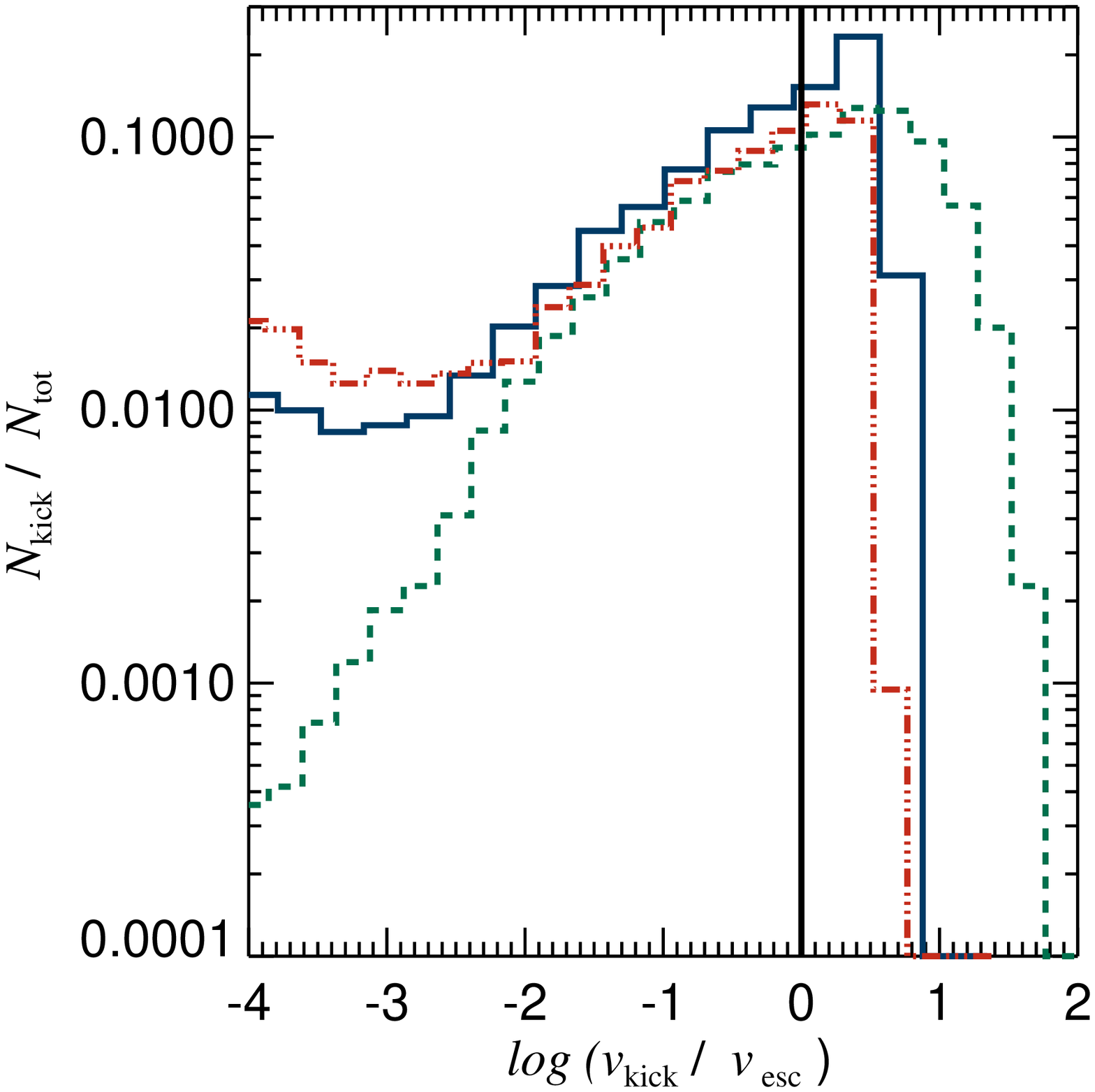,width=6.3truecm,height=6.2truecm}}}
\caption{Distribution of mass ratios (left-hand panel), kick velocities
(middle panel) and kick velocities divided by host halo escape velocities
(right-hand panel) for all BH mergers at $z \ge 6$ for simulations with the
three implementations of gravitational wave recoils. The red dot-dashed lines
are for simulation where recoil kicks are only due to the mass asymmetry of
merging BHs. The blue continuous lines are for the run where BH spins are
always aligned or anti-aligned, and the spin magnitude is kept fixed at
$0.5$. The green dashed lines represent the case where the spin orientations
are random and the spin magnitude is $0.9$. In all three cases BHs are seeded
in haloes with masses above  $10^{9} \,h^{-1} {\rm M}_\odot$. The gray region
in the middle panel denotes the range of escape velocities of haloes which
host merging BHs. The vertical arrows with matching colours and line styles
show the median value of the kick velocity distributions for the three cases
considered.}
\label{kick_dist}
\end{figure*}

In order to understand why BH kicks cannot stall the growth of massive BHs in
our model we have computed a number of BH properties at the moment when they
are about to undergo a merger. In Figure~\ref{kick_dist}, we show in the
left-hand panel the distribution of mass ratios of merging BH pairs over the
whole simulated timespan (from $z \sim 15$ when the first BHs are formed to $z
\sim 6$). Three different histograms (red dot-dashed lines: mass asymmetry
kicks,  blue continuous lines: parallel spins, and green dashed lines: random
spins) illustrate the three scenarios we have adopted for BH recoils. The
isolated peak to the right in the distribution corresponds to equal mass
mergers and is mainly due to BHs close to their seed mass merging with each
other. Apart from this feature, the distribution of mass ratios is broad and
peaks at the value of $1:10$, with mass ratios of $0.07$ being as common as
ratios of $0.2$.

In the middle panel of Figure~\ref{kick_dist}, we instead show the
distribution of kick velocities of remnant BHs. The vertical arrows with
matching colours and line styles denote the median values of the kick velocity
distributions for the three cases considered. While for all three scenarios
the distribution of kick velocities peaks at around $150-250\,{\rm
km\,s^{-1}}$, the maximum kick velocity is  smallest for the mass asymmetry
induced kicks, and largest for randomly oriented spins, as expected. The tail
of the distribution at low kick velocities is more pronounced in the case of
mass asymmetry kicks and for parallel spins, than for the run with random
spins.  This can be understood as follows. For the mass asymmetry kicks, the
kick velocity will be very low both when the mass ratio is close to one and
when it is very small, and will hence contribute to the tail of the
distribution. Instead, in the case of parallel spins, given that the two BH
spins can be aligned or anti-aligned with equal probability (recall that we
are choosing their alignment with respect to the orbital angular momentum
randomly), the contribution to the kick velocity will be zero when the
alignment of the spins occurs. On the other hand, for the case of random
spins, it is unlikely that the spins become mutually aligned and cause no
kick, as we do not impose any constraints on the spin orientation in this
case. The variety of possible spin orientations in this case causes kick
velocities that are higher on average. Finally, the shaded region in the
middle panel of Figure~\ref{kick_dist} indicates the range of escape
velocities from haloes when a BH merger occurs. This shows that in a number of
mergers the BHs actually get kicked out from their host halo.

In order to better constrain the fraction of kicked out BHs, we show the
distribution of the ratio of kick velocity to escape velocity from the host
halo at the moment of the merger in the right-hand panel of
Figure~\ref{kick_dist}. All the mergers that populate the distribution to the
right of the vertical black line can cause the remnant BH to get
expelled. Thus, in all three kick scenarios explored here, there are BHs which
experience kick-outs from their host halo, and the fraction of these events
varies from $22\%$ for mass asymmetry kicks, over $31\%$ for parallel
spin recoils, up to $36\%$ for the case of randomly oriented spins. Therefore,
the fraction of BHs kicked out from their host haloes is significant, ranging
from about $20\%$ if the BHs are not rapidly spinning to almost $40\%$ if we
assume that the whole population of BHs is instead characterized by a rather
large spin value of $0.9$.

\begin{figure*}\centerline{ \vbox{\hbox{
\psfig{file=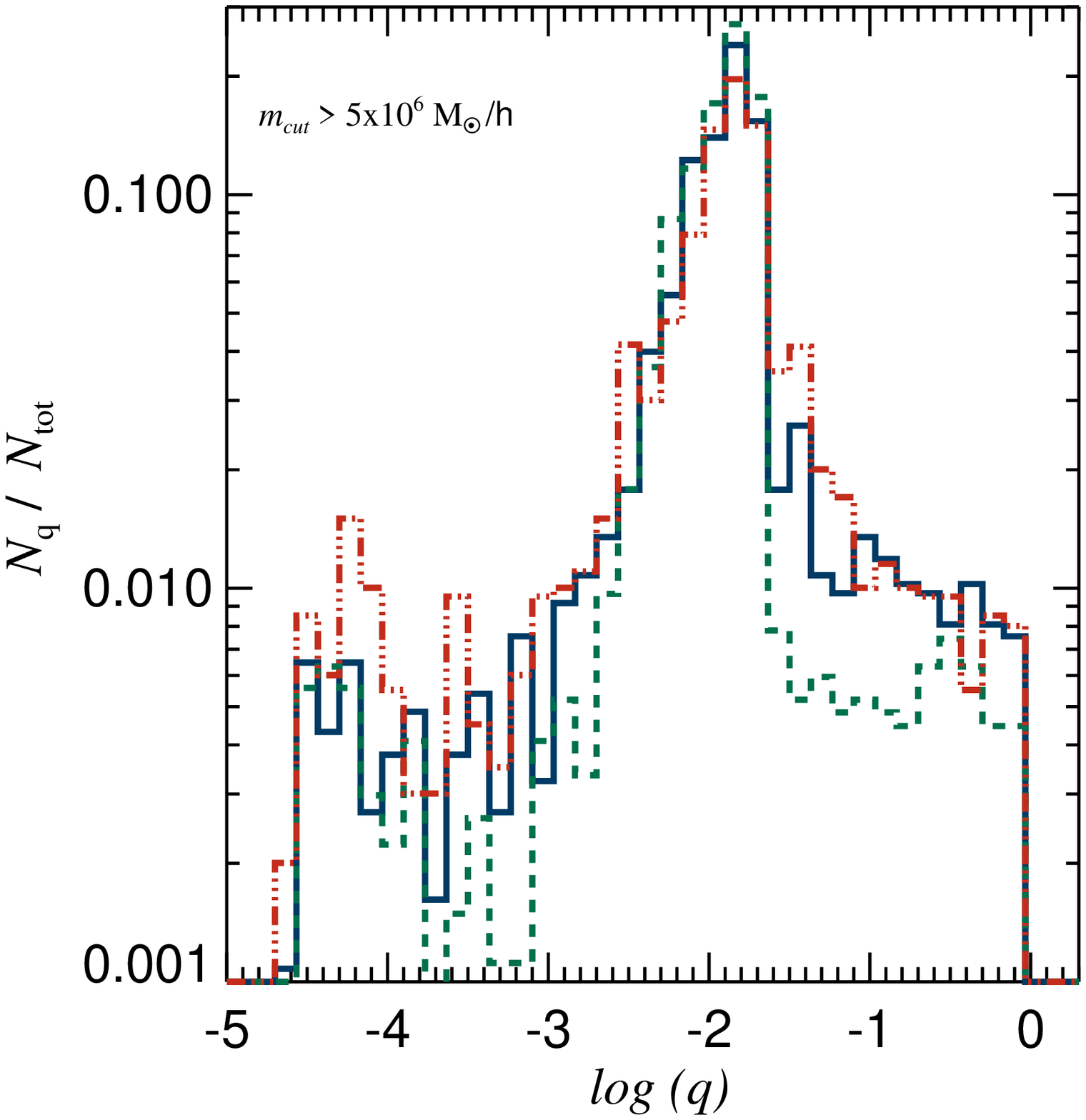,width=7.3truecm,height=7.truecm}
\psfig{file=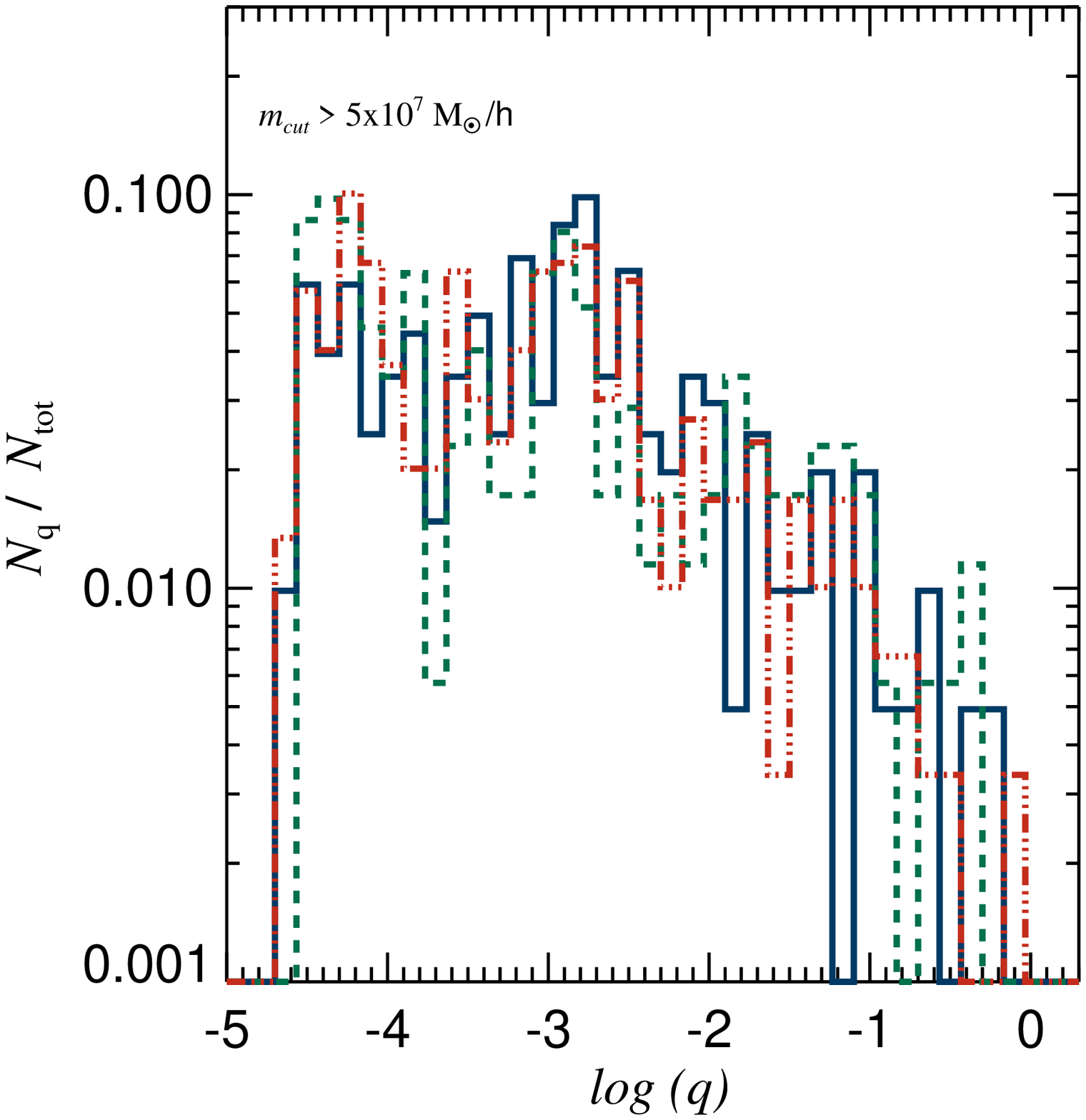,width=7.3truecm,height=7.truecm}}
\hbox{
\psfig{file=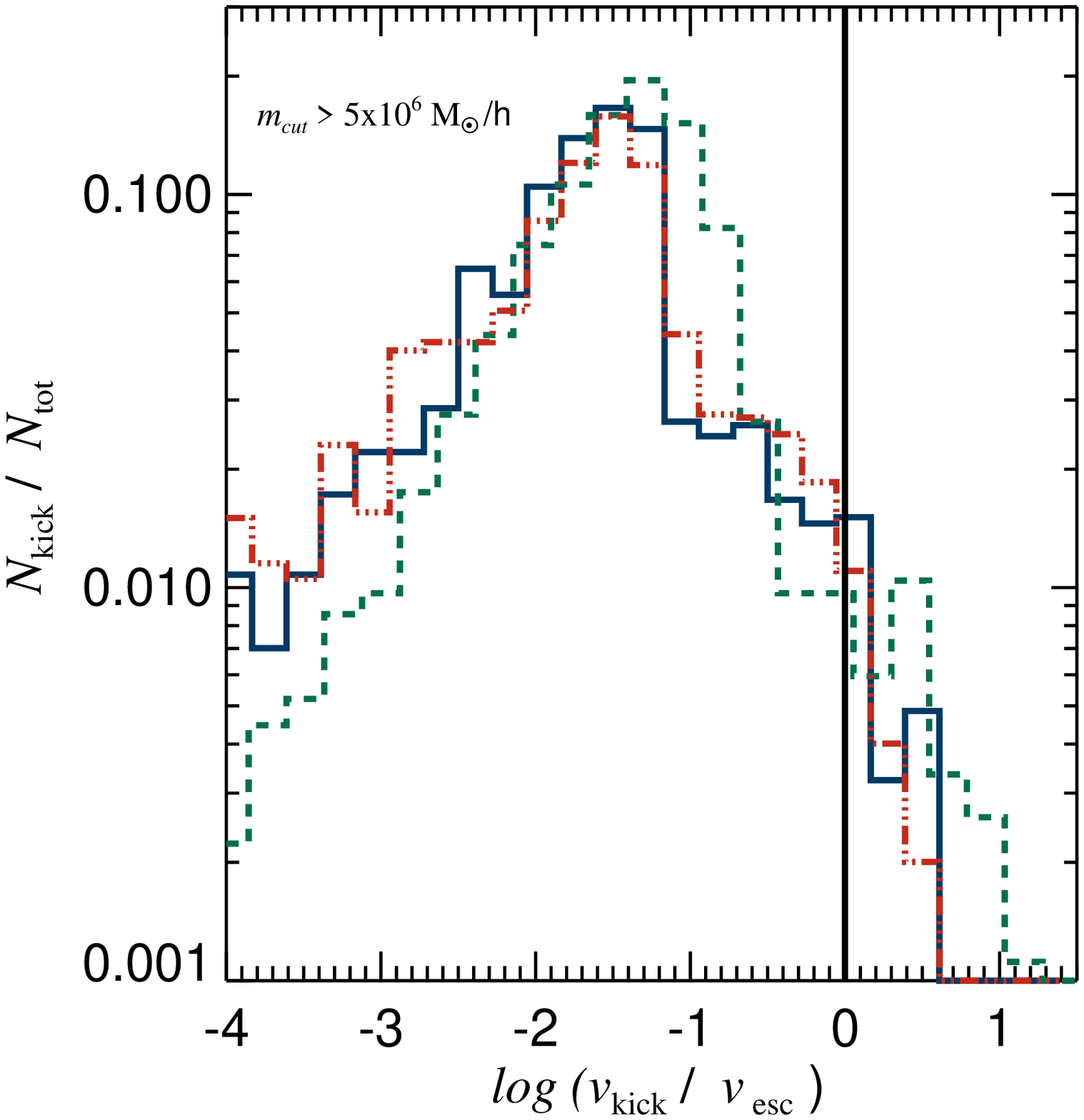,width=7.3truecm,height=7.truecm}
\psfig{file=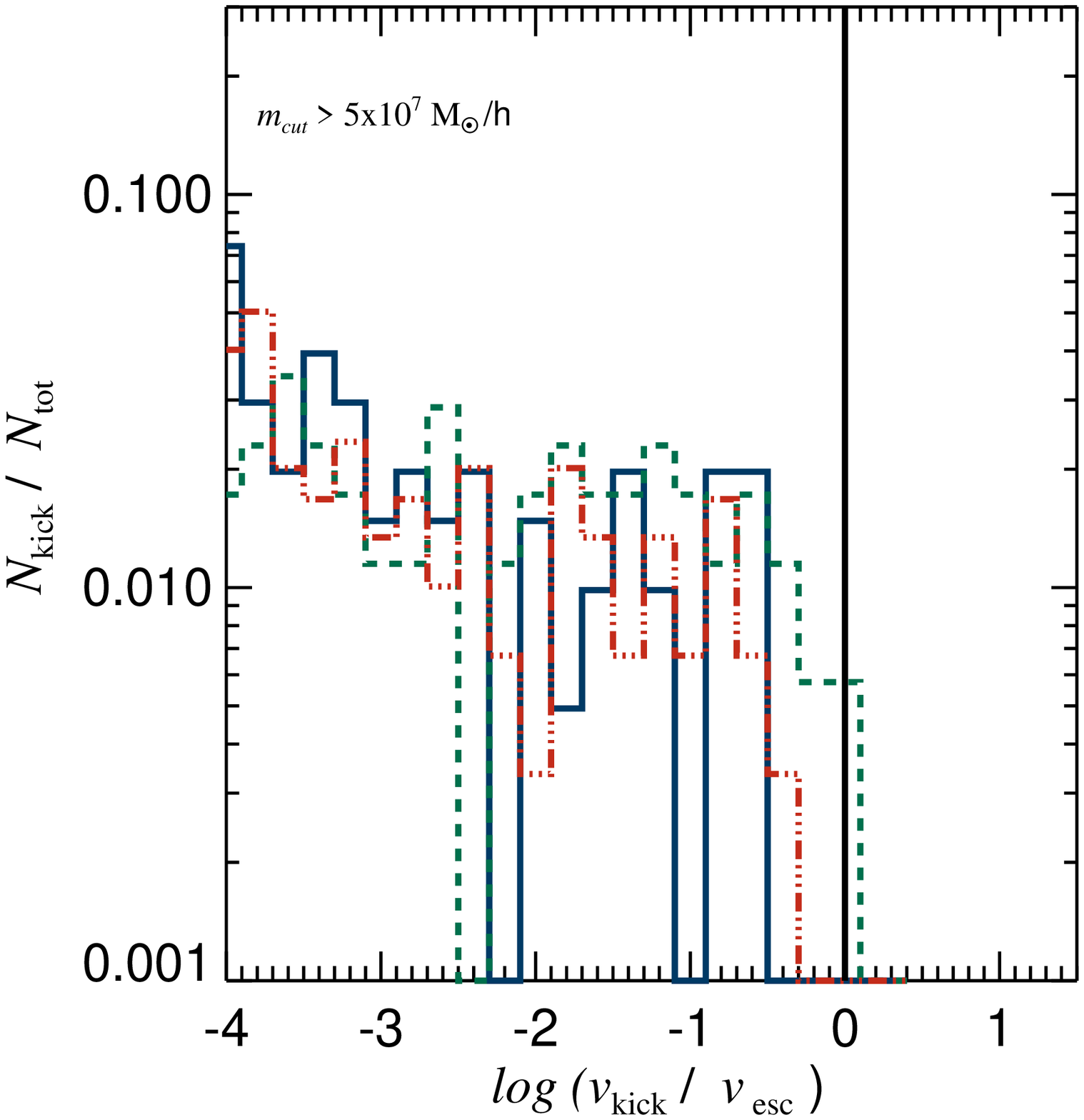,width=7.3truecm,height=7.truecm}}}}
\caption{Distribution of mass ratios (upper panels) and kick divided by escape
velocities (lower panels) for the same simulations as in
Figure~\ref{kick_dist}, using the same colour-coding. However, here the
distributions have been computed by considering that at least one of the two
merging BHs has to be more massive than $5 \times 10^{6} \,h^{-1} {\rm
M}_\odot$ (left-hand panel) and $5 \times 10^{7} \,h^{-1} {\rm M}_\odot$
(right-hand panel), respectively.}
  \label{kick_dist1}
\end{figure*}

However, the majority of these BHs that are expelled turn out to be fairly
small mass BHs. To demonstrate this, in Figure~\ref{kick_dist1} we show again
the distribution of mass ratios (top panels) and kick over escape velocities
(bottom panels), but this time imposing that at least one of the two merging
BHs needs to be more massive than $5 \times 10^{6} \,h^{-1} {\rm M}_\odot$
(left-hand panels) or $5 \times 10^{7} \,h^{-1} {\rm  M}_\odot$ (right-hand
panels). From these panels we can deduce that more massive BHs are less likely
to undergo a merger with a similar mass BH. While in the case of BHs more
massive than $5 \times 10^{6} \,h^{-1} {\rm M}_\odot$ there are still some
mergers that can cause the remnant to be kicked out of the host halo, for BHs
more massive then $5 \times 10^{7} \,h^{-1} {\rm  M}_\odot$ virtually no BH
remnant has a sufficient kick velocity to escape from its host halo. Three
factors contribute to this: similar mass mergers are quite rare, the
probability of having BH spins oriented such that high enough kicks are
produced is low, and as BHs grow in mass so do their host haloes and
consequently the escape velocity increases. Combined, these effects imply that
gravitational wave induced BH recoils {\it cannot} prevent the formation of
SMBHs at high redshifts. This result is in line with our previous finding that
most of the SMBH mass at $z=6$ is assembled by gas accretion, and that BH
mergers contribute only in a minor way to the hole's formation.

Nonetheless, we should stress that an important caveat for this result lies in
our seeding prescription. In scenarios where BH seeds are present also in much
smaller haloes than we have assumed and/or where BH seeding occurs at even
higher redshifts than $z \sim 15$, it is likely that BH mergers will
contribute much more to the mass assembly of early BHs. If in such scenarios
BHs happen to be sufficiently spinning, gravitational wave recoils could have
a much more damaging effect than we find here. Still, further direct modelling
by means of numerical simulations is necessary to establish if this would
indeed lead to a bottleneck for the early formation of SMBHs. Interestingly,
if it can be shown that this should really be the case, then the fact that
SMBHs {\em are observed} at $z=6$ can be used to argue against such scenarios
and be taken as indirect evidence that BHs at high redshift should have modest
values of BH spins and/or that the majority of BH seeds should be formed in
the centres of relatively massive dark matter hosts, as we have assumed in
this work.

\subsection{Spinning BHs with high radiative efficiency} \label{Spinning_quasars}

\begin{figure*}\centerline{ \hbox{
\psfig{file=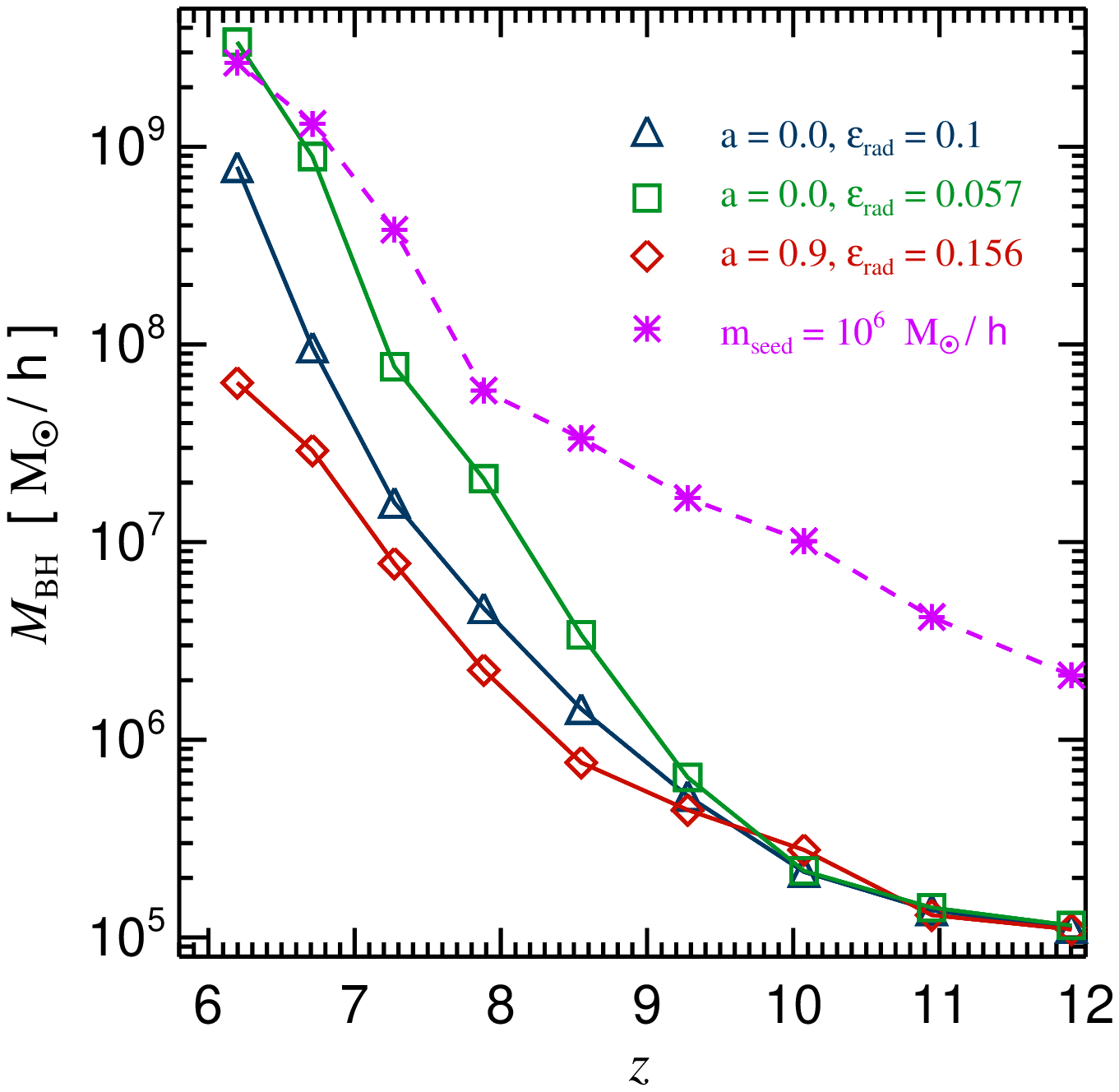,width=8.truecm,height=8.truecm}
\psfig{file=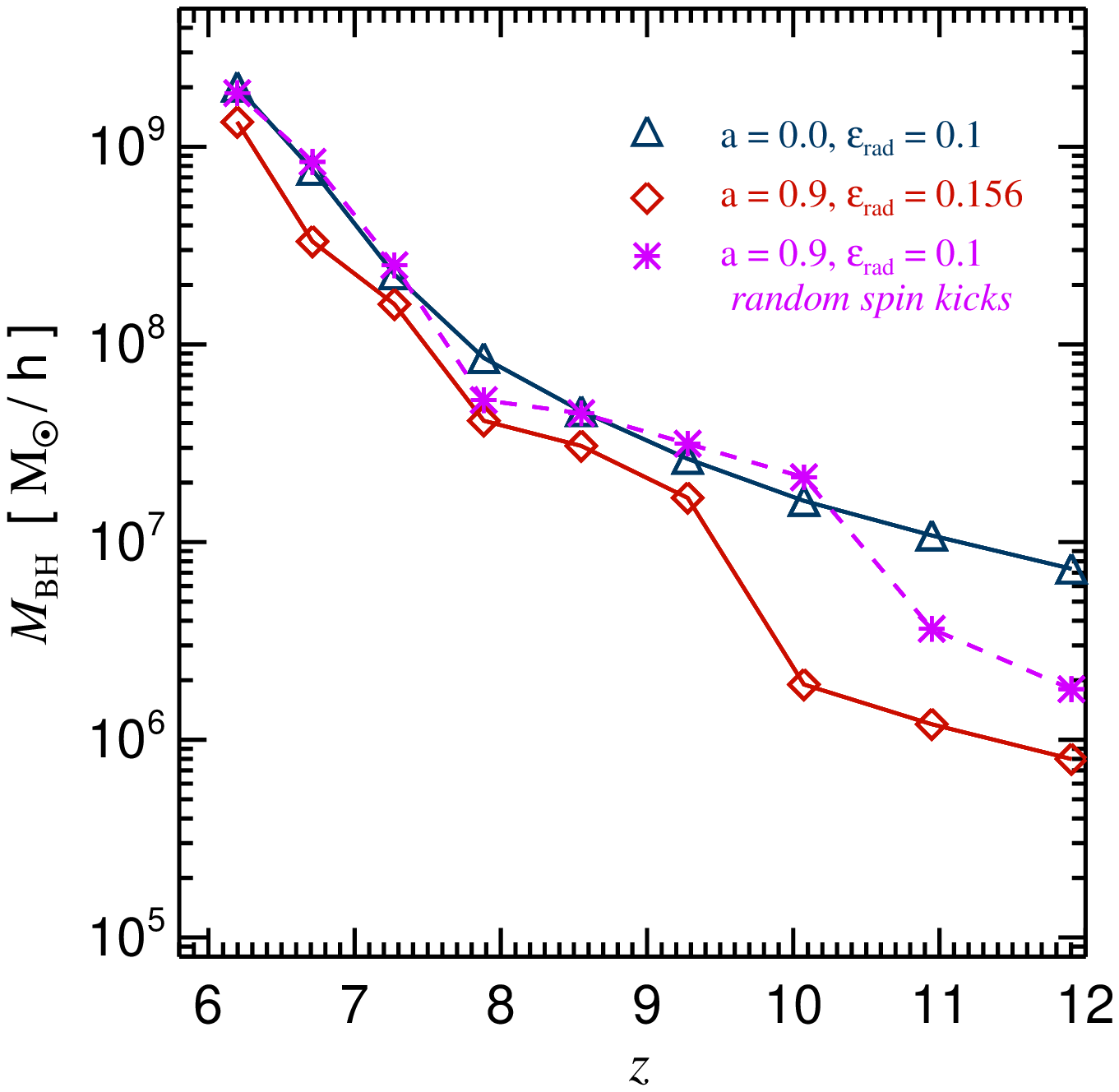,width=8.truecm,height=8.truecm}}}
\caption{Mass of the most massive BH in the simulated volume as a function of
redshift in the runs where BHs are seeded in $\sim 10^{10} \,h^{-1} {\rm
M}_\odot$ haloes (left-hand panel) or in $\sim 10^{9} \,h^{-1} {\rm M}_\odot$
haloes (right-hand panel). Blue triangle symbols denote the BH mass obtained
adopting our default model, where BHs are spinless and the radiative
efficiency is $0.1$. Green square symbols are for a run where BHs are spinless
as well, but the adopted radiative efficiency is $0.057$ (calculated from
equation~\ref{Bardeen}). Red diamond symbols represent the result in the case
where all BHs are rapidly spinning, characterized by a spin of $0.9$ and a
radiative efficiency computed according to equation~(\ref{Bardeen}). Finally,
magenta star symbols connected with dashed lines in the left-hand panel denote
the result of a simulation performed with our default BH model where initial
seeds were assumed to be more massive, having $10^{6} \,h^{-1} {\rm M}_\odot$.
The magenta symbols in the right panel show the case where gravitational wave
recoils with random spin orientations were switched-on.}
  \label{mbh_spin}
\end{figure*}

Above we have shown that BH recoils cannot stall the growth of the most
massive BHs in our simulations, even when they are characterized by a rather
large spin value. However, we have not yet taken into account that the radiative
efficiency is spin dependent. The spin dependent radiative efficiencies can
modify both the amount of material that a hole can accrete (here we assume
that the BH accretion is radiatively efficient, and not advection dominated)
and thus also the amount of thermal feedback. Highly spinning BHs are expected
to have radiative efficiencies approaching $0.42$ \citep{Bardeen1972,
Thorne1974}, and if indeed such a large fraction of the accreted rest mass
energy is lost to radiation this can have a major impact onto the mass
assembly of BHs.

In order to explore this possibility we have performed additional simulations
where we kept the spins of all BHs constant, but where we have computed the
radiative efficiencies from equation~(\ref{Bardeen}). In
Figure~\ref{mbh_spin}, we show the mass of the most massive BH in the
simulated volume as a function of redshift in the case where BHs where seeded
in $10^{10} \,h^{-1} {\rm M}_\odot$ haloes (left-hand panel) and in $10^{9}
\,h^{-1} {\rm M}_\odot$ haloes (right-hand panel). Blue connected triangles
denote the BH mass obtained from our default model. Green connected squares
(only in the left-hand panel) are instead for the run where the BHs are
spinless (as in our default model), but the radiative efficiency assumed is
somewhat lower than in our default model (given by
equation~\ref{Bardeen}). Finally, the red connected diamonds illustrate the
case where all BHs are rapidly spinning, with a spin of $0.9$, and the
radiative efficiency is $0.156$. From Figure~\ref{mbh_spin} it is evident that
for the rapidly spinning BHs it is much more difficult to become supermassive
by $z=6$. However, as shown in the right-hand panel of Figure~\ref{mbh_spin},
this problem is somewhat alleviated if BH mergers contribute more
significantly to the build-up of the SMBH at high redshifts, as is the case
when we seed smaller mass haloes with BHs.

It is interesting to ask whether the BH mass growth by accretion in the case
of high radiative efficiencies is reduced because a larger fraction of the
accreted mass goes into radiation combined with the fact that the Eddington
limit is lower, or because the BH thermal feedback is more powerful and tends
to shut off the accretion flow. We find that the main reason for the reduced
BH growth is the lower Eddington limit in the case of large radiative
efficiencies. We therefore conclude that the BHs seeded at $z \sim 12-15$,
which grow primarily in isolation and are characterized by large radiative
efficiencies, {\it cannot} easily become supermassive by $z=6$, unless they
accrete in a super-Eddington fashion.

Finally, in the left-hand panel of Figure~\ref{mbh_spin} we also show the mass
of the most massive BH in the simulated volume evolved with our default BH
model, but starting from somewhat larger initial seeds of $10^{6} \,h^{-1}
{\rm M}_\odot$ which were placed in all haloes above $10^{10} \,h^{-1} {\rm
M}_\odot$ (magenta star symbols connected with a dashed line). Whereas for $z
> 7$ the most massive BH in this run is always more massive than the
corresponding BH which grows from a $10^{5} \,h^{-1} {\rm M}_\odot$ seed, at
$z \sim 6$ -- due to the self-regulated feedback -- its final mass is
comparable. In fact, its mass evolution is rather similar to the mass
evolution of the most massive BH which grows from a $10^{5} \,h^{-1} {\rm
M}_\odot$ seed, but has been seeded in a $10^{9} \,h^{-1} {\rm M}_\odot$ halo,
as shown in the right-hand panel of Figure~\ref{mbh_spin}.

\subsection{BH spin evolution due to BH mergers} \label{Spin_evolution}

\begin{figure}\centerline{
\psfig{file=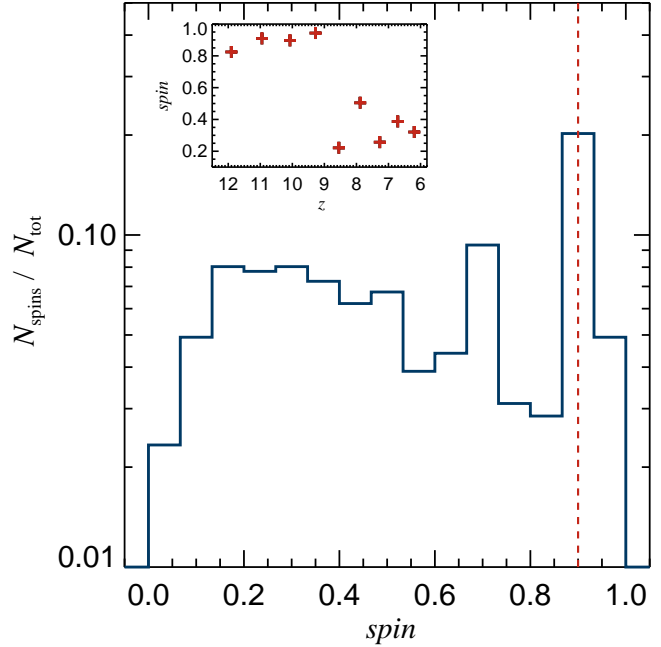,width=9.5truecm,height=9.5truecm}}
\caption{Distribution of BH spins at $z=6.2$ in the simulation where the
initial BH spin is $0.9$ and the spins change due to BH mergers. The inset
shows the spin of the most massive BH in the simulated volume as a function of
redshift.}
\label{spin_evo}
\end{figure}

We have seen in the previous section that BH mergers can contribute to the BH
mass assembly, which at least partially alleviates the problem of
highly-spinning BHs becoming supermassive. However, BH mergers also modify the
spin of the remnant BH and thus could additionally influence the BH mass
growth. In order to understand this issue in more depth we have computed the
final spin of the remnant BH for every BH merger, taking advantage of the
numerical relativity simulations of BH binary mergers \citep{Rezzolla2007a,
Rezzolla2007b, Rezzolla2008}. The expected distribution of BH spins at $z=6$
is illustrated in Figure~\ref{spin_evo}, where we have assumed that the BH
spins prior to a merger are parallel (aligned or anti-aligned to the orbital
angular momentum with equal probability) and initially characterized by a
large spin value of $0.9$.

The majority of BHs is spun-down after experiencing several mergers, and hence
the resulting BH spin distribution is broad at $z \sim 6$. There is also a
small fraction of BHs that are actually spun-up by mergers with other BHs,
lying to the right of the vertical dashed line. The peak in the BH spin
distribution at $a=0.9$ mostly comes from BH seeds (or BHs that have accreted
very little mass) that have not experienced any merger yet and thus have the
initial value of the spin. These BHs are about $15\%$ of the BH population
at $z=6$ in our simulated volume. The BHs that are spun-up amount to only
$5\%$ of the BH population. They result from mergers where both BH spins are
aligned with the orbital angular momentum and both are spinning rapidly if the
mass ratio is large (otherwise it is sufficient that the more massive BH has a
spin of $0.9$). Spin-ups can also occur if the less massive BH has a spin
orientation which is anti-aligned, but only in case of small mass ratios. For
other orientations, spin-downs are more likely, explaining why our BH spin
distribution ends up being biased towards spin values lower than the initial
spin \citep[see][for a detailed discussion]{Hughes2003}.

The above has interesting implications for the BH spin evolution of the
massive BHs at high redshift, as illustrated in the inset of
Figure~\ref{spin_evo}. Here the cross symbols denote the spin of the most
massive BH as a function of redshift. Initially the most massive BH is
characterized by a rather large spin value, but then after experiencing
several merger events it is spun-down. Thus, its radiative efficiency will get
lower as well, allowing it to grow rapidly. However, an important caveat in
this conclusion is that in our modelling we have neglected BH spin evolution
due to gas accretion itself, which can be expected to certainly play some
role. The question then is: Can extended episodes of gas accretion spin up the
BH significantly and thus ruin the beneficial effects of BH mergers?  Direct
numerical simulations of BH accretion flows are needed to give an answer to
this interesting question, something we plan to address in our future work.

\section{Evolution of the first bright quasars for  $z < 6$}\label{Beyond_z_6}

\subsection{Properties of the most massive BH and its host halo down to $z=2$}\label{Global_prop}

\begin{figure*}\centerline{
\psfig{file=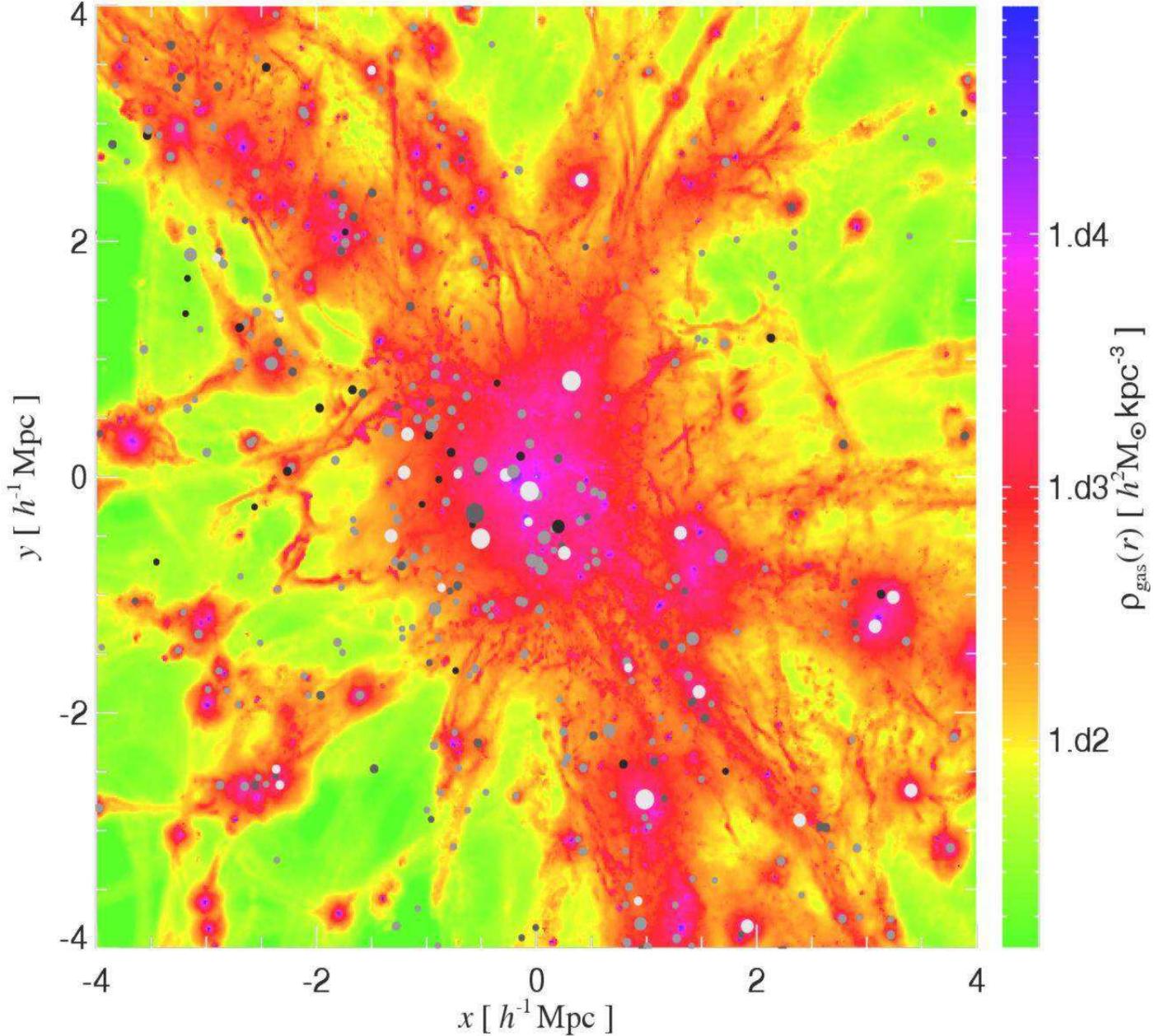,width=20.truecm,height=19.truecm}}
\caption{Projected mass-weighted gas density map at $z=2.24$ of the descendent
of the most massive halo at $z=6$. The halo has reached a mass of  $\sim
10^{14}  \,h^{-1} {\rm M}_\odot$, and is connected to a vast number of smaller
haloes by an intricate web of filaments. Numerous small scale features and
fine details of the ICM and the IGM are visible due to the very large spatial
and mass resolution achieved, with more than $10^7$ particles within the
virial radius of the halo. BHs within the projected slice are shown with
coloured dot symbols. The sizes of the dots encode information about the BH
mass, with smallest to biggest dots covering the following mass ranges:
$10^{6} - 10^{7}\,h^{-1} {\rm M}_\odot$, $10^{7} - 10^{8}\,h^{-1} {\rm
M}_\odot$, $10^{8} - 10^{9}\,h^{-1} {\rm M}_\odot$, $ > 10^{9} \,h^{-1} {\rm
M}_\odot$. The colours of the dots encode the bolometric luminosity of BHs,
from black over shades of grey to white, corresponding to the intervals: $<
10^{42} {\rm erg}\,{\rm s}^{-1}$, $10^{42} - 10^{43} {\rm erg}\,{\rm s}^{-1}$,
$10^{43} - 10^{44} {\rm erg}\,{\rm s}^{-1}$, and $> 10^{44} {\rm erg}\,{\rm
s}^{-1}$. Note that only two BHs have a bolometric luminosity greater than
$10^{45} {\rm erg}\,{\rm s}^{-1}$, one has a luminosity of $2.2 \times
10^{45}\, {\rm erg}\,{\rm s}^{-1}$, and the other $1.1 \times 10^{46}\, {\rm
erg}\,{\rm s}^{-1}$. The latter is the descendent of our $z=6$ quasar).}
\label{rhomap_z2}
\end{figure*}

Having a numerical model that can successfully reproduce the main properties
of high redshift quasars and their hosts it is interesting to study the
predictions of this model at lower redshifts. For this, we have used the
merger tree of the most massive $z=6$ Millennium halo  and extracted its
descendents both at redshifts $z=4$ and $z=2$. This gives us suitable target
haloes for resimulations that we start again from an initial redshift of
$z=127$. For these runs, we adopt the same BH model, and choose the zoom
factors of $5$ and $8$ (see Table~\ref{tab_simpar} for details). In
Figure~\ref{rhomap_z2}, we show a projected gas density map of the descendent
halo at $z=2.24$. The halo has already become fairly massive, with a virial
mass of $\sim 10^{14} \,h^{-1} {\rm M}_\odot$ (note that at $z=0$, the
descendant is the second most massive galaxy cluster in the whole volume of the
Millennium run). With a zoom factor of $5$, the halo at $z \sim 2$ has more
than $4 \times 10^7$ particles within the virial radius and is embedded within
a rich intergalactic environment, with many smaller haloes interconnected by a
complex filamentary structure. In Figure~\ref{rhomap_z2}, the BH particles
have been marked with coloured dot symbols. The size of each symbol encodes
the BH mass, while the colour-coding reflects the instantaneous bolometric
luminosities, where for simplicity we have assumed that all BHs are
radiatively efficient with a radiative efficiency of $0.1$ (see the figure
caption for more details). Besides the central BH, which has the largest mass
and the highest bolometric luminosity, a number of other BHs with a range of
masses have bolometric luminosities in excess of $10^{43} {\rm erg}\,{\rm
s}^{-1}$, indicating that their accretion rates are significant. We will come
back to this issue later on in this section.

\begin{figure} \psfig{file=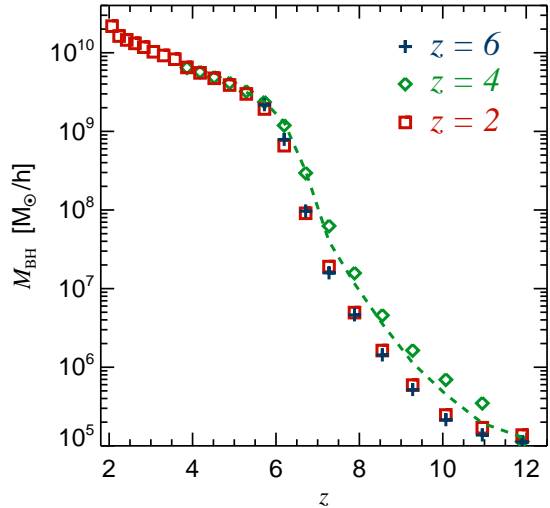,width=8.truecm,height=7.5truecm}
\caption{Mass of the most massive BH in the simulated volume as a function of
redshift. Blue crosses are for the run where the most massive halo of the
Millennium simulation has been selected at $z=6$ and then was resimulated at
$5^3$ times better mass resolution. The green diamonds are for the
resimulation of the main descendent of our $z=6$ halo selected at $z=4$, and
resimulated with the same zoom factor of $5$. The dashed green line is for the
same run but this time performed at even higher resolution adopting a zoom
factor of $8$. The red squares are for the case where the main descendent of
our  $z=6$ halo has been selected at $z=2$ and was resimulated with a zoom
factor of $5$. In all runs, the BHs were seeded in haloes with masses larger
than $10^{10} \,h^{-1} {\rm M}_\odot$. The BH mass growth is very similar in
all runs performed within the overlapping interval of redshifts and that it
does not depend on our choice for the resimulated region.}
\label{mbhz2}
\end{figure}

In order to verify that our resimulation technique gives similar results for
the BH mass assembly when we extract a descendent of the $z=6$ halo either at
$z=4$ or at $z=2$, in Figure~\ref{mbhz2} we compare the most massive BH in the
simulated volume as a function of redshift for both cases. Reassuringly, the
BH mass at a given epoch obtained from a resimulation done for the $z=6$ (blue
symbols) halo is very similar to the BH mass obtained from a resimulation done
for the $z=4$ (green diamonds) or $z=2$ (red squares) most massive
descendents. This confirms that these resimulations of the same object
extracted at different redshifts give us consistent estimates of the growth of
the most massive BH in overlapping redshifts intervals. Furthermore, the
green dashed line in Figure~\ref{mbhz2} shows the BH mass for the
resimulations where the host halo has been extracted at $z=4$ (same as green
diamonds), but the numerical resolution was higher, corresponding to a zoom
factor of $8$ (the other runs considered in the figure have a zoom factor of
$5$). Given that the mass of the most massive BH in this simulation starts to
be indistinguishable for $5 < z < 7$ from the analogous run performed with
lower resolution, we are confident that the numerical resolution of $5^3$
higher than that of the Millennium simulation is sufficient to resolve the
growth of the most massive BH reliably all the way to $z=2$. Moreover, we have
verified that the properties of the host halo (e.g.~its mass, gas and stellar
content, mean temperature, total SFR) are also recovered very well in all runs
performed.

\begin{table*} \bc
\begin{tabular}{p{0.8cm}@{\quad}p{1.cm}@{\quad}p{1.cm}@{\quad}p{1.cm}@{\quad}p{0.9cm}@{\quad}p{0.8cm}@{\qquad}p{0.8cm}@{\qquad}p{0.8cm}@{\qquad}p{0.8cm}@{\qquad}p{0.7cm}@{\qquad}p{1.cm}@{\qquad}p{0.8cm}@{\qquad}p{0.8cm}}
\hline \hline \multicolumn{13}{c}{MAIN HALO PROPERTIES AT $z=3.9$}\\ \hline Run &
$N_{\rm 200}$ & $N_{\rm 200,DM}$ & $N_{\rm 200,gas}$ & $R_{\rm 200}$ & $M_{\rm
200}$ & $M_{\rm 200,DM}$ & $M_{\rm 200,gas}$ & $M_{\rm 200,*}$ & $T_{\rm 200}$
& $SFR$ & $M_{\rm BH}$ & $M_{\rm Edd}$\\ & & & & [${\rm kpc}\,/h$] & [${\rm
M}_\odot\,/h$] & [${\rm M}_\odot\,/h$] & [${\rm M}_\odot\,/h$] & [${\rm
M}_\odot\,/h$] & [${\rm K}$] & [${\rm M}_\odot\, /{\rm yr} $] & [${\rm
M}_\odot\,/h$] & \\     \hline  with BHs & $6240576$ & $2772900$ &
$1413716$ & $719.6$ & $2.17 \times 10^{13}$ & $1.87 \times 10^{13}$ & $1.57
\times 10^{12}$ & $1.36 \times 10^{12}$ & $1.2 \times 10^7$ & $1105$ & $6.51
\times 10^9$ & 0.06 \\ \hline \hline
\end{tabular}
\begin{tabular}{p{0.8cm}@{\quad}p{1.cm}@{\quad}p{1.cm}@{\quad}p{1.cm}@{\quad}p{0.9cm}@{\quad}p{0.8cm}@{\qquad}p{0.8cm}@{\qquad}p{0.8cm}@{\qquad}p{0.8cm}@{\qquad}p{0.7cm}@{\qquad}p{1.cm}@{\qquad}p{0.8cm}@{\qquad}p{0.8cm}}
\hline \hline \multicolumn{13}{c}{MAIN HALO PROPERTIES AT $z=2.1$}\\ \hline Run &
$N_{\rm 200}$ & $N_{\rm 200,DM}$ & $N_{\rm 200,gas}$ & $R_{\rm 200}$ & $M_{\rm
200}$ & $M_{\rm 200,DM}$ & $M_{\rm 200,gas}$ & $M_{\rm 200,*}$ & $T_{\rm 200}$
& $SFR$ & $M_{\rm BH}$ & $M_{\rm Edd}$\\ & & & & [${\rm kpc}\,/h$] & [${\rm
M}_\odot\,/h$] & [${\rm M}_\odot\,/h$] & [${\rm M}_\odot\,/h$] & [${\rm
M}_\odot\,/h$] & [${\rm K}$] & [${\rm M}_\odot\, /{\rm yr} $] & [${\rm
M}_\odot\,/h$] & \\     \hline  with BHs & $46051219$ & $19457020$ &
$12593848$ & $1387.9$ & $1.55 \times 10^{14}$ & $1.31 \times 10^{14}$ & $1.47
\times 10^{13}$ & $9.27 \times 10^{12}$ & $3.7 \times 10^7$ & $2512$ & $2.18
\times 10^{10}$ & 0.01 \\ \hline \hline
\end{tabular}
\caption{The main properties of the descendents of the most massive Millennium
halo at $z=6$, resimulated to $z=3.9$ and $z=2.1$, respectively.  BH seeds
were introduced in haloes with masses larger than $10^{10} \,h^{-1} {\rm
M}_\odot$, and adopted numerical resolution corresponds to the zoom factor of
$5$. The second to fourth columns indicate the total, dark matter, and gas
particle numbers within the virial radius (fifth column). The total, dark
matter, gas, and stellar mass of the halo are listed in columns six to
nine. The mean mass-weighted temperature and the total star formation rate
within the virial radius are given in columns ten and eleven. The last two
columns give the mass of the central BH and the accretion rate in Eddington
units.
\label{tab_haloparz2}} \ec
\end{table*}

Similarly to the properties of the $z=6$ halo that are listed in
Table~\ref{tab_halopar}, the properties of its $z=4$ and $z=2$ descendent are
summarised in Table~\ref{tab_haloparz2}. From $z\sim 6$ to $z\sim 4$ and from
$z \sim 4$ to $z \sim 2$, the BH increases its mass by similar factors of
order $\sim 3.3$ (even though the first redshift interval corresponds to
$0.73\,{\rm Gyr}$, while the second one is much longer, amounting to
$1.6\,{\rm Gyr}$), while the BH's host halo (both in the DM and baryonic
components) experiences quite different relative growth, increasing its mass
by  factors of $\sim 4.5$ and $\sim 7$, respectively. At redshifts higher than
$6$, the BH increases its mass by a  huge factor of $\sim 2 \times 10^4$ from
$z \sim 12$ to $z \sim 6$, while the host halo mass grows from $\sim 1.5
\times 10^{10} \,h^{-1} {\rm M}_\odot$ to $\sim 4.9 \times 10^{12} \,h^{-1}
{\rm M}_\odot$. Therefore, the mass assembly of the host halo and its central
BH do not go ``hand-in-hand'', and if there is a tight relationship between BH
mass and host halo mass \citep{Ferrarese2002} it needs to evolve with redshift
over the interval considered here \citep[see also e.g.][]{Wyithe2006,
Shankar2007, Colberg2008, Croton2009}. This also holds true if instead of the
host halo dark matter mass we consider the total stellar mass within $R_{\rm
200}$. Finally, the fastest growth of the most massive BH at high redshifts,
can be viewed as a signature of so-called `downsizing', as we discuss next in
more detail.

\begin{figure*}\centerline{ \hbox{
\psfig{file=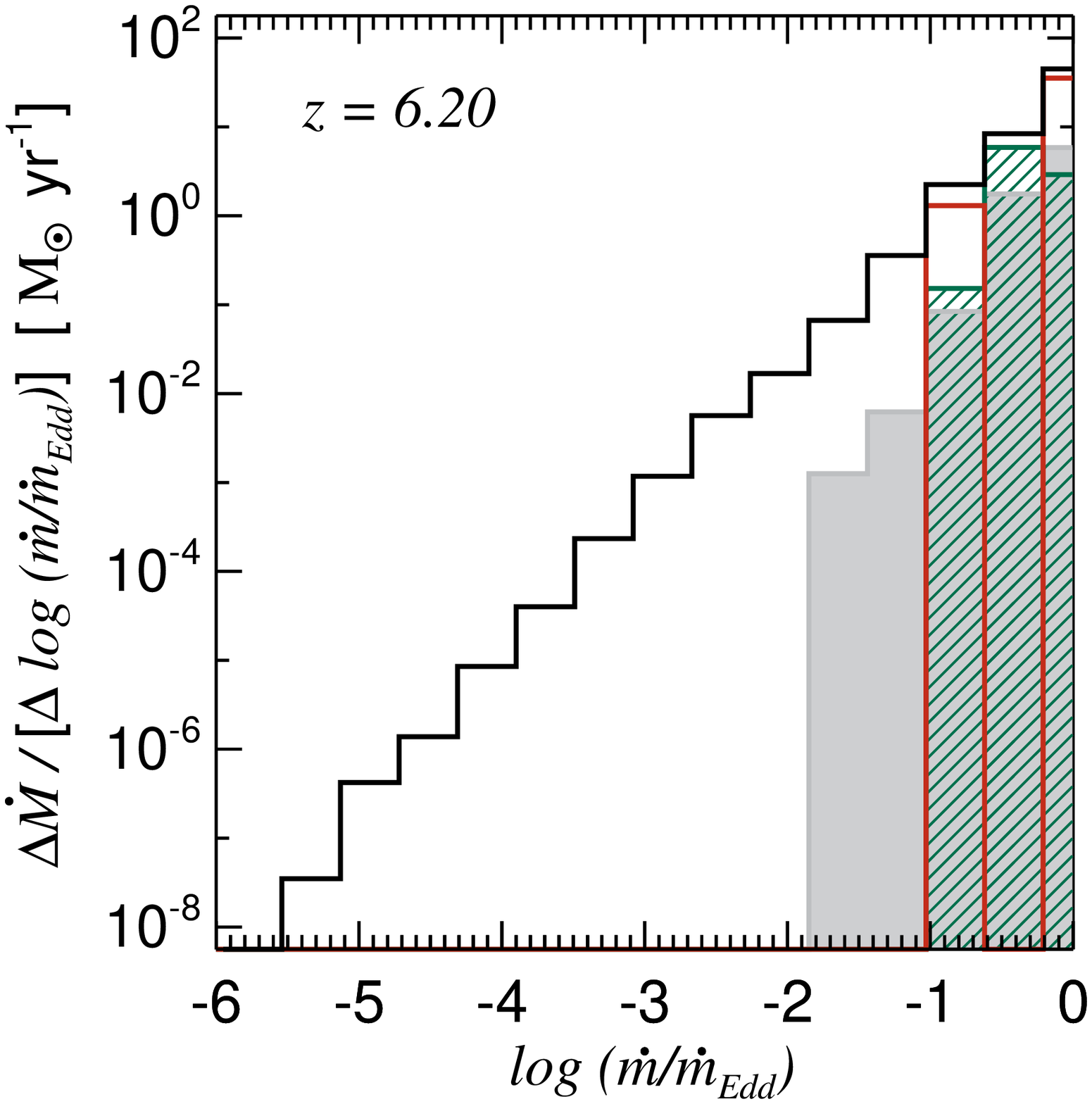,width=6.3truecm,height=6.2truecm}
\hspace{-0.5truecm}
\psfig{file=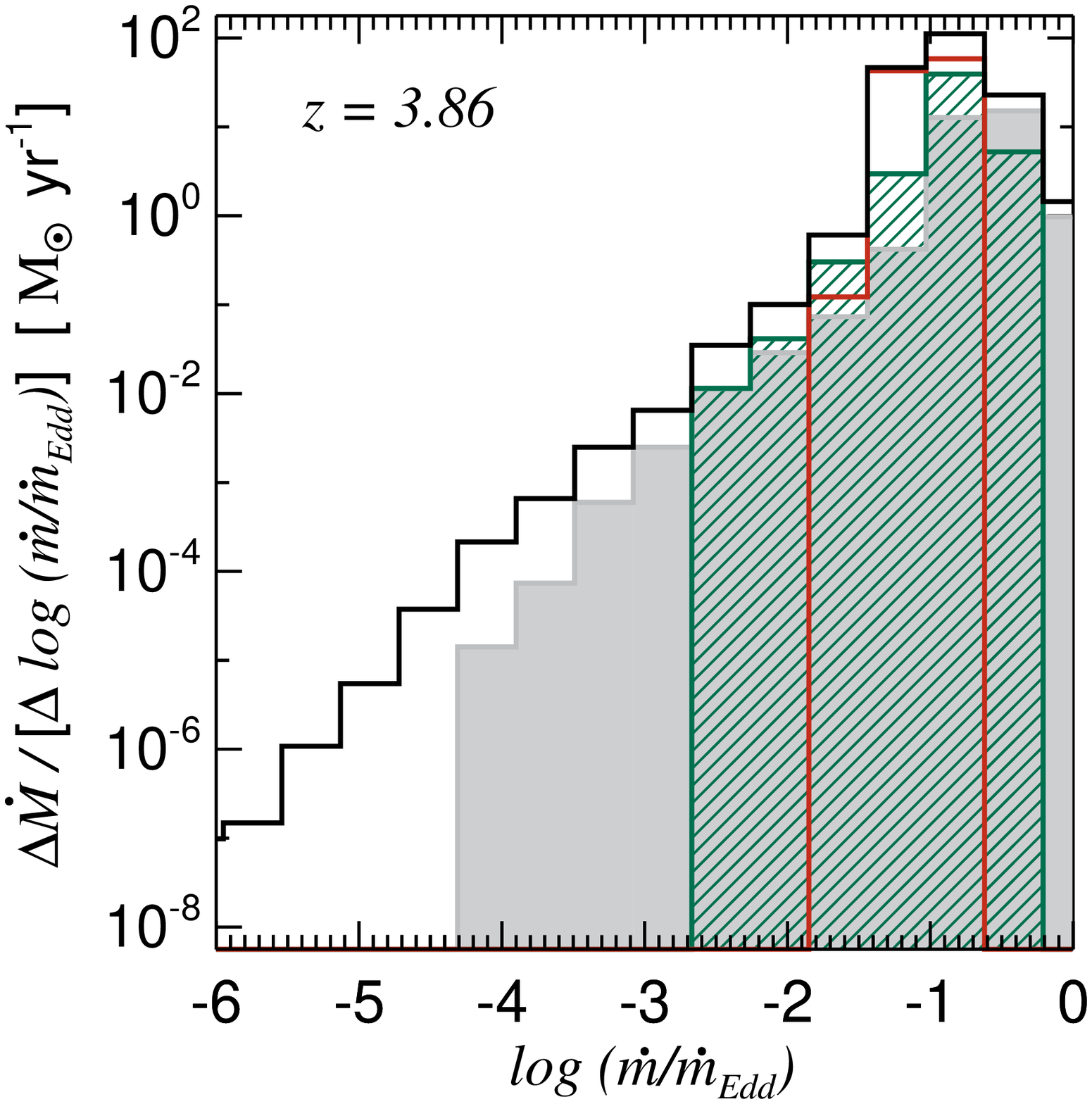,width=6.3truecm,height=6.2truecm}
\hspace{-0.5truecm}
\psfig{file=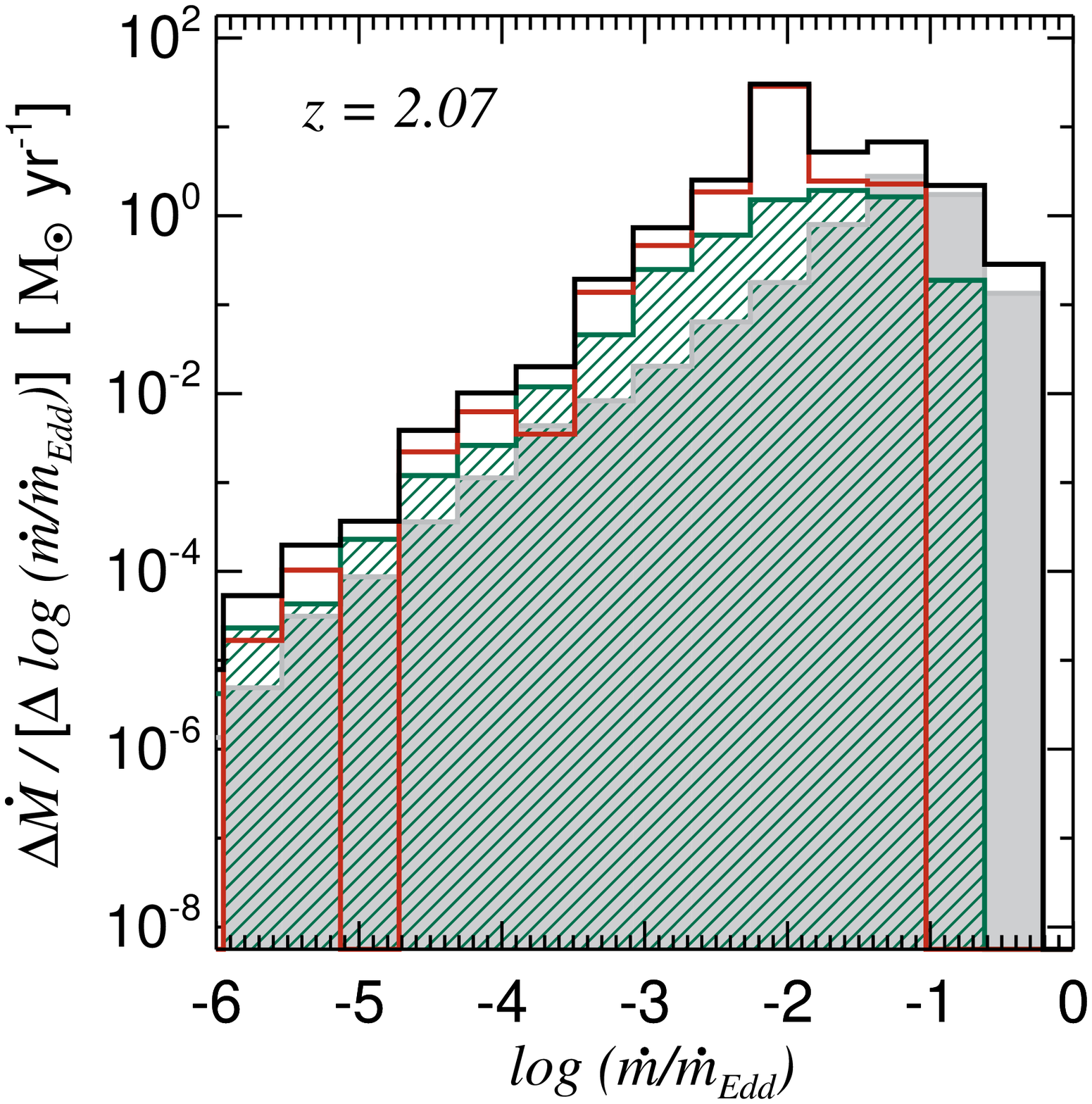,width=6.3truecm,height=6.2truecm}}}
\caption{The distribution of BH accretion rate counts (in ${\rm M}_\odot
\,{\rm yr}^{-1}$) in logarithmic intervals of BH accretion rate expressed in
Eddington units, at three different redshifts as indicated on the panels. The
black histogram is for all BHs; The other histograms have been computed
taking into account only BHs belonging to a certain mass range: grey
histograms are for $10^{6} \,h^{-1} {\rm M}_\odot < {\rm M}_{\rm BH} \le
10^{7} \,h^{-1} {\rm M}_\odot$, green hatched histograms are for $10^{7}
\,h^{-1} {\rm M}_\odot < M_{\rm BH} \le 10^{8} \,h^{-1} {\rm M}_\odot$, while
the red histograms are for $10^{8} \,h^{-1} {\rm M}_\odot < M_{\rm BH}$. The
most massive BHs accrete most efficiently at the highest redshift considered
and their BH accretion rate distribution evolves fastest with redshift with
respect to the two other mass bins. This can be viewed as a clear signature of
``downsizing''.}
\label{bhar_hist}
\end{figure*}

\subsection{The cosmological downsizing of the BH mass accretion}

In Figure~\ref{bhar_hist} we examine how efficiently our simulated BHs grow by
accretion at different epochs. To this end we show the distribution of BH
accretion rates (in ${\rm M}_\odot \, {\rm yr}^{-1}$) per logarithmic bin of
accretion rate in Eddington units at $z=6.2$ (left-hand panel), $z=3.86$
(central panel), and $z=2.07$ (right-hand panel). We both consider the whole
population of BHs (black histograms), as well as BHs belonging to specific
mass ranges: grey histograms denote the lowest, green hatched histograms the
intermediate, and red histograms the highest mass range considered. A number
of interesting features can be seen from these plots: (a) for all mass ranges
considered, the distribution of Eddington ratios broadens at lower redshifts,
implying that the fraction of BH accretion that occurs in the sub-Eddington
regime is more significant at later cosmic times; (b) the peak of the BH
accretion rate distribution shifts to the left most strongly for the highest
mass range considered, indicating that the high-mass end of the BH population
experiences faster evolution in the average accretion rate; (c) the same 
holds true for the intermediate mass range BHs compared with the lowest mass
range BHs, thus confirming that there is a systematic trend in the evolution
of the peak BH accretion rate with BH mass. All these features are clear
signatures of `downsizing' in the BH mass growth, i.e.~of a shift of the main
activity from high to low masses with cosmic time, running in a opposite sense
to the hierarchical build up of cosmic structures \citep[e.g.][]{Barger2005}.

\begin{figure}
\psfig{file=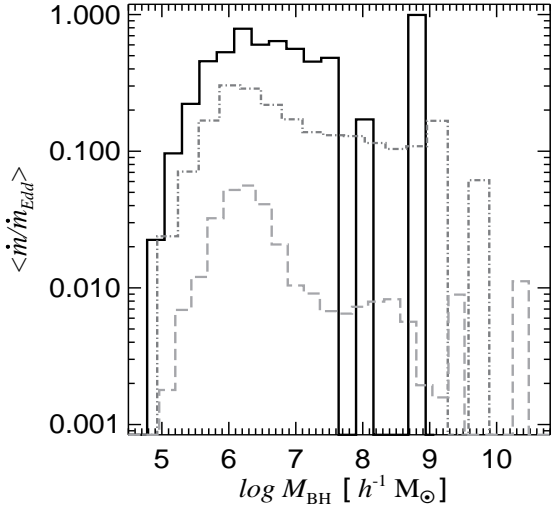,width=8.truecm,height=7.5truecm}
\caption{The distribution of the mean BH accretion rates in Eddington units
computed per logarithmic BH mass bin at three different redshifts: $z=6.2$
(black continuous line), $z=3.86$ (grey dot-dashed line), and $z=2.07$ (light
grey dashed line). For each mass bin the mean BH accretion rate measured in
Eddington units increases with increasing redshift, and the relative increase
is strongest for the most massive BHs.}
\label{bhar_histzall}
\end{figure}

To see this effect in perhaps even clearer way, we plot in
Figure~\ref{bhar_histzall} the distribution of mean Eddington ratios as a
function of BH mass at the same three epochs considered above. For a given BH
mass bin, the average Eddington ratio increases with increasing redshift, and
the relative increase is most pronounced for the high mass end of the BH mass
function.

\begin{figure}
\psfig{file=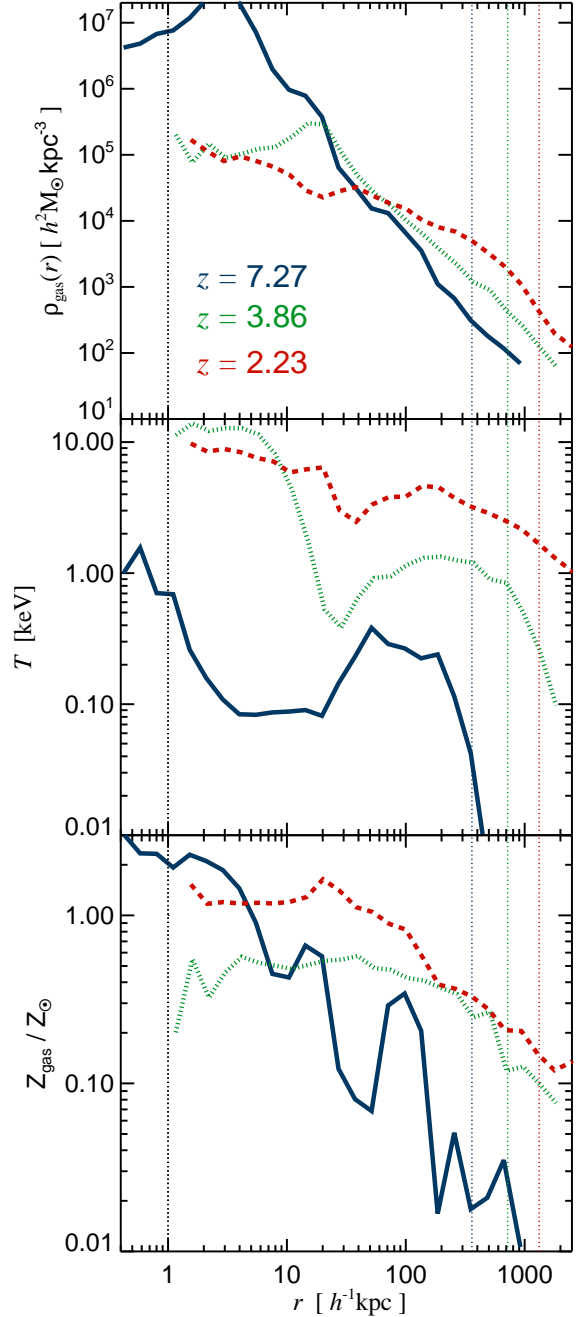,width=8.truecm,height=18truecm}
\caption{Radial profiles of gas density, mass-weighted temperature and
mass-weighted gas metallicity of the resimulated $z=2$ descendent of the most
massive massive halo in the Millennium simulation at $z=6$, at three different
redshifts: $z=7.27$  (blue continuous lines), $z=3.86$ (green dotted lines)
and $z=2.23$ (red dashed lines). The simulation has been performed with a zoom
factor of $5$, and the BHs were seeded in haloes above $10^{10} \,h^{-1} {\rm
M}_\odot$. The vertical dotted lines denote the adopted gravitational
softening length and the virial radius of the halo at these three redshifts
(in comoving units), using the same colour-coding.}
\label{profiles_evo}
\end{figure}

\subsection{Evolution of the intracluster gas properties}

We now analyze how the central SMBH affects the properties of the
intracluster gas down to low redshift.  In Figure~\ref{profiles_evo}, we show
the change of the gas density, mass-weighted temperature, and gas metallicity
(in Solar units) at $z=7.27$ (blue continuous lines), $z=3.86$ (green dotted
lines), and $z=2.23$ (red dashed lines). The vertical lines denote the
gravitational softening length and the virial radii at these three
epochs. While there is some cold dense gas in the innermost regions present at
$z=7.27$, due to the powerful AGN feedback the gas in the central regions is
heated at lower redshifts. Furthermore, while the gas metallicity is rather
patchy at the highest redshift considered, it becomes much more uniform
throughout the hot ICM at later times, which is also caused by AGN
feedback. Interestingly, the gas metallicity in the very centre at $z=7.27$
reaches super-Solar values, after which it declines at $z=3.86$ to then rise
again to super-Solar metallicities at $z=2.23$. This turns out to be
caused by BH feedback as well: highly enriched gas present at early times
is transported and mixed by AGN generated outflows with lower metallicity gas
further away. Of course, some of the highly enriched gas gets also swallowed
by the central BH itself. At later epochs, new generations of stars form in
the halo's central galaxy, and together with metal enriched gas of the
infalling satellites, causes the ICM gas metallicity to rise in the central
regions again.

\subsection{Constraints on the BH growth to $z=0$ and the presence of ultra-massive BHs}

As can be seen from Figure~\ref{mbhz2} and Table~\ref{tab_haloparz2}, the most
massive BH in our simulated volume at $z=2$ reaches a mass of $\sim 2 \times
10^{10} \,h^{-1} {\rm M}_\odot$. Several questions immediately arise from
this: What causes the BH to become so massive at $z=2$? What would be its
final mass at $z=0$, and would this be compatible with current observational
constraints on the high-mass end of the BH population?

\begin{figure*}\centerline{ \hbox{
\psfig{file=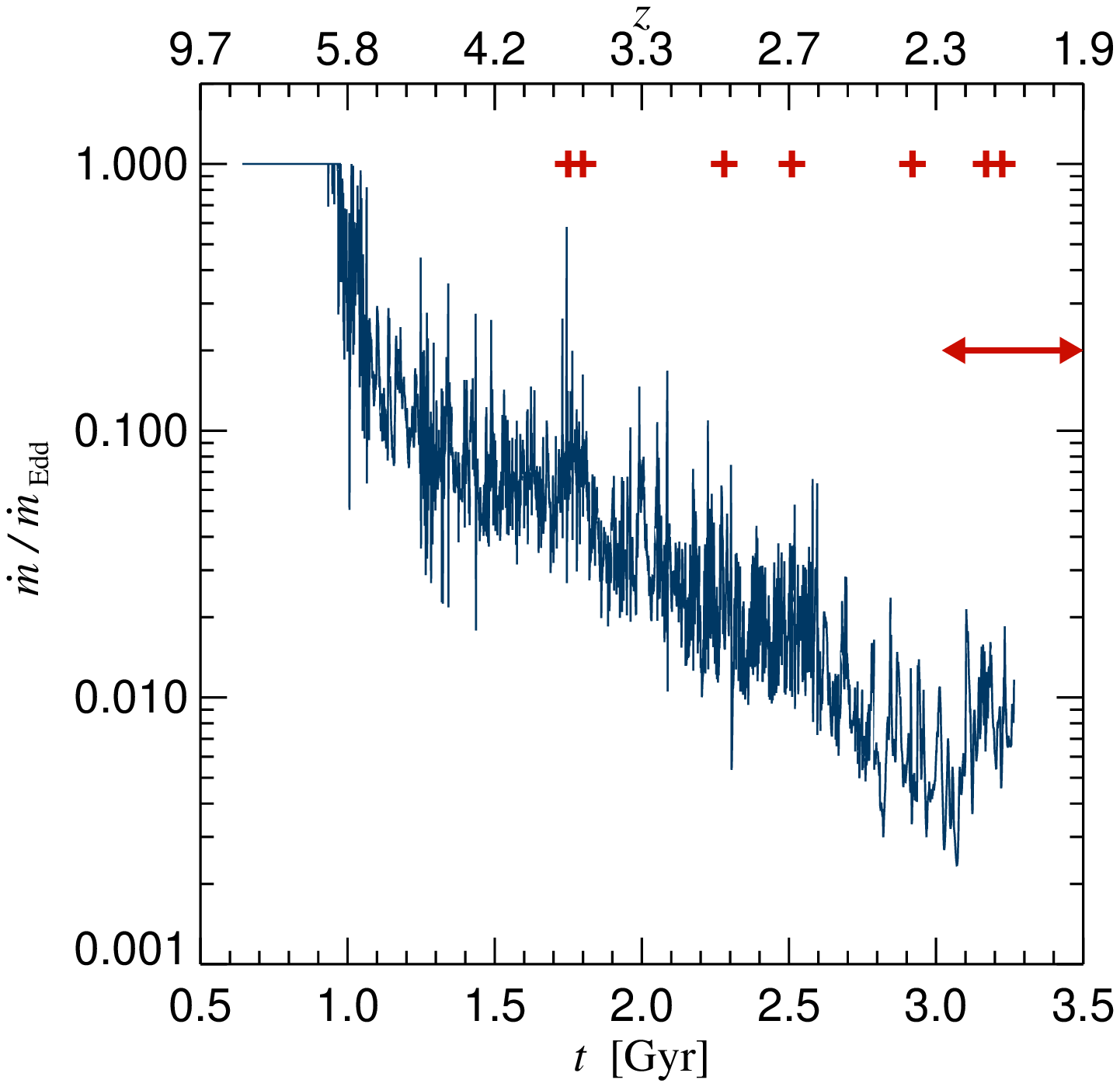,width=8.8truecm,height=8.8truecm}
\psfig{file=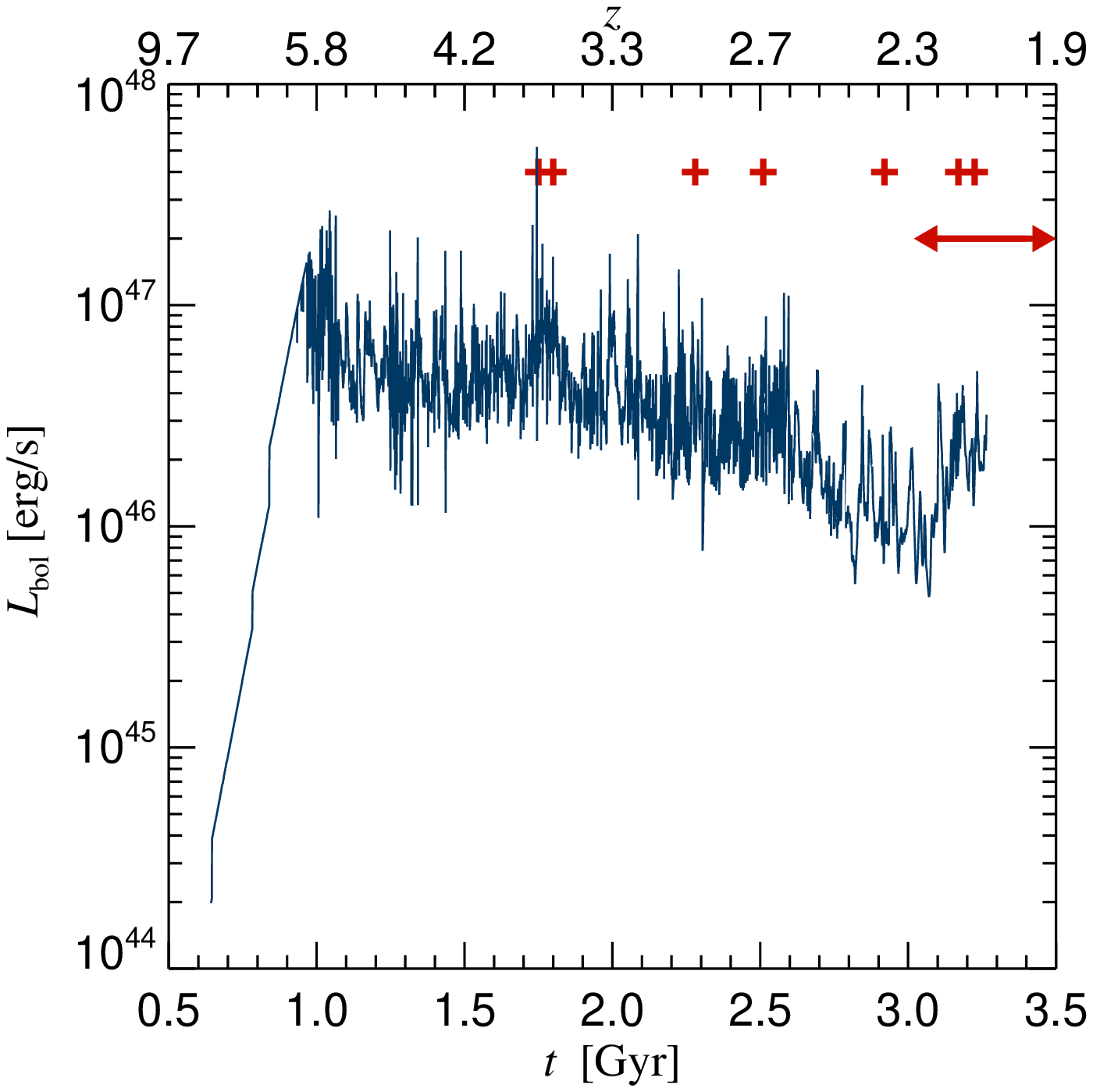,width=8.8truecm,height=8.8truecm}}}
\caption{The BH accretion rate in Eddington units (left-hand panel) and the
bolometric luminosity (right-hand panel) of the main progenitor of the most
massive BH at $z=2$. At $z < 6$ both quantities start decreasing. The red
crosses on both panels indicate mergers of the main progenitor with another BH
with mass larger than $10^{8} \,h^{-1} {\rm M}_\odot$. There is a close
correspondence between these mergers and the intervals of enhanced
accretion. The red arrow denotes the time interval of the most important major
merger the host halo undergoes (of mass ratio 0.8), from the redshift where
the two FoF haloes first `merge' (or rather touch), to the moment when the
surviving core of the secondary halo merges with the core of the primary.}
\label{medd_z2}
\end{figure*}

In order to obtain tentative answers to these questions, we have constructed
the merger tree of the whole BH population present in our simulated volume and
extracted the history of the main progenitor of the most massive BH at
$z=2$. For the BH's main progenitor, we have also accumulated information
about the instantaneous BH mass and accretion rate at each simulation
timestep.  This data is shown in Figure~\ref{medd_z2}, where we plot the
accretion rate in Eddington units (left-hand panel) and the bolometric
luminosity (right-hand panel), assuming a constant radiative efficiency of
$0.1$. After an initial essentially Eddington limited growth, the accretion
rate onto the BH's main progenitor is progressively declining for $z<6$, both
in relative and in absolute terms. The red crosses in both panels indicate
when the main progenitor undergoes a merger with a massive BH of mass larger
than $10^{8} \,h^{-1} {\rm M}_\odot$. The red arrow denotes the time-span
during which the host halo of the main BH progenitor experiences a major
merger with an almost equal mass halo (the merger mass ratio is $0.8$). This
shows that there is a relationship between the merging history of the host
halo and the activity of the central BH. Note that the peaks in bolometric
luminosity do not strictly correspond to the instants of BH mergers, given
that the accretion rate is also enhanced by gas which is funnelled towards the
central region during halo mergers. This often happens with a time offset
relative to the merger event of the BHs.

In Section~\ref{Global_prop}, we mentioned that the most massive BH increases
its mass by a similar factor of $\sim 3.3$ over the redshift ranges $z=6-4$
and $z=4-2$. Given that the BH accretion rate is declining with time, this
means that the contribution of BH mergers has to be more significant towards
lower redshift. In fact, analysis of the merger tree shows that from $z=6$ to
$z=4$, BH mergers contribute $\sim 20\%$ of the main progenitor mass at $z=4$,
while the BH mergers from $z=4$ to $z=2$ contribute up to $\sim 40\%$ of the
main progenitor mass at $z=2$.

We do not have a cosmological simulation at this high resolution all the way
to $z=0$, but we can, nevertheless, try to use our results to estimate a rough
upper limit for the mass the most massive BH would attain by the present
day. For this, we need to consider contributions both from gas that is locally
accreted and also from BHs that are likely to merge with the descendant of the
most massive $z=2$ BH. As for the contribution from accretion, we have fitted
a linear relation to the logarithm of the BH accretion rate as a function of
cosmic time, and then integrated this result from $t=3.3\,{\rm Gyr}$ to
$t=13.6\,{\rm Gyr}$. This extrapolation yields a total mass of about $\sim 2
\times 10^{9} \,h^{-1} {\rm M}_\odot$ that is expected to be accreted locally
by the BH. In order to estimate the contribution to the growth by BH mergers,
we have tracked back in time the merger history of the host halo from $z=0$,
based on the Millennium merger trees. We have then assumed that all haloes
that merge with our target halo contain a massive BH in their centre, and that
all these BHs will merge with our target BH. To estimate the cumulative mass
involved in these mergers, we have associated BH masses with dark matter
haloes based on the following relation: $M_{\rm BH}/10^8 {\rm M}_\odot = 0.1
(M_{\rm DM}/ 10^{12} {\rm M}_\odot)^{1.65}$ \citep{Ferrarese2002}. For the
dark matter masses $M_{\rm DM}$ of haloes we have taken the values measured
when the haloes were still in separate FoF groups for the last time. With this
procedure we conclude that the total mass of all BHs which can potentially
merge with the descendent of our $z=2$ BH is of the order of $\sim 9 \times
10^{9} \,h^{-1} {\rm M}_\odot$\footnote{After undergoing a major merger at
  $z=2$ the host cluster has a much more quiet
merging history with only one important merger happening at $z=0.9$, of
mass-ratio 0.2. This further justifies our assumption that the BH accretion
rate will continue to decrease given that not much fresh gas for accretion can
be supplied by merging satellites.}. Therefore, as an upper limit we are quite
confident that our BH will not increase its mass from $z=2$ to $z=0$ by more
than a factor of $2$.

Is a mass of $3-4 \times 10^{10} \,h^{-1} {\rm M}_\odot$ reasonable for the
present day descendent of one of the most massive BHs at $z=6$?
Observationally we have no constraints yet on what the duty cycle of the SDSS
quasars is, but if the BHs have grown substantially in an Eddington limited
accretion regime as in our simulations, the duty cycle has to be close to
unity. The local volume for which we can measure BH masses reasonably well is
then significantly smaller than that probed by either the SDSS survey or the
Millennium simulation. Estimating the mass of the most massive BH expected in
such a volume requires therefore a significant extrapolation of the local
scaling relations between BH mass and galaxy luminosity (or stellar velocity
dispersion) and an estimate of the likewise uncertain space density of rather
rare galaxies \citep[see also][]{Natarajan2009}.

Recently, an interesting issue regarding a potential mismatch between BH
masses estimated from the $M_{\rm BH}-\sigma$ versus the $M_{\rm BH}-L$
relation at the high-mass end has been raised by \cite{Lauer2007}. Their
findings imply that the masses of the most massive BHs are probably
underestimated if one uses the $M_{\rm BH}-\sigma$ relationship, and that at
the high mass end of the BH mass function there should be BHs with mass of
the order of $\sim 10^{10} {\rm M}_\odot$, which however should be very rare
and reside in the most massive galaxy clusters today. \cite{Lauer2007} give an
estimate of the cumulative number of BHs above a given mass, which for the
mass threshold of $10^{10} {\rm M}_\odot$ yields that there should be only a
handful of these ultra-massive BHs in the Millennium simulation volume. This
is exactly in line with our findings, which furthermore predict that the most
massive BHs today reside in the most massive and very rare haloes, and have
powered in their past the most luminous quasars.

However, we caution that one should take our estimate of the mass of the most
massive BHs at low redshifts with a grain of salt, in the sense that this is
likely to be somewhat at the high end within our model. This is due to several
reasons: first, the contribution of BH mergers to the BH mass assembly starts
to be more important at lower redshifts, especially for the BH sitting in the
centre of the brightest cluster galaxy \citep[see
also][]{Ruszkowski2009}. Given that our prescription for BH mergers is very
efficient (and additionally given that here we have not considered 3-body BH
effects and gravitational recoils), our estimate for the contribution of
merging BHs to the mass assembly is an upper limit. Additionally, we have not
considered that gravitational wave emission carries away part of the rest-mass
energy (albeit this effect is probably not very significant), and that
non-vanishing BH spins can potentially reduce the amount of material that a
hole can accrete (which however could build some tension with the growth of
BHs at $z > 6$). Nevertheless, we clearly find that the existence of rare
ultra-massive BHs of mass $\sim 1-2 \times 10^{10} {\rm M}_\odot$ in our local
Universe is a very interesting and plausible possibility. At the same time,
this highlights how special most luminous quasars are.

\section{Discussion and Conclusions}  \label{Discussion}

We have studied the growth of BHs at high redshifts with  state-of-the-art
numerical simulations in the full cosmological context that take into account
feedback effects from BH gas accretion. The main objective was to investigate
which are the most plausible formation scenarios of the SMBHs at $z=6$ whose
existence is inferred from observations of SDSS quasars. For this purpose, we
have selected the most massive dark matter halo from the Millennium simulation
at $z=6$, and resimulated it at a much higher mass and spatial resolution,
including gas physics, star formation and feedback processes, and a model for
BH seeding, growth and feedback. We have first thoroughly tested our model and
confirmed that numerical convergence in the SFR and BH accretion rate has been
reached. We have then explored the BH growth at high redshift, systematically
varying different physical assumptions in the simulations that could
potentially help or prevent early BH assembly.

We have found that it is possible in our default BH model to grow SMBHs by
$z=6$ which have a space density, mass, and bolometric luminosity consistent
with the findings from the SDSS quasar observations. This is a non-trivial
result given that the same model for BH growth and feedback also reproduces
the observed BH mass density, the observed relations between BH mass and host
galaxy at low redshifts, and alleviates overcooling in massive ellipticals
\citep[see][]{Springel2005b, Sijacki2007}. Regardless of the two different
seeding prescriptions we have considered (using halo mass thresholds of
$10^{9} \,h^{-1} {\rm M}_\odot$ or $10^{10} \,h^{-1} {\rm M}_\odot$) the SMBH
we form at $z=6$ gains most of its mass by gas accretion, undergoing extended
episodes of Eddington-limited accretion. We have checked whether starburst
powered galactic winds can expel a sufficient amount of the central gas supply
and thus stall the formation of the SMBH. This is not the case in our
simulations. If this mechanism operates in reality, then galactic winds would
need to have rather large mass-loading factors and velocities ($\sim
1000\,{\rm km\,s^{-1}}$), and would have to blow away a substantial fraction
of the innermost dense gas which is the reservoir for BH accretion.

We have extended our default model by incorporating prescriptions for
gravitational wave induced BH recoils, based on recent numerical relativity
simulations of merging BH binaries. We considered both non-spinning and
spinning BHs. We have found that a large number of BHs experience
gravitational recoils that kick them out of their host haloes ($20\%$-$40\%$,
depending on the model adopted). The vast majority of these BHs have low mass,
and the probability of expelling more massive BHs is not very high. In our
simulations a SMBH still forms even in the presence of strong gravitational
wave emission recoils, with very similar mass as in the case without recoils.

However, we would like to point out a potential caveat in our modelling. Since
we are primarily interested in how many BHs get kicked out from their hosts,
we have only imparted kick velocities to the remnant BHs if the estimated
recoil velocity is larger than the escape velocity of the host halo. Otherwise
we have neglected the recoil, assuming that the remnant BH stays in the host
halo and quickly sinks again to the centre of the halo. This treatment should
be adequate as our simulations do not have sufficient resolution to accurately
calculate the dynamical friction processes that would bring a displaced BH
back to the centre of a halo. Depending on how much material a displaced
remnant BH can carry with it and how long it takes the BH to return to the
centre, minor recoils could potentially also produce periods of stagnation in
the BH mass growth. This is a very interesting possibility that we have
neglected here but plan to investigate in forthcoming work.

The introduction of BH spin in our modelling also allows us to investigate the
impact of spin-dependent radiative efficiencies on the BH mass growth. A
rapidly spinning BH (with a constant spin of $a \ge 0.9$) seeded at $z\simeq
12-15$ will fail to become supermassive by $z=6$ unless it grows substantially
by mergers with other BHs, or can accrete at a super-Eddington rate. On the
other hand, if an initially rapidly spinning BH experiences several mergers
with other BHs (which are also initially rapidly spinning), and if we compute
at each merger the spin of the BH remnant according to the numerical
relativity findings, then it appears very likely that the BH will get
spun-down and end up with low spin. This suggests that highly spinning BHs
which grow in ``isolation'' (not experiencing many mergers) have the largest
difficulty in becoming massive enough by $z=6$. This conclusion remains valid
even when we start from fairly massive BH seeds. We caution however that
super-Eddington accretion or the effect of gas accretion on the spin evolution
(that we have not considered here) could in principle alleviate this problem.

Finally, we wanted to understand the kinds of objects the first SMBHs in our
simulations evolve into, how their properties change with time and whether
they are consistent with what we know about the demographics of low redshift
BHs. To this end, we have identified the descendents of our $z=6$ host halo at
$z=4$ and $z=2$, and resimulated them with the same mass and spatial
resolution as before. We have then tracked  the accretion rate of the
descendent of the most massive BH at $z=6$ all the way to $z=2$. The accretion
rate becomes systematically more sub-Eddington, and even in absolute terms the
accretion rate decreases. At $z \sim 6$ accretion of gas onto the most massive
BH changes from Eddington-limited to being limited by the thermal feedback of
the AGN implemented in the simulation. Why does this change occur at $z \sim
6$? This can be best understood as a selection effect. The mass of the host
haloes necessary to reproduce the space density of SDSS quasars assuming a
duty cycle of order of unity and the inferred mass of the BHs are close to
the threshold where the AGN feedback in our simulations becomes efficient in
shutting off the fuel supply.

Comparing the accretion rates in Eddington units for different mass ranges and
at different epochs, we found a clear signature of downsizing of BH
accretion rates. While we had seen a hint of this effect in
\cite{Sijacki2007}, we have here extended and confirmed this finding all the
way to very high redshifts, where very rare and massive BHs are accreting
efficiently and driving the trend. In this respect, the simulations performed
in this study turned out to be ideal to highlight the issue of the downsizing
of the BH mass growth by accretion.
   
At the high resolution achieved here it is too costly to use the same
simulation technique down to redshift $z=0$, but our results up to $z=2$ can
still be used to estimate the mass finally reached at $z=0$ by the descendant
of our most massive BH.  We found that the mass should reach a few times
$10^{10} \,h^{-1} {\rm M}_\odot$, taking into account both gas accretion and
further BH mergers. This may appear as a rather large mass (and in part is due
to the very efficient BH mergers in our model), but note that present
observations already suggest that there should be a handful of ultra-massive
BHs with masses of order of $1-2 \times 10^{10} \,h^{-1} {\rm M}_\odot$ in the
local Universe. These should be mostly dormant, low activity remnants of the
most luminous $z=6$ quasars which are situated today in the central galaxies
of rich galaxy clusters.

It is encouraging that the cosmological simulations examined here provide such
a successful simultaneous description of the build-up of SMBHs at high and low
redshift, despite the simplicity of our BH growth model that glosses over much
of the detailed small-scale physics of accretion flows. It will be very
interesting to refine this picture in future with ever more sophisticated
cosmological simulations. Among other improvements, this will make it
hopefully possible to account more consistently for the origin of the BH
seeds, and the spin evolution of BHs due to gas accretion.

\section*{Acknowledgements} 
DS acknowledges a Postdoctoral Fellowship from the UK Science and Technology
Funding Council (STFC). The simulations were performed on the Cambridge High
Performance Computing Cluster DARWIN in Cambridge (http://www.hpc.cam.ac.uk).

\bibliographystyle{mnras}

\bibliography{paper}

\end{document}